\documentclass[12pt]{article}

\usepackage{latexsym,epsfig,amssymb,amsmath,subfigure,color,pictex,fullpage,amsfonts}
\usepackage[margin=1in]{geometry}
\usepackage{cite} 
\usepackage{authblk}
\usepackage{hyperref}  % For clickable equation, citation and figure numbers in pdf
\usepackage{longtable} % For tables that span multiple pages

\definecolor{maroon}{rgb}{0.5, 0, 0}

\newcommand{\ie}{{\it i.e.}\ }
\newcommand{\eg}{{\it e.g.}\ }
\newcommand{\etal}{{\it et al.}\ }
\newcommand{\dd} \partial

\newcommand{\non} \nonumber

\newcommand{\be}{\begin{eqnarray}}
\newcommand{\ee}{\end{eqnarray}}
\newcommand{\bes}{\begin{eqnarray*}}
\newcommand{\ees}{\end{eqnarray*}}
\newcommand{\bet}{\begin{equation}} % command for standard equation environment used with \tag
\newcommand{\eet}{\end{equation}}   % as \tag does not work with eqnarray
\newcommand{\ds}{\displaystyle}

\newcommand{\nl}{n^{(l)}_{{\rm ETL}}}
\newcommand{\pl}{p^{(l)}}
\newcommand{\pr}{p^{(r)}_{{\rm HTL}}}
\newcommand{\nr}{n^{(r)}}

\newcommand{\jg}{j_{\rm gen}}
\newcommand{\nap}{n_{\textnormal{ap}}}      % Replacing nec
\newcommand{\nec}{n_{\textnormal{ec}}}      % True ectypal factor 
\newcommand{\nmec}{\bar{n}_{\textnormal{ec}}}% Measured ectypal factor 

\newcommand{\Voc}{V_{\textnormal{oc}}}
\newcommand{\voc}{V_{\textnormal{oc}}}
\newcommand{\Vbi}{V_{\textnormal{bi}}}
\newcommand{\vbi}{V_{\textnormal{bi}}}
\newcommand{\nel}{n_{\textnormal{el}}}
\newcommand{\neli}{n_{\textnormal{el}_i}}
\newcommand{\VDC}{V_{\textnormal{DC}}}

\newcommand{\QDC}{Q_{\textnormal{DC}}}

\newcommand{\jrec}{j_{\textnormal{rec}}}

\newcommand{\super}[1]{\textsuperscript{#1}}

\title{A new ideality factor for perovskite solar cells and an analytical theory for their impedance spectroscopy response}
\author[1]{Laurence J. Bennett}
\author[2]{Antonio J. Riquelme}
\author[3]{Nicola E. Courtier} 
\author[2]{Juan A. Anta}
\author[1,*]{Giles Richardson}
\affil[1]{Mathematical Sciences, University of Southampton, Southampton, SO17 1BJ, UK}
\affil[2]{Área de Química Física, Universidad Pablo de Olavide, E-41013, Seville, Spain}
\affil[3]{Department of Engineering Science, University of Oxford, OX1 3PJ, Oxford, UK}
\affil[*]{\tt G.Richardson@soton.ac.uk}

\begin{document}

\maketitle 

\begin{abstract}
Impedance spectroscopy (IS) is a relatively straightforward experimental technique that is commonly used to obtain information about the physical and chemical characteristics of photovoltaic devices. However, the non-standard physical behaviour of perovskite solar cells (PSC), which are heavily influenced by the motion of mobile ion vacancies, has hindered efforts to obtain a consistent theory to interpret PSC impedance data. This work rectifies this omission by deriving a simple analytic model of the impedance response of a PSC from the underlying drift-diffusion model of charge carrier dynamics and ion vacancy motion. Extremely good agreement is shown between the analytic model and the much more complex drift-diffusion model in regimes (including maximum power point) where the applied voltage is close to the open circuit voltage  $\Voc$. Both models show good qualitative agreement to experimental IS data in the literature and  predict many of the observed anomalous features found in impedance measurements on PSCs, such as `the giant low frequency capacitance` and `inductive arcs' in the Nyquist plots. Where the physical properties of the PSC are already known the analytic model can be used to predict the recombination current $\jrec$ and the high and low frequency resistances and capacitances of the cell, $R_{HF}$, $C_{HF}$, $R_{LF}$ and $C_{LF}$.
In scenarios where the physical properties of the cell are unknown the analytic model can also used to extract physical parameters from experimental PSC impedance data. 
{A novel physical parameter of particular significance to PSC physics is identified. This is termed the electronic ideality factor, $\nel$, and (as opposed to the standard ideality factor) can be used to deduce the dominant source of recombination in a PSC, independent of its ionic properties.}

{\bf keywords: perovskite solar cell, drift-diffusion model, impedance spectroscopy, ideality factor, {drift-diffusion} model.}

\end{abstract}

% -------------------------------------- Introduction -------------------------------------- % 
\section{Introduction}

% Intro to PSCs.
Over the last decade, metal halide perovskite solar cells (PSCs) have emerged as a promising new photovoltaic technology. Certified power conversion efficiency  of PSC cells  of 25.5\% and 29.5\%, for single junction and tandem configurations, respectively, are comparable to those of silicon \cite{green2020solar,jeong2020stable,nrelchart2020}. However, concerns over their long term stability, and an incomplete description of their fundamental physics, present barriers to widespread commercialisation \cite{berhe2016organometal,unger2020perovskite,kim2020advanced}.

% Intro to IS and why it's relevant/useful, particularly for PSCs and addressing the problems identified. 
Impedance spectroscopy (IS) is a characterisation method that probes the physical and chemical properties of (photo-)electrochemical devices under operation \cite{zhang2006eis,guillen2011electron}. Measurement of impedance involves applying a DC voltage $\VDC$ with an additional small sinusoidal voltage perturbation, of frequency $\omega$, to the system and measuring the current response. The applied voltage for an impedance measurement takes the form
\be
V(t) = \VDC + V_{p}\cos(\omega t).  \label{eq:V(t)_IS}
\ee
By design, the amplitude of the voltage input is small enough to ensure that the current response is linear. Although the resulting current response varies sinusoidally with the same frequency as the voltage input, its magnitude and phase lag depend on the frequency $\omega$ and the DC voltage. Typically the relationship between current response and the input voltage is represented in the form of a complex impedance $Z(\omega)$, such that
\be
Z(\omega) = |Z(\omega)|e^{-i \theta(\omega) },
\ee
where the magnitude, $|Z(\omega)|$ = $V_p/J_p(\omega)$ and the phase, $\theta(\omega)$ are obtained from the current response which takes the form
\be
J(t) = J_{DC} + J_{p}(\omega) \cos(\omega t + \theta).  \label{eq:J(t)_IS}
\ee
% This current response is obtained by applying the following voltage
% \be
% V(t) = \VDC + V_{p}\sin(\omega t). 
% \ee
The impedance is typically expressed in terms of the real and imaginary components as follows
\be
\textnormal{Re}(Z(\omega)) = R(\omega), \hspace{5mm} \textnormal{Im}(Z(\omega)) = X(\omega)
\ee
where the real component, $R$ is termed the resistance and the imaginary component, $X$ is termed the reactance.
% \ma{where the input voltage is denoted by
% \be
% V=\VDC+ V_p  \mbox{Re} (e^{i \omega t}), \label{vinput}
% \ee
% where $V_1  \mbox{Re} (e^{i \omega t})$ is the small sinusoidal perturbation to the DC voltage,
% the output current can be written in the form
% \be
% I=I_0(\VDC) +V_p \mbox{Re} \left( \frac{ e^{- i \omega t}}{Z(\omega,\VDC)} \right). \label{ioutput}
% \ee
% Here $I_0(\VDC)$ is the DC response while the final term is the small oscillatory current response. 
% Note that $ \mbox{Re}(\cdot)$ signifies the real part of the quantity in brackets.}
Impedance spectra are determined, at fixed DC voltage, by measuring the response to voltage perturbations over a  range of frequencies $\omega$. More information about the cell's properties can be obtained by collecting impedance spectra at different DC voltages $\VDC$, illumination intensities and temperature.

Impedance spectroscopy measurements are relatively simple  to perform and require equipment that is commonly found in most labs. With the correct interpretation, IS will offer significant insight into the  physics and characteristics of PSCs. In particular, it will allow ionic properties and interfacial potentials \cite{klotz2019detecting,riquelme2020identification} to be probed, both of which have been proposed as factors that strongly influence chemical reactivity and degradation \cite{aranda2019impedance,rizzo2019understanding,guerrero2016interfacial}, in addition to allowing the interrogation of the recombination pathways \cite{riquelme2020identification}. One of the main advantages of IS over other characterisation techniques is that it can be performed on complete devices under working conditions. Notably many other characterisation techniques are performed on half cells, or use destructive methods, to physically access regions of the cell, such as the transport layer interfaces \cite{domanski2016not,weber2018formation}.
% Monitor device stability 
The fact that IS is cheap and non-destructive allows it to be used to monitor device stability over time \cite{klotz2019detecting,hailegnaw2020impedance} which, when paired with the accurate spectral interpretation techniques outlined here, will allow makers of devices to better understand and diagnose the causes of degradation.  

\begin{figure}[htbp] 
\includegraphics[trim=0.0cm 0cm 0cm 0cm, clip, width=0.95\textwidth]{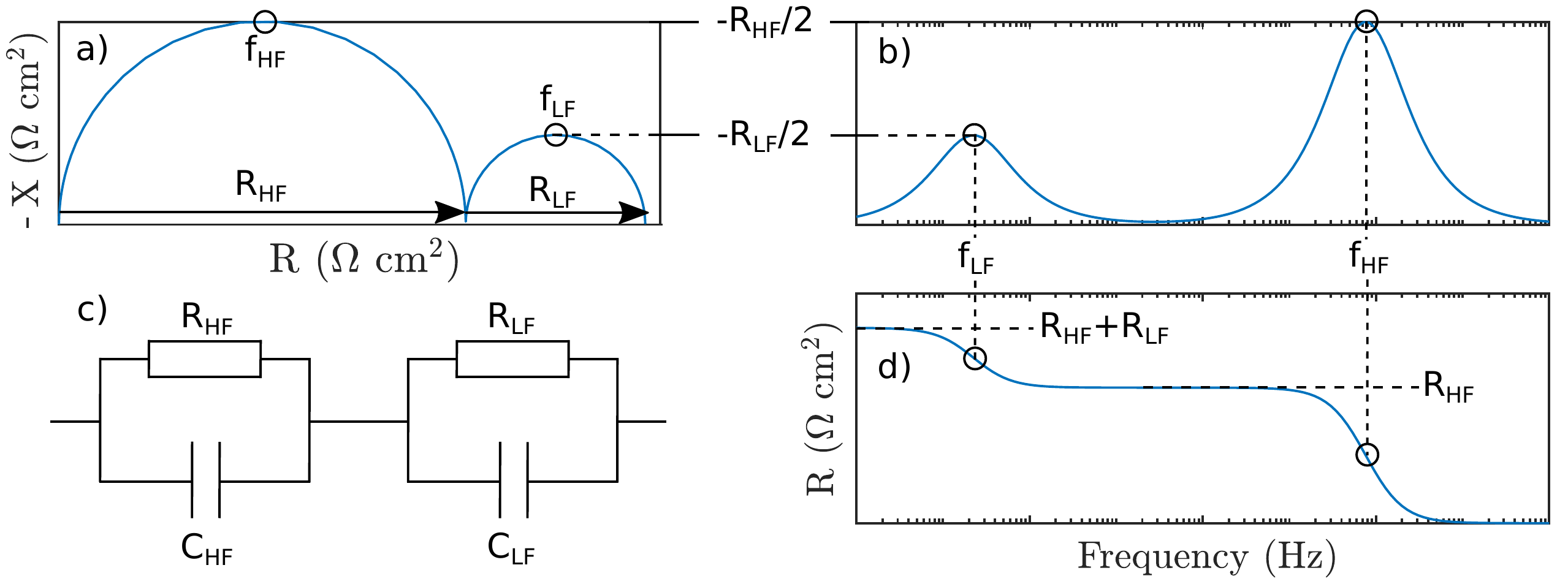}
\centering
\caption{a) Nyquist, b) and d) frequency plots with key impedance parameters labelled for a typical PSC impedance spectra exhibiting two features. c) An $RC$-$RC$ equivalent circuit used to extract resistances and capacitances. The labelled frequencies are related to the angular frequencies via $f = \omega/2\pi$. The low frequency semicircle can lie below the axis on the Nyquist plot, resulting in negative LF resistance and capacitance values \cite{alvarez2020negative}} \label{fig:Nyq_Bode_labelled}
\end{figure}

% What is typically/universally observed in IS of PSCs?
An impedance spectrum for a PSC typically exhibits two (or more) time constants that result in semicircular features on a Nyquist diagram (a plot of $-\mbox{Im}(Z(\omega))$ versus $\mbox{Re}(Z(\omega))\,$). A low frequency (LF) feature is observed at around 10$^{-2}$-10$^{1}$ Hz and a high frequency (HF) feature is seen around 10$^4$-10$^6$ Hz \cite{pockett2017microseconds,contreras2019impedance,wang2019kinetic}. Intermediate frequency features, such as loops or `bumps' are sometimes seen that correspond to additional time constants \cite{pockett2017microseconds,ravishankar2019perovskite,guerrero2016properties,garcia2019influence}. The presence and magnitude of these features varies significantly between cells of different compositions and under different experimental conditions. Figure \ref{fig:Nyq_Bode_labelled} shows a diagram of a typical Nyquist and frequency plot with two characteristic features. An ad hoc `$RC$-$RC$' equivalent circuit, is also shown, which is commonly employed to extract resistances and capacitances from PSC impedance spectra. In the presence of an additional series resistance $R_s$ (such as may be attributed to contact resistances) the semicircles in the Nyquist plane are shifted along the $R$-axis an amount $R_s$.

The impedance spectra of real PSCs show some diversity, however Contreras-Bernal \etal identify the following commonly observed characteristics \cite{contreras2019impedance} (which are based in part on fitting to the $RC$-$RC$ equivalent circuit shown in Figure \ref{fig:Nyq_Bode_labelled}c): 
\begin{itemize}
    \item[i)]  Two (or more) features, corresponding to time constants that are visible as peaks in frequency plots or semicircles/arcs in Nyquist plots.
	\item[ii)] The characteristic frequency of the LF feature is independent of DC voltage, whereas the characteristic frequency of the HF feature increases exponentially with DC voltage.  
    \item[iii)] Both the low and high frequency resistances ($R_{LF}$ and $R_{HF}$ respectively) decrease exponentially with DC voltage. Only the slope of $\ln(R_{HF})$ gives the same `ideality factor' as that obtained from $\Voc$ versus light intensity. 
    \item[iv)] The capacitance associated with the HF feature is independent of DC voltage, for low voltages, while the capacitance associated with the LF feature increases exponentially with open-circuit voltage. This is implicit given ii-iii) and that $\tau = RC$. 
\end{itemize}

% Why models are important - identify the landscape that our model fits into
The main difficulty in extracting useful physical information from IS measurements conducted on PSCs is that the physics of these cells is, to a large extent, determined by the motion of large numbers of slow-moving ion vacancies and so is markedly different to that of other photovoltaic devices. Intuition brought to the  perovskite field by experts in IS, who previously worked on other types of device, is therefore often useless because it is confounded by the unusual physics of these perovskite devices. Useful interpretation of IS results should therefore be based on a PSC model that captures the effects of ion motion. We note that IS results for PSCs are commonly fitted to equivalent circuit models in an attempt to extract useful information from the spectra, but that it is doubtful whether this approach can yield sensible conclusions {unless all the elements in the equivalent circuit can be related to real physical processes occurring in the device.}
In this context we note three recent works that have compared results from IS experiments to drift-diffusion simulations, which incorporate the effects of ion vacancy motion, namely \cite{moia2019ionic,jacobs2018two,riquelme2020identification}. {These works all require that the IS response is simulated computationally from a drift-diffusion model of the device that incorporates the motion of both ions and charge carriers. Such models are based on a large set of device parameters that must be specified before the numerical results can be compared to the experimental ones.} This leads to a costly fitting process in which the device parameters are repeatedly adjusted, and the simulations re-run, until the simulated IS results mimic the experimental ones. {Nevertheless, drift-diffusion models have the advantage that all parameters (diffusion coefficients, carrier concentrations, lifetimes, etc.) have a clear physical meaning.} The aim of this work is to avoid this cumbersome fitting process by seeking approximate (yet highly accurate) analytic solutions to the drift-diffusion model that can be directly compared to the experimental IS results, and thereby used to extract device parameters from experimental data without the need for large numbers of computational simulations.

{One of the peculiarities of PSCs is that, unlike other photovoltaic cells, recombination cannot be readily inferred from the ideality factor determined from standard measurements. These measurements, such as the Suns-$\Voc$ or the dark $J$-$V$ methods, lead to predictions of an `apparent' ideality factor that is voltage dependent, highly sensitive to ion concentration (see, for example, \cite{riquelme2020identification}) and has a value that spans a range less than 0.9 to greater than 5 \cite{tress2018interpretation,correa2017changes,kiermasch2018revisiting,pockett2015characterization}. Evidently, this is not concordant with interpretations formed from a naive analysis of recombination based on the Shockley diode equation. 
As pointed out in \cite{courtier2020interpreting} this is hardly surprising as the physics of no other common photovoltaic is as strongly influenced by ion vacancy as is that of PSCs.} {Instead, in  \cite{courtier2020interpreting}, it is suggested that the ideality factor obtained via the standard techniques should instead be interpreted as an `ectypal factor', and which we term the `apparent ideality factor'.}

The analytic approach we adopt here leads directly to an alternative to the apparent ideality factor, which we term the electronic ideality factor $\nel$, that is {independent of the parameters governing ion motion and close to being constant over a wide range of applied voltages. Moreover, it serves as an analogue to the ideality factor that is typically used for conventional photovoltaic devices and can be used, in a similar fashion, to identify the dominant recombination mechanism in a PSC.} 

{To summarise our approach, we employ a coupled electronic-ionic {drift-diffusion model} to describe the operational physics  of a PSC. A systematic, and highly accurate, approximation of this {drift-diffusion model} \cite{courtier2018systematic,courtier2019transport}, termed the surface polarisation model, is used to reduce the complexity of the drift-diffusion equations. This approximate model is able to accurately describe the evolution of the mobile ion vacancies and the electric potential and so can be used to predict the results of impedance measurements. However, in order to arrive at tractable analytic expressions (\ie a set of transcendental equations) for the impedance of the device, the electron and hole distributions are also assumed to be Boltzmann distributed throughout the device. This, as we shall demonstrate, is a suitable approximation to the charge carrier distributions where the applied voltage lies in vicinity of the open-circuit voltage and the maximum power point. This approach results in analytic expressions, in terms of the fundamental physical properties of the cell, for an appropriately defined (electronic) ideality factor and for the high- and low-frequency resistances and capacitances typically extracted from the impedance spectra of PSCs. These analytic expressions 
are validated against numerical solutions to the full {drift-diffusion model} with different recombination mechanisms, as shown in Figures \ref{fig:RlNyquist}, \ref{fig:R_combined_Nyquist_point1sun} and \ref{fig:RallNyquist_point1sun}. From a pragmatic perspective these results provide a practical tool with which to interpret, and to extract useful information from, experimental impedance spectra of perovskite solar cells and other related devices with substantial ionic motion.  }

%%%%%%%%%%%%%%%%%%%%%%%%%%%%%%%%%%%%%%%%%%%%%%%%%%%%%%%%%%%%%%%%%%%
%%%%%%%%%%%%%%%%%%%%%%%%%%%%%%%%%%%%%%%%%%%%%%%%%%%%%%%%%%%%%%%%%%%
%%%%%%%%%         MODEL
%%%%%%%%%%%%%%%%%%%%%%%%%%%%%%%%%%%%%%%%%%%%%%%%%%%%%%%%%%%%%%%%%%%
%%%%%%%%%%%%%%%%%%%%%%%%%%%%%%%%%%%%%%%%%%%%%%%%%%%%%%%%%%%%%%%%%%%

\section{Analytic model derivation} 

{In this section we outline the {\it surface polarization}, or  {\it ionic capacitance}, model of a PSC. This is a systematically derived approximation (see \cite{courtier2018systematic,courtier2019transport}) to the standard  coupled drift-diffusion model for charge carrier transport across the cell and the motion of a single ion species in the perovskite absorber \cite{calado2016,calado2020,courtier2019transport,moia2019ionic,neukom2017,neukom2019,richardson2016can}.\footnote{Note that approximations of this sort have wider validity and can be used to describe  situations in which, for example, there is more than one ion species.} Further, on assuming that the charge carriers (electrons and holes) are Boltzmann distributed, we can directly derive simple analytic formulae for the the cell's impedance response. We use these formulae to reproduce the impedance spectra (both Nyquist and frequency plots) and show that they compare extremely favourably to IS simulations conducted using the standard coupled drift-diffusion model for ion vacancy motion and charge transport (\eg \cite{courtier2019transport,neukom2017,neukom2019,richardson2016can}), {which in turn, reproduce experiment remarkably well}. Throughout this work we focus primarily on cells in which charge carrier recombination takes place via a single recombination mechanism and only briefly consider cells where more that one recombination mechanism is important, noting that the analysis of such cells is broadly similar but somewhat more complex.}

Unlike other photovoltaics the response of PSCs is determined not only by the motion of the charge carriers (electrons and holes) but also by that of mobile ion vacancies. This makes their physics more complex than that of other solar cells. As has been noted by Eames \etal \cite{eames2015ionic} ion vacancies occur at much higher densities than those of the charge carriers, and furthermore move much more slowly than the charge carriers. Since the perovskite is a crystal composed of at least three ionic species, for example MAPbI$_3$ is composed of methylammonium ions (CH$_3$NH$_3^+$), iodide ions (I$^-$) and lead ions (Pb$^{2+}$), multiple species of vacancy may play a role in the cell's response. In the case of MAPbI$_3$  it is known, both from experiment \cite{senocrate2017,yang2015} and from {\it ab-initio} molecular calculations \cite{eames2015ionic}, that there is significant motion of the relatively mobile iodide ion vacancies and, it is suspected, that the much less mobile methylammonium ion vacancies may also influence the physics over much longer timescales, of the order of several hours \cite{domanski2017}. The comparatively  large density of mobile charged ion vacancies in the perovskite crystal structure means that the internal electric field within the device is, to a very good approximation, determined almost solely by the ion vacancy distributions, and is almost completely unaffected by  those of the charge carriers \cite{courtier2018systematic}. In addition, mobile ion vacancies are believed to occur at sufficiently high densities to result in the formation of narrow Debye (or double) layers at the interfaces between the perovskite and the transport layers (see, for example, \cite{eames2015ionic,weber2018formation}). Indeed it seems that a Debye layer is a requirement for simulations to reproduce the behaviours characteristic of PSCs  \cite{courtier2019transport,neukom2017,neukom2019,richardson2016can,moia2019ionic,jacobs2018two}.

\begin{figure}[htbp] % trim=left botm right top  \fbox{ whole include graphics bit for box around it to crop}
\includegraphics[trim=0cm 0cm 0cm 0cm, clip, width=.99\textwidth]{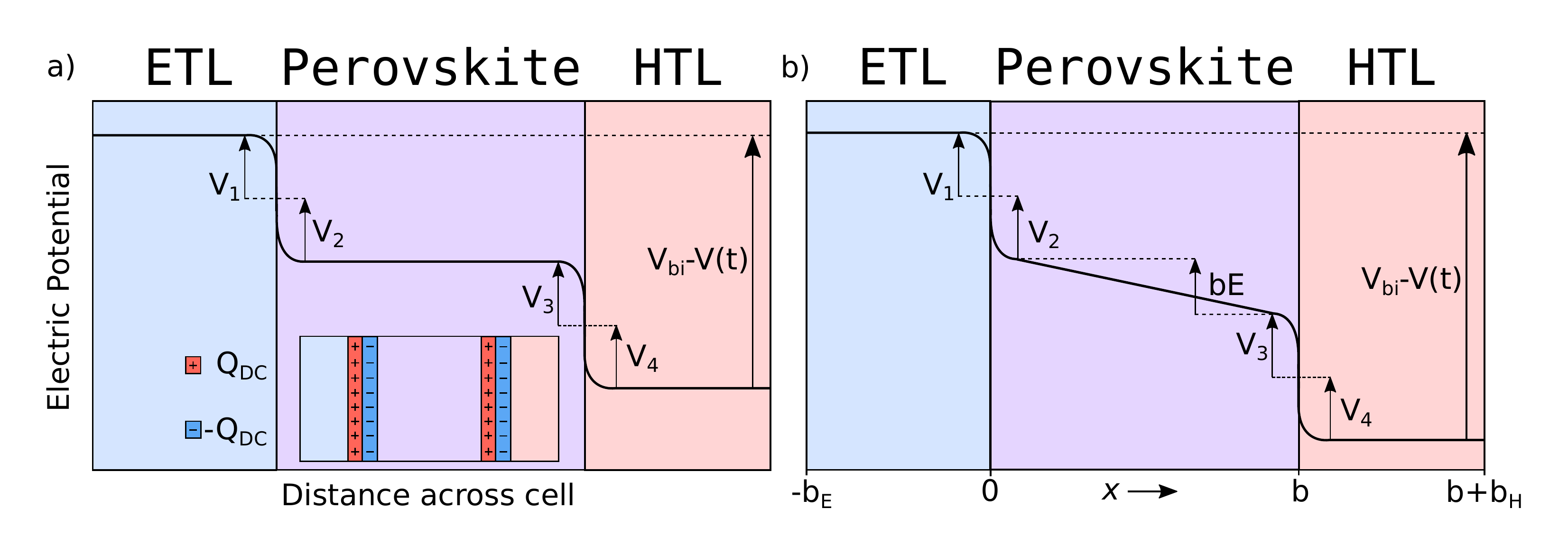}
\centering
\caption{a) Diagram illustrating the potential drops $V_1$-$V_4$ across a perovskite solar cell at steady-state. Inset shows the charge contained within the perovskite Debye layers at steady-state. b) Diagram illustrating the potential across a PSC with a non-zero bulk electric field after a rapid reduction to the applied voltage.}
\label{fig:V1-4}
\end{figure}

The total potential drop across the cell $\Vbi - V(t)$ is composed of a built-in potential $\Vbi$, arising from differences in the Fermi levels of the two transport layers adjacent to the perovskite, and the applied voltage $V(t)$, see figure \ref{fig:V1-4}. It is usually a reasonable assumption that the energy levels of the transport layers and their contacts line up, so that there is no significant potential drop at the outer edges of the ETL and the HTL, and furthermore, that there is some level of doping within the transport layers so that the internal electric field in the transport layers is minimal. Then, as shown in \cite{courtier2019transport}, the total potential drop across the cell occurs predominantly within the perovskite layer, in the form of a uniform electric field $E(t)$, and across the two Debye layers that form at the interfaces between the perovskite and the electron transport layer (ETL) and between the perovskite and the hole transport layer (HTL), as illustrated in Figure \ref{fig:V1-4}(b). The recombination losses within the cell, and therefore its behaviour, depends sensitively on how the total potential drop  $\Vbi - V(t)$ is distributed across the cell \cite{courtier2020interpreting}. This motivates the sub-division of the potential drop $\Vbi - V(t)$ into five  component drops $V_1$, $V_2$, $bE(t)$, $V_3$ and $V_4$ occurring across the different regions of the cell such that $\Vbi - V(t)=V_1(t)+V_2(t)+b E(t)+V_3(t)+V_4(t)$.   Here $V_1(t)$ is the potential drop across the portion of left-hand Debye layer lying within the ETL; $V_2(t)$ is the potential drop across the portion of left-hand Debye layer lying within the perovskite; $bE(t)$ is the potential drop occurring across the central region of the perovskite; $V_3(t)$ is the potential drop across the portion of right-hand Debye layer lying within the perovskite; and, $V_4(t)$ is the potential drop across the portion of right-hand Debye layer lying within the HTL. Typical electric potential profiles across the device, at steady-state and at non-steady-state are illustrated in Figure \ref{fig:V1-4}(a) and \ref{fig:V1-4}(b), respectively.

\subsection{{The surface polarisation model/ionic capacitance model}}
For both the steady-state and non steady-state configurations capacitance relations may be determined from the underlying {drift-diffusion model} of the device. This gives $V_{1-4}$ in terms of the  charge $Q$ (per unit area) contained in the Debye layers. The exact functional forms of $V_1(Q)$, $V_2(Q)$, $V_3(Q)$ and $V_4(Q)$ are contingent on the physics of the device. In the widely considered scenario where there is a single mobile positively charged ion species (typically a halide vacancy), confined to the perovskite layer, and where both electron and hole transport layers are moderately doped these capacitance relations have been derived in \cite{courtier2019transport} from the corresponding, and extensively used, version of the {drift-diffusion} model \cite{calado2016,calado2020,courtier2019transport,moia2019ionic,neukom2017,neukom2019,richardson2016can}. Since, in \S \ref{sec:results}, we will compare our approximate expressions for the high and low frequency resistances and capacitances against those obtained from impedance spectra generated by this version of the {drift-diffusion model}, we restate these capacitance relations below
\be
\begin{array}{lll}
\ds V_1(Q) = -\mathcal{V}(-\Omega_E Q ), & \ds V_2(Q) = -\mathcal{V}(-Q), & \ds V_3(Q) = \mathcal{V}(Q),  \\*[4mm]
\ds V_4(Q) = -\mathcal{V}(-\Omega_HQ ), &\ds \Omega_E = \sqrt{\frac{\varepsilon_pN_0}{\varepsilon_Ed_E}}, &\ds \Omega_H = \sqrt{\frac{\varepsilon_pN_0}{\varepsilon_Hd_H}},
\label{eq:VsandOmegas}
\end{array}
\ee
where $\mathcal{V}(Q)$ is the inverse of the function
\begin{equation}
Q(\mathcal{V}) = \sqrt{qN_0\varepsilon_p V_T}  ~\mbox{sign}(\mathcal{V})\sqrt{2}\left(e^{\mathcal{V}/V_T}-\mathcal{V}/V_T-1\right)^{\frac{1}{2}}, \label{eq:charge}
\end{equation}
and the material parameters that appear in these equations are defined in Table \ref{tab:params}.
Although the version of the capacitance model described above in \eqref{eq:VsandOmegas}-\eqref{eq:charge} applies only to the specific drift-diffusion model discussed above (and stated in full in \cite{courtier2019transport}), it will often be possible to derive analogous capacitance relations for other versions of the drift-diffusion model, which describe a modified physical picture of a PSC (for example, one in which both anion and cation vacancies are mobile, or one in which the charge carriers in the transport layers obey degenerate statistics).
 
Under non-steady-state conditions, the Debye layer charge $Q(t)$ (per unit area) evolves in response to the motion of ion vacancies driven across the perovskite by the internal electric field and, in the case of a single positively charged ion vacancy species \cite{richardson2016can}, evolves according to the ordinary differential equation (ODE)
\be
\frac{d Q}{d t}= \frac{q D_+ N_0}{V_T} E(t), \label{Qeq}
\ee
in which $D_+$ is the vacancy diffusivity, $q$ is the charge on a vacancy and $V_T$ is the thermal voltage (equal to $k_BT/q$ $\approx$ 26 mV). Since the ion motion is relatively slow, the charge within the Debye layers lags behind the changes in the applied voltage leading to a non-negligible internal internal electric field that is determined by the relation
\be
E(t)=\frac{1}{b} \left( \Vbi - V(t) - \left\{V_1(Q(t))+V_2(Q(t))+V_3(Q(t))+V_4(Q(t)) \right\} \right) \label{eq:E(t)}
\ee

\subsection{Evolution of internal potential drops during an IS experiment}
In the case of the IS measurements that we are interested in, small perturbations to the steady-state result from the application of a small oscillating potential on to a steady fixed potential difference across the cell, so that the applied potential takes the form
\begin{equation}
  V(t) = \VDC + V_{p}\cos(\omega t).  \tag{\ref{eq:V(t)_IS} reprinted}  
\end{equation}
in which the magnitude of the perturbation $V_p$ is small relative to (or at worst comparable to) the thermal voltage $V_T$.

In order to investigate the impedance response of the cell we first consider its steady-state behaviour subjected to the constant applied voltage $\VDC$. In this configuration, we denote the excess charge (per unit area) contained within the left- and right-hand Debye layers, lying within the perovskite, as $-\QDC$ and $+\QDC$, respectively (see Figure \ref{fig:V1-4}a). At steady-state the bulk electric field within the central portion of the perovskite is zero such that the charge density within the Debye layers is found, on appealing to \eqref{eq:E(t)}, to satisfy the transcendental equation
\be
V_1(\QDC)+V_2(\QDC)+V_3(\QDC)+V_4(\QDC)=\Vbi - \VDC, \label{eq:pot_drops}
\ee
from which it is possible to determine $\QDC$ from the applied voltage $\VDC$.
{Henceforth, non-steady-state conditions are denoted as functions of time (e.g. $Q(t)$), while steady-state conditions are expressed as functions of $\QDC$.} The response of the internal electric field of the cell to the impedance input voltage \eqref{eq:V(t)_IS} can be readily calculated from the simplified capacitance model simply by substituting \eqref{eq:V(t)_IS} into \eqref{Qeq}-\eqref{eq:E(t)}, which on solution of the first order ODE leads to a complete time course for the quantities $V_1(Q(t))$-$V_4(Q(t))$ and $E(t)$. {For the capacitance model described here it is not possible to determine an analytic solution for the steady-state ionic surface charge density $\QDC$ in terms of $\VDC$. Therefore, a numerical root finding algorithm is used to determine $\QDC$ for a given applied potential $\VDC$. This is the only part of the solution that requires numerical evaluation, the rest is analytic.} Once these quantities have been obtained the charge carrier concentrations, and hence also the recombination losses and the current, can be found by solving a {\it linear} {drift-diffusion model} for the charge carriers, as described in \S B of the ESI to \cite{courtier2019transport}. However, in order to arrive at a tractable analytic result, from which the influence of the parameters within the model on the impedance response of the cell can be readily inferred, we make the additional assumption, which is appropriate when the cell is held at an applied voltage in the vicinity of $\Voc$, that the charge carriers (holes and electrons) lie in approximate quasi-equilibrium, \ie that they are Boltzmann distributed. {This assumption can be justified provided that the carrier densities are large enough such that the electron and hole currents do not lead to significant band bending. The approximation works well close to $\Voc$, where there is significant build up of carrier concentrations, but can break down as the voltage is reduced towards short-circuit and the carrier densities are much reduced. We confirm the validity of this assumption by comparing  the carrier densities obtained with this approximation against the full results from the numerical drift-diffusion model, see figure \ref{fig:npcomptoIM}.}

\paragraph{Summary of our approach.} This is based upon the observation that the internal electric potential $\phi(x,t)$ within the perovskite (as determined by the drift-diffusion model) is determined almost solely by the evolving distribution of high concentrations of ion vacancies \cite{richardson2016can}, in comparison to which those of the charge carriers in the perovskite layer are negligible. This allows us to approximate: 1) the internal electric potential within the device by a model of the form \eqref{Qeq}-\eqref{eq:E(t)}, an approach that has been validated in the case of a single ion vacancy species in  \cite{courtier2019transport,courtier2018systematic,richardson2016can}; 2) the time-dependent perturbation to the applied voltage during an impedance spectroscopy measurement eqn.\eqref{eq:V(t)_IS} is small in comparison to the thermal voltage $V_T$, so that the model can be linearised about the steady-state; and 3) that the charge carriers are in approximate quasi-equilibrium throughout the device, so that they are Boltzmann distributed (an approximation that has been previously adopted in \cite{courtier2020interpreting}).

\begin{table}[ht!]
\small
\centering
\begin{tabular}{|l l l l|} 
\hline
\textbf{Symbol} & \textbf{Description} & \textbf{Value} & \textbf{Source} \\ [0.5ex] 
\hline 
\multicolumn{4}{|l|}{\rule{0pt}{1.0\normalbaselineskip}\textbf{Constants}}  \\
%\rule{0pt}{1.0\normalbaselineskip}$q$ & charge of proton  & $1.60\times10^{-19}$ C & NA \\
$q$             & Elementary charge             & $1.60\times10^{-19}$ C                         &  \\
$\varepsilon_0$ & Permittivity of free space    & $8.85\times10^{-12}$ F m\super{-1}            &  \\ [0.5ex]
\multicolumn{4}{|l|}{\textbf{MAPbI$_3$ properties}}  \\
$b$             & Width                         & 300 nm                                        & \cite{kim2013mechanism,bag2015kinetics} \\ %\cite{lee2012efficient, pockett2015characterization}\\
$\alpha$        & Absorption coefficient        & $1.3\times10^7$ m\super{-1}                   & \cite{loper2015complex}\\
$\varepsilon_p$ & Permittivity                  & $24.1\varepsilon_0$                           & \cite{brivio2014relativistic}\\
$E_C$           & Conduction band level         & -3.8 eV                                       & \cite{schulz2014interface}\\
$E_V$           & Valence band level            & -5.4 eV                                       & \cite{schulz2014interface}\\
$g_c$           & Conduction band DoS           & $8.1\times10^{24}$ m\super{-3}                & \cite{brivio2014relativistic}\\
$g_v$           & Valence band DoS              & $5.8\times10^{24}$ m\super{-3}                & \cite{brivio2014relativistic}\\
$D_n$           & Electron diffusion coefficient& $1.7\times10^{-4}$ m\super{2}s\super{-1}      & \cite{stoumpos2013semiconducting}\\
$D_p$           & Hole diffusion coefficient    & $1.7\times10^{-4}$ m\super{2}s\super{-1}      & \cite{stoumpos2013semiconducting}\\
$D_+$           & Ionic vacancy diffusion coefficient& $1 \times 10^{-16}$ m\super{2}s\super{-1}& \cite{eames2015ionic, richardson2016can}\\
$N_0$           & Ionic vacancy density         & $1.6\times10^{25}$ m\super{-3}                &\cite{walsh2015self}\\
% $\tau_n$        & Electron pseudo-lifetime      & $3\times10^{-12}$ s                         &\cite{richardson2016can}\\
% $\tau_p$        & Hole pseudo-lifetime          & $9\times10^{-10}$ s                         & \cite{richardson2016can}\\ [0.5ex]
\multicolumn{4}{|l|}{\textbf{ETL properties (compact-TiO$_2$)}}  \\
$b_E$           & Width                         & 100 nm                                        & \\
$d_E$           & Effective doping density      & $5\times10^{24}$ m\super{-3}                  & \\
$\varepsilon_E$ & Permittivity                  & $10\varepsilon_0$                             & \\
$D_E$           & Electron diffusion coefficient& $1\times10^{-5}$ m\super{2}s\super{-1}        & \\
$g_{cE}$        & Effective conduction band DoS & $1\times10^{26}$ m\super{-3}                  & \\ [0.5ex]  % \cite{schulz2014interface}
$E_{cE}$        & Conduction band minimum       & -4.1 eV                                       & \cite{fujisawa2017comparative}\\
\multicolumn{4}{|l|}{\textbf{HTL properties (Spiro)}}  \\
$b_H$           & Width                         & 200 nm                                        & \\
$d_H$           & Effective doping density      & $5\times10^{24}$ m\super{-3}                  & \\
$\varepsilon_E$ & Permittivity                  & $3\varepsilon_0$                              & \\
$D_H$           & Hole diffusion coefficient    & $1\times10^{-6}$ m\super{2}s\super{-1}        & \\
$g_{vH}$        & Effective valence band DoS    & $1\times10^{26}$ m\super{-3}                  & \\ 
$E_{vH}$        & Valence band maximum          & -5.1 eV                                       & \cite{schulz2014interface} \\
\multicolumn{4}{|l|}{\textbf{Other}}  \\
$T$             & Temperature                   & 298 K                                         & \\ 
$V_T$           & Thermal voltage (at 298 K)    & 25.7 mV                                       & \\
$F_{ph}$        & Incident photon flux          & $1.4\times10^{21}$ m\super{-2} s\super{-1}    & \\ %\cite{loper2015complex,richardson2016can}
$\Vbi$        & Built-in voltage              & 0.85 V                                        & \\ [1ex]
\hline
\end{tabular}
\caption{ Parameter definitions and their values used in this work.}
\label{tab:params}
\end{table}

\subsection{The current response during an IS experiment} 
The total current density flowing across the cell $J(t)$ is comprised of three components: the current generated by the incident radiation $\jg$, the current loss due to charge carrier recombination $-\jrec$ and, at high frequencies, the displacement current $j_d$.  It can thus be represented by the formula
\be
J(t) = \jg - \jrec + j_d. \label{eq:J(t)}
\ee
Here $\jg$ the current density generated by the incident radiation depends upon the thickness of the perovskite absorber layer $b$ through the Beer-Lambert law 
\be
\jg = qF_{ph}\left(1-e^{-\alpha b}\right). \label{eq:js}
\ee
where $F_{ph}$ is the intensity of the incident radiation and $\alpha$ is the absorption coefficient and $b$ (more details of the cell parameters are given in Table \ref{tab:params}).
%where $F_{ph}$ is the incident photon flux, $\alpha$ is the absorption coefficient and $b$ the width of the perovskite layer.
The displacement current is given by
\be
j_d(t) = \varepsilon_p \frac{d E}{d t}, \label{eq:j_d} 
\ee
where {$\varepsilon_p$} is the perovskite permittivity. The approximate surface polarisation/ionic capacitance model (as described above) can be used to compute $J(t)$ (the current flow) when an oscillatory voltage, of the form \eqref{eq:V(t)_IS}, is applied across the device. This is accomplished by first solving equations \eqref{Qeq}-\eqref{eq:E(t)} for the evolution of the electric potential in the device;\footnote{As noted in \cite{courtier2019transport,courtier2018systematic,richardson2016can} this gives excellent agreement to the full {drift-diffusion} model.} and the using the solutions of these equations to obtain the potential drops $V_1(t)$-$V_4(t)$ and the internal electric field $E(t)$. Then, on making the assumption that the carriers are Boltzmann distributed, we can  deduce expressions for the electron and hole densities $n(x,t)$ and $p(x,t)$ and, in turn, use these to compute the current loss due to charge carrier recombination $-j_{\rm rec}$ and hence obtain $J(t)$. 

\subsubsection*{Approximate carrier distributions.} 
\begin{figure}[htbp] % trim=left botm right top  \fbox{ whole include graphics bit for box around it to crop} 
\includegraphics[trim=0.4cm 0.4cm 0.4cm 0.4cm, clip, width=0.7\textwidth]{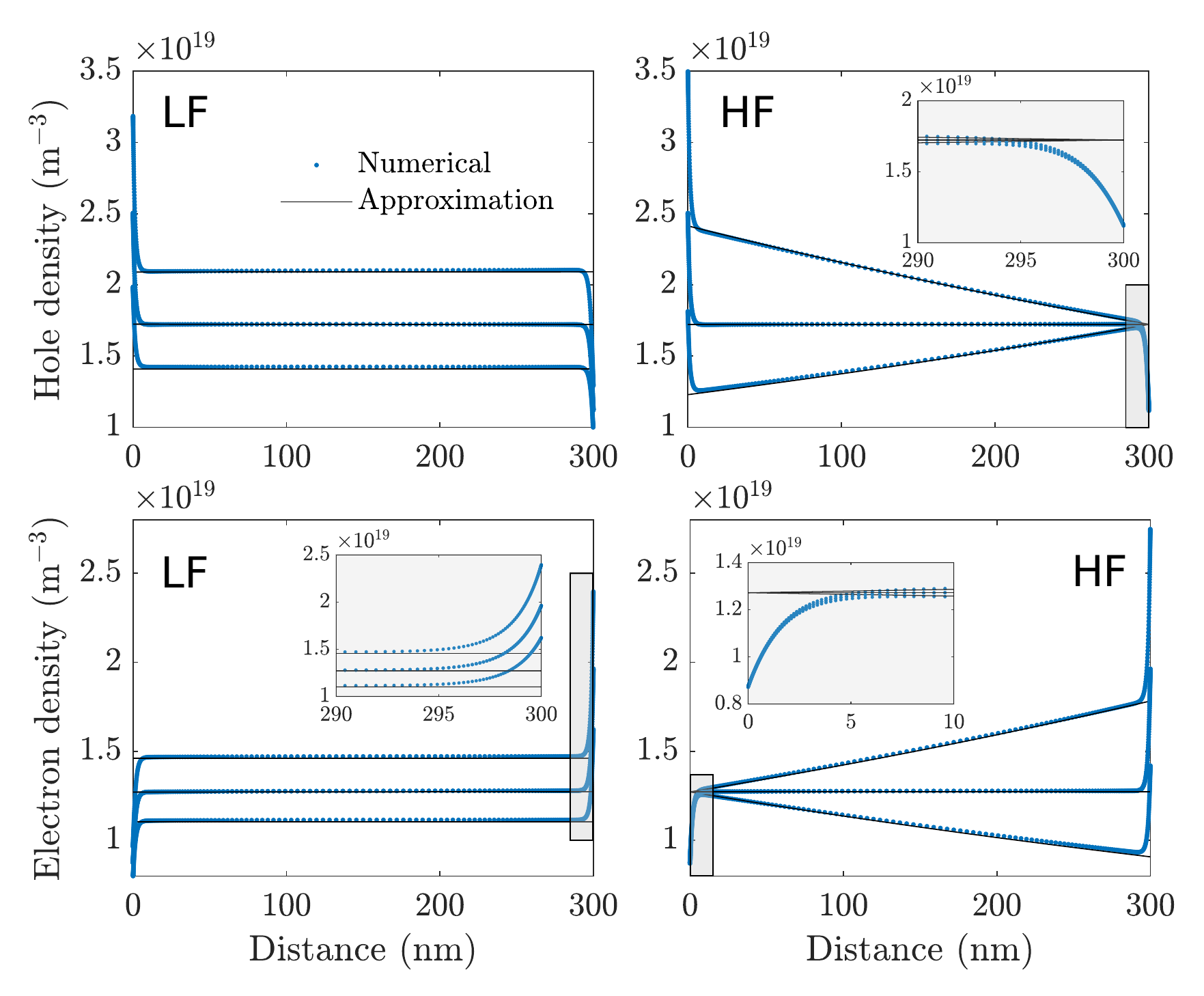}
\centering
\caption{{Comparison between solutions obtained by using the Boltzmann approximation to evaluate electron and hole densities and solutions of the the full drift-diffusion model. In both cases $V_{DC}=V_{OC}$. Left: carrier density at equally spaced intervals over a low frequency period (1 mHz). The right is the equivalent but over an intermediate/high frequency period (25 kHz). The parameters used are detailed in Table \ref{tab:params}, under 0.1-sun equivalent illumination and with recombination at the ETL/perovskite interface ($R_l$) from Table \ref{tab:recomb}.}}
\label{fig:npcomptoIM}
\end{figure}

Recombination at the interfaces between the perovskite and the transport layers (TLs) is known to be a significant source of energy loss within the cell and, in order to account for this, we need not only to compute the electron and hole densities $n(x,t)$ and $p(x,t)$ within the central bulk region of the perovskite, but also at the interfaces between the perovskite and TLs.  At the ETL/perovskite interface (on $x=0$) we denote the electron density within the ETL by $\nl$ and the hole density within the perovskite by $\pl$. Correspondingly, at the perovskite/HTL interface ($x=b$) we denote the electron density within the perovskite by $\nr$ and the hole density within the HTL by $\pr$. On assuming Boltzmann distributions of the carriers the expressions for electron and hole densities within the perovskite layer, and at the interfaces between the transport layers and the perovskite,  read as follows
\be
\left.\begin{array}{l}
n(x,t)=  k_Ed_E\exp\left(-\frac{V_1 + V_2 + xE}{V_T}\right) \\
p(x,t)=  k_Hd_H\exp\left(-\frac{V_3 + V_4 + (b-x)E}{V_T}\right)
\end{array}
\right\} \qquad \mbox{in} \qquad 0<x<b,~~~~~~~~~~~~~~~~~~ \label{eq:n_p_Boltz}\\
\left.\begin{array}{l}
\nl(t)=  d_E\exp\left(-\frac{V_1}{V_T}\right) \\
\pl(t)=  k_Hd_H\exp\left(-\frac{V_2 + V_3 + V_4 + bE}{V_T}\right)
\end{array}
\right\} \quad \mbox{on the ETL/perovskite interface}, \label{eq:n_l_p_l} \\
\left.\begin{array}{l}
\nr(t)=  k_Ed_E\exp\left(-\frac{V_1 + V_2 + V_3 + bE}{V_T}\right)\\
\pr(t)=  d_H\exp\left(-\frac{ V_4 }{V_T}\right)
\end{array}
\right\} \quad \mbox{on the perovskite/HTL interface,}  \label{eq:n_r_p_r}
\ee
Here, the ratios $k_E$ (and $k_H$) between the electron (and hole) densities across the ETL/perovskite (and perovskite/HTL) boundaries are given by
\be
k_E = \frac{g_c}{g_{cE}} \exp \left( \frac{E_{cE}-E_c}{V_T}\right), \hspace{5mm} k_H = \frac{g_v}{g_{vH}} \exp \left( \frac{E_{v}-E_{vH}}{V_T}\right), \label{eq:kE_kH}
\ee
respectively, in which the parameters appearing in this equation are defined in Table 1. An example of the extremely good agreement obtained between this approximation for the carrier densities and the full drift-diffusion model is shown in Figure \ref{fig:npcomptoIM} which is computed for a IS voltage input of the form \eqref{eq:V(t)_IS} with $V_{DC}=V_{OC}$ and $V_p=10$mV.

\subsubsection*{The recombination current.}
Given the carrier distributions the recombination current can determined from the bulk recombination rate $R_\textnormal{bulk}(n,p)$, and the interfacial rates $R_l(\nl,\pl)$ (on the interface with the ETL) and $R_r(\nr,\pr)$ (on the interface with the HTL) as follows:
\be
j_{\rm rec}= qR_l \left(\nl(t),\pl(t) \right) + qR_r \left(\nr(t),\pr(t) \right)  + q\int_0^b R_\textnormal{bulk}\big(n(x,t),p(x,t)\big) dx. \label{eq:jrec_int}
\ee
In the case where there is a single dominant recombination mechanism, as in the examples given in Table \ref{tab:recomb}, the recombination current takes the form
\begin{align}
        \jrec(t) = j_{R_i}\exp\left(-\frac{F_i(V_1(Q(t)),V_2(Q(t)),V_3(Q(t)),V_4(Q(t)))}{V_T} - \frac{b}{\nel V_T}E(t)\right).    \label{eq:j_rec_general}
\end{align}
Here $F_i(V_1,V_2,V_3,V_4)$ (the potential barrier to recombination for recombination type $i$, where $i=b,p,n,l$ or $r$), $\nel$ (the \textit{electronic} ideality factor) and $j_{R_i}$ (the recombination current prefactor) all depend on the dominant recombination mechanism and, for the recombination mechanism stated in Table \ref{tab:recomb},  are presented in Table \ref{tab:F_i(V)andjR_iforR_i}. Notably, $\nel$, the \textit{electronic} ideality factor, takes either the value 1 or 2, depending on the recombination mechanism, and can thus be used as a diagnostic tool (just as the ideality factor is in conventional photovoltaics) in order to distinguish between different possible sources of recombination within the cell. Notably all the ionic effects are contained in the function $F_i$ so that $\nel$ depends only upon purely electronic effects. 

Further discussion of the electronic ideality factor can be found in Sections \ref{sec:elec_id_fac} and \ref{sec:n_id_n_ec_comp}, along with details of how it may be obtained experimentally.

\begin{table}[htbp]
\begin{center}
\begin{tabular}{ |c|c|c|c|} 
 \hline
  \textbf{Recombination Type} & \textbf{Full Form}  & \textbf{Approximation} & \textbf{Parameter Values}  \\
 \hline
  Bimolecular, $R_b$ &  $R_{\textnormal{bulk}} = \beta\left(np - n_i^2\right)$ & $R_{\textnormal{bulk}} \approx \beta np$ & $ \beta = 10^{-12}$ m$^3$s\super{-1}\\ 
 \hline
  \begin{tabular}{@{}c@{}}Hole-limited  \\ SRH, $R_p$\end{tabular} &  $R_{\textnormal{bulk}} = \frac{np - n_i^2}{\tau_np + \tau_pn + k_1}$ & $R_{\textnormal{bulk}} \approx \frac{p}{\tau_p} $ &  \begin{tabular}{@{}c@{}}$\tau_n = 3\times10$\super{-10} s \\ $\tau_p = 3\times10$\super{-8}s\end{tabular}  \\ 
   \hline
  \begin{tabular}{@{}c@{}}Electron-limited \\  SRH, $R_n$ \end{tabular}&  $R_{\textnormal{bulk}} = \frac{np - n_i^2}{\tau_np + \tau_pn + k_2}$ & $R_{\textnormal{bulk}} \approx \frac{n}{\tau_n} $ & \begin{tabular}{@{}c@{}}$ \tau_n = 3\times10$\super{-8} s \\ $\tau_p = 3\times10$\super{-10} s\end{tabular} \\ 
   \hline
  \begin{tabular}{@{}c@{}}ETL/Perovskite \\ interfacial SRH, $R_l$\end{tabular}  &  $R_l = \frac{\nl \pl - n_i^2/k_E}{\pl/v_{n_E} + \nl/v_{p_E} + k_3}$ & $R_l \approx v_{p_E} \pl$  &  \begin{tabular}{@{}c@{}}$v_{n_E}$ = 10$^{5}$ ms\super{-1}, \\ $v_{p_E}$ = 5 ms\super{-1}\end{tabular} \\ 
   \hline
  \begin{tabular}{@{}c@{}}Perovskite/HTL \\  interfacial SRH, $R_r$ \end{tabular}&  $R_r = \frac{\nr \pr - n_i^2/k_H}{\pr/v_{n_H} + \nr/v_{p_H} + k_4}$ & $R_r \approx v_{n_H}\nr $  & \begin{tabular}{@{}c@{}}$v_{n_H}$ = 5 ms\super{-1}  \\ $v_{p_H}$ = 10$^{5}$ ms\super{-1}\end{tabular}\\ 
  \hline
\end{tabular}
\end{center}
\caption{List of all recombination types considered in this study, including their approximations and relevant parameter values. The intrinsic carrier density within the perovskite, given by {$n_i = \sqrt{ g_cg_v }\exp\left(\left(E_v-E_c\right)/2V_T\right)$}, is negligible relative to the bulk carrier densities under illumination. Assuming trap state energies lie close to the centre of the band-gap, the parameters $k_{1-4}$ are also negligible. The interfacial recombination rate equations are dependent on the carrier densities at the left and right interfaces defined by equations (\ref{eq:n_l_p_l}) and (\ref{eq:n_r_p_r}) respectively. 
\label{tab:recomb}} % See Nelson pg 110 ish and https://ecee.colorado.edu/~bart/book/recomb.htm
\end{table}
\begin{table}[ht!]
\begin{center}
\begin{tabular}{ |c|c|c|c|c|c|c| }  
 \hline
  \textbf{Label} $\boldsymbol{R_i}$ & \textbf{Recombination} & $\boldsymbol{F_i(V_1,V_2,V_3,V_4)}$ & $\boldsymbol{j_{R_i}}$ & $\boldsymbol{n_{\textnormal{\textbf{el}}}}$\\
 \hline 
  $R_b$ & $R_{\textnormal{bulk}} = \beta np$  & $V_1 + V_2 + V_3 + V_4 $ & $ qb\beta k_Ed_Ek_Hd_H$ & 1 \\ 
 \hline
  $R_p$ &$R_{\textnormal{bulk}} =  p/\tau_p$  & $ V_3 + V_4 $ & $\frac{qbk_Hd_H}{\tau_p}$ & 2  \\ 
   \hline
  $R_n$ & $R_{\textnormal{bulk}} = n/\tau_n $  & $V_1 + V_2 $ & $ \frac{qbk_Ed_E}{\tau_n} $ & 2 \\ 
   \hline
  $R_l$ & $R_l =  v_{p_E}\pl$  & $V_2 + V_3 + V_4 $ & $qk_Hd_Hv_{p_E}$ & 1 \\ 
   \hline
  $R_r$ & $R_r = v_{n_H} \nr$  & $V_1 + V_2 + V_3 $ & $qk_Ed_Ev_{n_H}$ & 1 \\ 
  \hline
\end{tabular}
\end{center}
\caption{Recombination types with labelling convention and corresponding values for the electronic ideality factor. $F_i(V_1,V_2,V_3,V_4)$ is the potential barrier to recombination for recombination of type $R_i$, where $i=b,p,n,l,r$. This notation enables the recombination current (eq.\eqref{eq:j_rec_general}) and impedance parameters (eq.\eqref{eq:RandC_HF} and eq.\eqref{eq:RandC_LF}) to be written in a general form. The total potential drop across the cell at steady state is given by $V_1 + V_2 + V_3 + V_4 = \Vbi-\VDC$. \label{tab:F_i(V)andjR_iforR_i} }
\end{table}

%%%%%%%%%%%%%%%%%%%%%%%%%%%%%%%%%%%%%%%%%%%%%%%%%%%%%%%%%%%%%%%%%%%
%%%%%%%%%%%%%%%%%%%%%%%%%%%%%%%%%%%%%%%%%%%%%%%%%%%%%%%%%%%%%%%%%%%
%%%%%%%%%         RESULTS
%%%%%%%%%%%%%%%%%%%%%%%%%%%%%%%%%%%%%%%%%%%%%%%%%%%%%%%%%%%%%%%%%%%
%%%%%%%%%%%%%%%%%%%%%%%%%%%%%%%%%%%%%%%%%%%%%%%%%%%%%%%%%%%%%%%%%%%

\section{Results} \label{sec:results}
In scenarios where the magnitude of the voltage perturbation $V_p$ is small (in comparison to $V_T$) it is possible to linearise the current response of the cell to the IS voltage input \eqref{eq:V(t)_IS}. In turn, this allows the current response to be written in the form \eqref{eq:J(t)_IS} where $J_p$ and the phase shift $\theta$ are linearly related to the voltage perturbation $V_p \cos(\omega t)$ via  the impedance $Z(\omega)$ (as described in \S1). In turn the impedance thus obtained can be related to the equivalent RC-RC circuit  depicted in figure \ref{fig:Nyq_Bode_labelled}(c) which allows us to identify the high frequency (HF) and low frequency (LF) resistances, $R_{HF}$ and $R_{LF}$, and capacitances, $C_{HF}$ and $C_{LF}$, of the cell. Details of this derivation are given in the Supplementary Information \ref{sec:derivation}.
The resistances and capacitances are most usefully expressed per unit area of cell, so that $R_{HF}$ and $R_{LF}$ are actually the inverse of  the conductances per unit area, and have units V A\super{-1} m$^2$ while $C_{HF}$ and $C_{LF}$ are capacitances per unit area, and have units A\,s V\super{-1} m$^{-2}$. On adopting this convention
\begin{align}
    R_{HF} &=  \frac{V_T\nel}{\jrec  (\VDC) }  &&C_{HF}= \frac{\varepsilon_p}{b}  \label{eq:RandC_HF}\\
    R_{LF} &=  \frac{V_T}{\jrec (\VDC) }\left( \nap(\VDC) - \nel\right) && C_{LF} = { \frac{\nap(\VDC) \jrec (\VDC) }{G_+V_T \nel \left( {\nap(\VDC)} - \nel \right)}\left(-\frac{d\QDC}{d\VDC}\right)}. \label{eq:RandC_LF}
\end{align}
Here $\jrec (\VDC)$ is the steady-state recombination current density, $\varepsilon_p$ is the perovskite layer permittivity, $\nap$ is the apparent ideality factor determined by the standard techniques used to obtain the ideality factor (such as the Suns-$\Voc$ or dark-$JV$ methods), $\nel$ is the electronic ideality factor,  and the parameter $G_+$ quantifies the ionic conductance per unit area of the perovskite layer and is given by
\be 
G_+  = \frac{qD_+N_0}{V_Tb}, \label{eq:G_+}
\ee
where $D_+$ and $N_0$ are the ionic vacancy density and diffusion coefficient in the perovskite layer.  The final term in \eqref{eq:RandC_LF} is found by  solving \eqref{eq:pot_drops} to obtain an expression for $\QDC$ as a function of $\VDC$ and then differentiating this function with respect to $\VDC$.  Plots of {$\QDC$} as a function of $\VDC$ are made in figure \ref{fig:Q_0_vs_V_0_and_dQbydV}(a) while those of its derivative $d\QDC/d \VDC$, which appears in the expression for $C_{LF}$, can be found in figure \ref{fig:Q_0_vs_V_0_and_dQbydV}(b). {Notably $(-d \QDC/d \VDC)$ is the ionic capacitance of the cell (per unit area) and this fact together with the formula for the ionic conductance \eqref{eq:G_+} allows us to rewrite equations \eqref{Qeq}, \eqref{eq:E(t)} and \eqref{eq:pot_drops} for the evolution of the ionic charge in the device in the form of an evolution equation for the potential drop $V_{\rm bulk}=b E(t)$ across the centre of the perovskite layer, in which $C_{LF}$ is a key parameter,
\be
\left.\left[ C_{LF} \frac{\nel(\nap-\nel)}{\nap \jrec} \right] \right|_{\VDC=V(t)+V_{\rm bulk}(t)} \left( \frac{d V_{\rm bulk} }{d t}+\frac{d V}{d t} \right) =-V_{\rm bulk}.
\ee }

\begin{figure}[htbp] % trim=left botm right top  \fbox{ whole `includegraphics{}' bit for box around it to crop}
\includegraphics[trim=3cm 11.5cm 4cm 12.5cm, clip, width=0.6\textwidth]{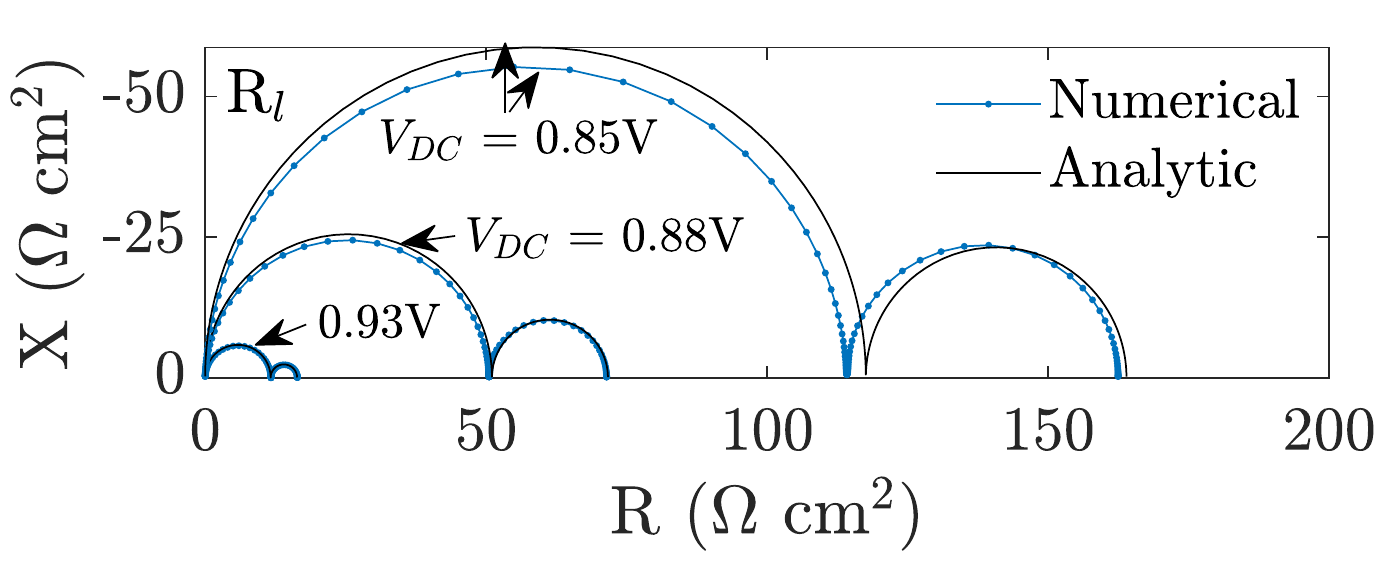}
\centering
\caption{Simulated impedance spectra for a PSC with recombination at the ETL/perovskite interface ($R_l$) under 0.1-Sun equivalent illumination with a perturbation amplitude of 10 mV. Spectra for three DC voltages are shown, including at open-circuit where $\VDC = \Voc$=0.93V. The cell parameters used in the numerical and analytic model are listed in Table \ref{tab:params}. The recombination parameters for the approximate rates (as used for the analytic solution) and full rates (as used for the numerical solution) are listed in Table \ref{tab:recomb}.}
\label{fig:RlNyquist}
\end{figure}

\begin{figure}[htbp] % trim=left botm right top  \fbox{ whole `includegraphics{}' bit for box around it to crop}
\includegraphics[trim=3cm 9.0cm 4cm 9.5cm, clip, width=0.6\textwidth]{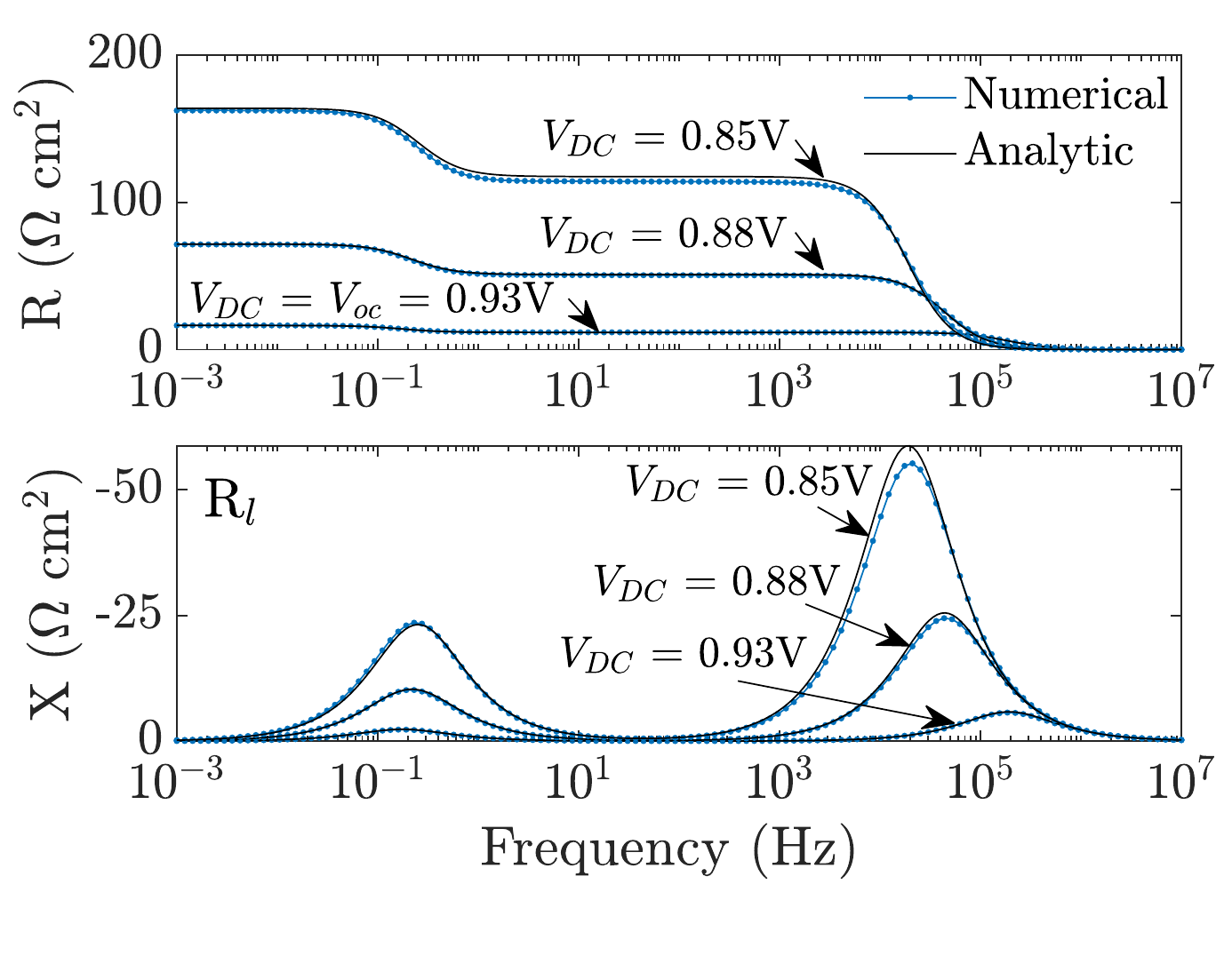}
\centering
\caption{Real (top) and imaginary (bottom) components of impedance versus frequency for the spectra presented in Figure \ref{fig:RlNyquist}.}
\label{fig:Rlfreqplot}
\end{figure}

\begin{figure}[htbp] % trim=left botm right top  \fbox{ whole include graphics bit for box around it to crop}
\includegraphics[trim=3.5cm 11.3cm 4.5cm 10.5cm, clip, width=0.7\textwidth]{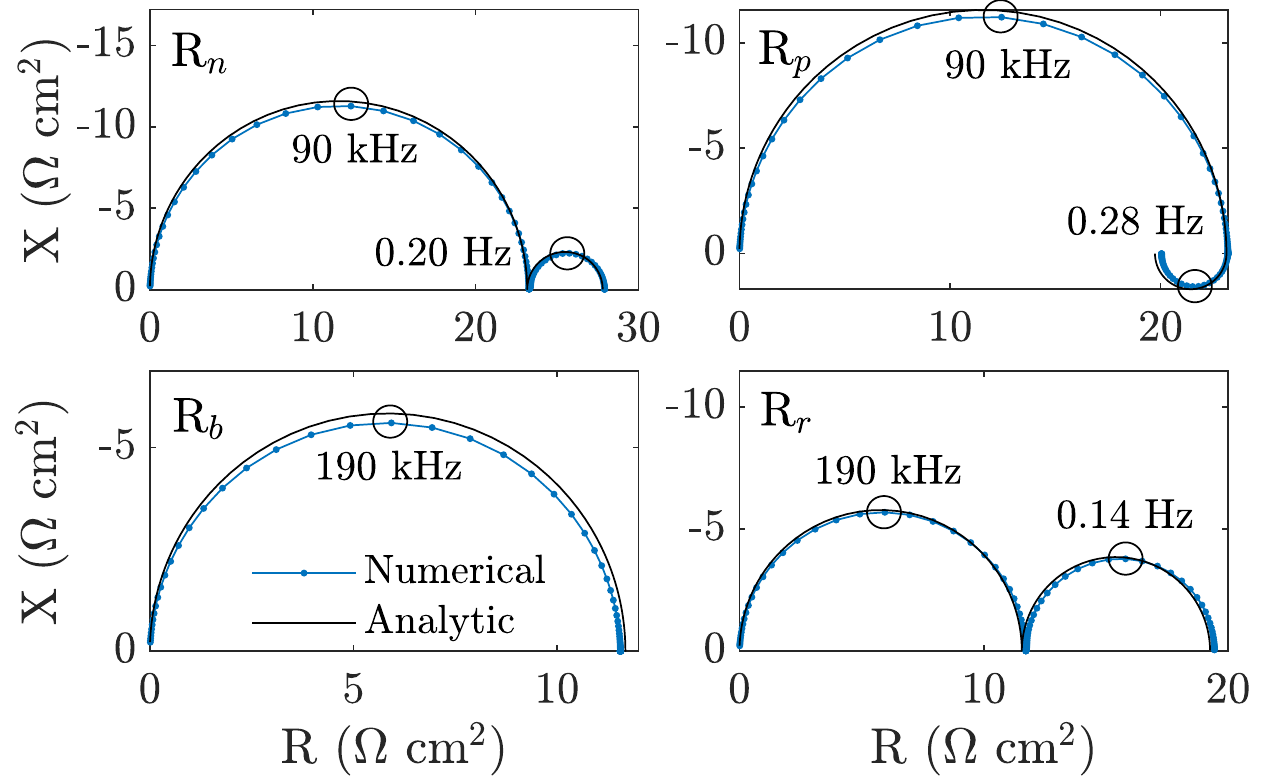}
\centering
\caption{Simulated impedance spectra at open-circuit with different recombination mechanisms. Clockwise from top left: electron-limited bulk SRH ($R_n$: V$_{oc}$=0.94 V), hole-limited bulk SRH ($R_p$: V$_{oc}$=0.92 V), perovskite/HTL interfacial ($R_r$: V$_{oc}$=0.95 V) and bimolecular bulk recombination ($R_b$: V$_{oc}$=0.95 V). Cell and recombination parameters are listed in Tables \ref{tab:params} and \ref{tab:recomb} respectively.}
\label{fig:R_combined_Nyquist_point1sun}
\end{figure}

\begin{figure}[htbp] % trim=left botm right top  \fbox{ whole `includegraphics{}' bit for box around it to crop}
\includegraphics[trim=3.5cm 10.0cm 4cm 9.5cm, clip, width=0.65\textwidth]{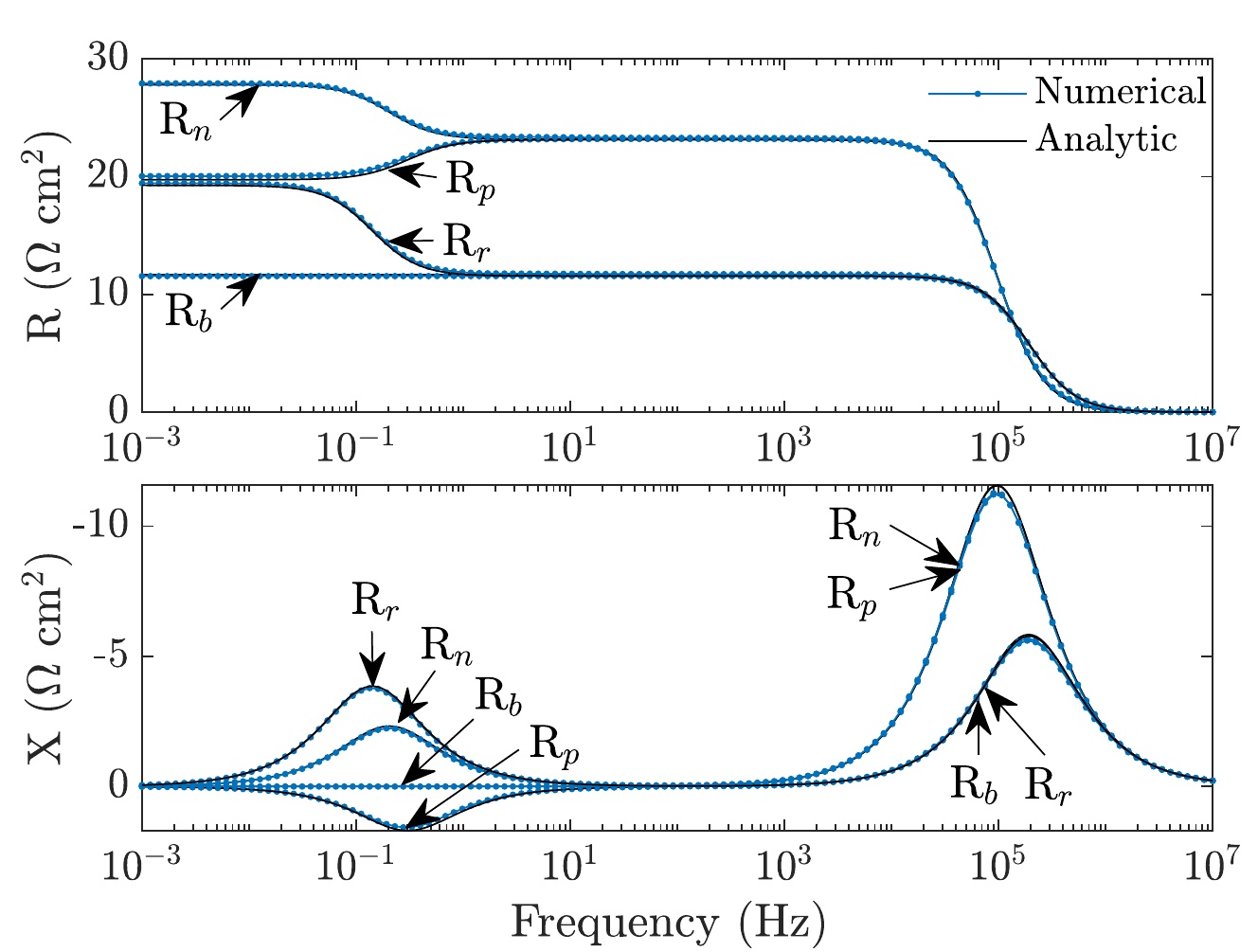}
\centering
\caption{Real (top) and imaginary (bottom) components of impedance versus frequency for the spectra presented in Figure \ref{fig:R_combined_Nyquist_point1sun}.}
\label{fig:R_combined_Bode_point1sun}
\end{figure}

{\paragraph{Interpretation of the results.} Equations \eqref{eq:RandC_HF} and \eqref{eq:RandC_LF} illustrate how to interpret the impedance response of a PSC. In particular, they show that the impedance spectrum of the drift-diffusion model of a PSC is associated with two dominant features, a low- and a high-frequency one. Furthermore, since this response is the same as that of the RC-RC circuit  depicted in figure \ref{fig:Nyq_Bode_labelled}(c), it can be associated with two arcs in a Nyquist plot, as  illustrated in figures \ref{fig:RlNyquist} and  \ref{fig:R_combined_Nyquist_point1sun}, and equivalently two peaks in the Bode plots, as illustrated in figures \ref{fig:Rlfreqplot} and \ref{fig:R_combined_Bode_point1sun}. However, in contrast to a physical RC-RC circuit, the low frequency capacitance and resistance are not guaranteed to be positive. In particular, in scenarios where $\nel > \nap$  both $R_{LF}$ and $C_{LF}$ are negative (since $d \QDC/d \VDC$ is always negative while all the other terms in these formulae are positive). It is these negative capacitances and resistances that give rise to {\it so-called} inductive arcs in the Nyquist plot that appear below the axis, as for example in figure \ref{fig:R_combined_Nyquist_point1sun}(b).}

{An alternative viewpoint is provided by starting from experimental IS data and using these data to obtain fits for $R_{HF}(\VDC)$, $C_{HF}(\VDC)$, $R_{LF}(\VDC)$ and $C_{LF}(\VDC)$. These experimentally derived expressions for the cell resistances and capacitances can, in turn, be used to infer many of the cell's properties.}
%Further discussion on the interpretations of these results is deferred to \S \ref{sec:interprettingIS}.

\paragraph{The apparent ideality factor $\nap$.} The unconventional physics of PSCs means that the value of $\nap$ cannot be related straightforwardly to the recombination mechanism, as is the case for a conventional solar cell. Instead the apparent ideality factor determined for PSCs using standard techniques should be interpreted in terms of the model by the following asymptotic expression, which is exact for $\VDC = \Vbi$, (see \cite{courtier2020interpreting}, where $\nap$ is termed the ectypal factor, for further details):
{\be
\nap(\VDC) \sim \frac{\vbi-\VDC}{ F_i\left(V_1(\QDC),V_2(\QDC),V_3(\QDC),V_4(\QDC)\right).} \label{eq:ectypal_approx_def}
\ee
Here we make use of the fact that $V_1(\QDC) + V_2(\QDC) + V_3(\QDC) + V_4(\QDC)=\vbi-\VDC$ and use the functional relation between the charge density and the applied voltage $\QDC=\QDC(\VDC)$ that is obtained by inverting \eqref{eq:pot_drops}. Notably $\nap$ gives the ratio of the total potential drop across the cell $\vbi-\VDC$ to the potential barrier for recombination $F_i$. Furthermore,  $\nap$ is inherently dependent on the applied voltage, the ionic vacancy density and transport layer properties, via $\Omega_E$ and $\Omega_H$ (as defined in \eqref{eq:VsandOmegas}). In scenarios where $\nap$ = $\nel$ the low frequency arc of the Nyquist plot disappears, as can be seen from \eqref{eq:RandC_LF}. This situation, \ie $\nap$ = $\nel$, occurs where there are no mobile ions (see \S \ref{sec:elec_id_fac} for further details) and a conventional (non-ideal) diodic response is observed. However, it can also occur even in the presence of mobile ions, as for example when recombination occurs solely in the perovskite via a purely bimolecular mechanism (see also \S \ref{sec:LF}), as illustrated in  figure \ref{fig:R_combined_Nyquist_point1sun}(c). Further properties of the apparent ideality factor are discussed \S \ref{sec:elec_id_fac} and \S \ref{sec:n_id_n_ec_comp}.}

\paragraph{The recombination current $\jrec$} {It is notable that $\jrec$ depends both on the steady-state voltage $\VDC$ and on how the potential difference across the cell is divided between the potential drops $V_{1}-V_{4}$; it is thus sensitive to parameters, such as the transport layer doping densities, that alter the relative distribution between the potential drops. In steady-state, the recombination current can be related to the photo-generated current $\jg$ and the  steady-state current output of the cell $J(\VDC)$ by the expression
\be
\jrec(\VDC) = \jg - J(\VDC). \label{eq:j_rec0_from_measurable_currents} \label{jrec}
\ee 
This allows the steady-state recombination current to be estimated from experimental measurements by making the assumption that $\jg \approx J(0)$.}

The full numerical model is detailed in \cite{courtier2019transport} and a description of its use to simulate IS measurements is given in \cite{riquelme2020identification}.

% Describe what the full model solves for and the experimental conditions for the simulated spectra
\paragraph{Numerical solutions to the drift-diffusion model.} These are obtained by using the open-source PSC simulation tool \texttt{IonMonger} \cite{courtier2019ionmonger}. This solves the fully coupled ionic-electronic drift-diffusion equations for a planar PSC with a single positively charged mobile ion vacancy species and mobile charge carriers in the perovskite layer (as described in \cite{courtier2019transport}). Impedance spectra are obtained by using \texttt{IonMonger} to solve the {drift-diffusion} model multiple times, over a range of frequencies $\omega$,  with the voltage input given by \eqref{eq:V(t)_IS} in which the amplitude, $V_p$, of the sinusoidal perturbations to the steady-state applied voltage $\VDC$ is small. A detailed description of this process is provided in  \cite{riquelme2020identification}. Comparison of the resulting output current to \eqref{eq:J(t)_IS}  allows the complex impedance $Z(\omega,\VDC)$ to be obtained as a function of the frequency $\omega$. Further details of the method are provided in Riquelme \etal \cite{riquelme2020identification}. Impedance spectra are calculated, in this way, about a specified steady-state voltage $\VDC$ (for example $\VDC=\Voc$) using the full forms of the recombination mechanisms specified in Table \ref{tab:recomb} and compared to the corresponding analytic spectra reconstructed from the analytic expressions for the low and high frequency resistances and capacitances, \eqref{eq:RandC_HF}-\eqref{eq:RandC_LF}.
The numerical and analytic impedance spectra presented in this work (unless otherwise stated) are simulated at open-circuit under monochromatic (520 nm) illumination with intensity that produces a photocurrent equivalent to 0.1-Suns at AM1.5. Impedance spectra are composed of 128 and 256 frequencies for the numerical and analytic solutions respectively over a range of 10$^{-3}$-10$^7$ Hz. The voltage perturbation amplitude is 10 mV throughout.

\paragraph{Comparison between impedance spectra computed using the drift-diffusion and analytic models.}
In Figures \ref{fig:RlNyquist}-\ref{fig:R_combined_Bode_point1sun} comparison is made between PSC impedance spectra reconstructed from the approximate analytic model \eqref{eq:RandC_HF}-\eqref{eq:RandC_LF} and those reconstructed from numerical solutions to the full drift-diffusion model. {In practice impedance data from a numerical solution to the drift-diffusion model is fitted to an RC-RC circuit (as depicted in figure \ref{fig:Nyq_Bode_labelled}(c)). In most cases this can be accomplished by measuring the radii of the arcs in the Nyquist plane to extract $R_{HF}$ and $R_{LF}$ and by noting the two frequencies, $\omega_{LF}$ and $\omega_{HF}$, at which $X$ (the imaginary component of the impedance $Z(\omega)$) has a local maximum (see, for example, figure \ref{fig:R_combined_Bode_point1sun}(b)). The two capacitances are related to these frequencies via the standard formulae
\be
C_{LF}=\frac{1}{R_{LF} \omega_{LF}}, \qquad C_{HF}=\frac{1}{R_{HF} \omega_{HF}}. \label{capac}
\ee 
The only other bit of information that needs to be extracted from the numerical simulations of the drift-diffusion model is the recombination current $\jrec(\VDC)$ which can be evaluated from \eqref{eq:j_rec0_from_measurable_currents}.}

The results plotted in Figures \ref{fig:RlNyquist}-\ref{fig:R_combined_Bode_point1sun} show examples of all five recombination mechanisms described in Table \ref{tab:recomb} and are all simulated under 0.1-Sun equivalent illumination at open-circuit with $\VDC=\Voc$; in all cases the other parameter values are taken from Table \ref{tab:params}. {Additional plots showing simulated IS at $\VDC=\Voc$ and at 1-Sun illumination can be found in the SI in figures \ref{fig:R_combined_Nyquist_1Sun}-\ref{fig:R_combined_Bode_1Sun} while plots of the spectra at $\VDC=V_{MPP}$ (\ie the maximum power point) and at 0.1-Sun illumination can be found in figure \ref{fig:Nyquists_at_MPP}.}
It is clear that there is a extremely good agreement between the impedance spectra predicted by the analytic model and those predicted by the full {drift-diffusion model}. Furthermore, both approaches can be used to illustrate how different recombination mechanisms impact the shape and features  of the impedance spectra. The ability of the analytic model to closely reproduce the results of the numerical model validates the use of the surface polarisation model \cite{courtier2019transport,courtier2018systematic,richardson2016can} and the use of the Boltzmann approximation in the computation of carrier densities and recombination rates.



\pagebreak
% State usefulness of relations for showing trends
{The key advantages of using the (approximate) analytic model, as opposed to the drift-diffusion model, to interpret impedance spectra are that (I) `trends' (such as the dependence of the spectra on illumination, open-circuit voltage and steady-state voltage) observed in real spectra can be much more easily understood in terms of the analytic model than from numerical simulations of the drift-diffusion model and (II) key physical parameters of the device may be obtained much more easily from impedance data by comparing to the explicit formulae for $\nap$, $R_{HF}$, $C_{HF}$, $R_{LF}$ and $C_{LF}$, generated by the analytic model, than by repeatedly solving the drift-diffusion model until a parameter set is found that matches the data.} 

In terms of the analytic model the characteristic low and high frequencies, $\omega_{LF}$ and $\Omega_{HF}$, which give the locations of the maxima in $X(\omega)$ (see Figure \ref{fig:R_combined_Bode_point1sun}), are found by substituting \eqref{eq:RandC_HF}-\eqref{eq:RandC_LF} into \eqref{capac} to obtain
\be
    \omega_{LF} = G_+ \frac{\nel}{\nap(\VDC)}\left({-\frac{d \QDC}{d \VDC}}\right)^{-1}, \label{eq:omega_LF} \quad 
    \omega_{HF} = \frac{b\jrec(\VDC)}{\varepsilon_p V_T \nel}, \label{eq:omega_HF}    
\ee
where it should be noted that ${{d \QDC}/{d \VDC}}$ is negative for all $\VDC$, as shown in Figure \ref{fig:Q_0_vs_V_0_and_dQbydV}. Furthermore $\omega_{LF}$, the characteristic frequency of the LF feature, is proportional to $G_+$, the parameter that quantifies the ionic conductance and which is defined in \eqref{eq:G_+}. In particular, $\omega_{LF}$ increases with increased anion vacancy density, or mobility.

Given that $1/(\nap {{d \QDC}/{d \VDC}})$ is not strongly temperature dependent, the activation energy extracted from the LF time constant is the result of the temperature dependence of $G_+(T)$ \cite{pockett2017microseconds,contreras2019impedance,garcia2019influence,meggiolaro2019formation}.

\subsection{The electronic ideality factor} \label{sec:elec_id_fac}

% Introduce the electronic ideality factor 
{Analysis of the PSC drift-diffusion, conducted here and in \cite{courtier2020interpreting}, demonstrates that the ideality factor as conventionally defined, and determined by the Suns-$\Voc$ or dark-$JV$ methods, is not independent of the applied voltage (a consequence of ion motion) and is thus not a particularly useful measure of PSC properties. 
%This is a direct consequence of the unusual physics of these devices. 
We have therefore termed this version of the ideality factor the apparent ideality factor, $\nap$. However, a deeper understanding of the drift-diffusion models that describe PSC behaviour has led to the identification of an alternative dimensionless constant, which we term the \textit{electronic ideality factor} $\nel$. This factor plays an analogous role in PSCs to that played by the traditional ideality factor in conventional photovoltaics and, in particular, can be used as a tool to deduce the dominant form of recombination taking place in the cell.}

% Detailing its value and definition
{For the recombination mechanisms studied in this work (see Table \ref{tab:recomb}) the electronic ideality factor takes a value of either 1 or 2, depending on the dominant recombination mechanism. Specifically, it takes a value of 2 where recombination occurs via an electron-, or hole-, limited SRH mechanism within the perovskite absorber layer; and takes a value of 1 where bimolecular recombination  in the perovskite is the dominant loss mechanism, or recombination occurs (via an SRH mechanism) on the interfaces with the transport layers. These results are stated in full in Table \ref{tab:F_i(V)andjR_iforR_i}. From an experimental perspective the electronic ideality factor can be determined from the high frequency resistance and the recombination current (see (\ref{eq:RandC_HF}a)) via the formula
\be
\nel = \frac{R_{\textnormal{HF}}(\VDC)\jrec(\VDC)}{V_T}, \label{eq:n_id} \label{nel}
\ee
where $\jrec$ may be estimated from (\ref{eq:j_rec0_from_measurable_currents}). This result is key to analysing the behaviour of real cells from from experimental data. In particular, it is straightforward to obtain  $\nel$ since both $R_{\textnormal{HF}}(\VDC)$ and $\jrec(\VDC)$ are readily measured. It is also the way that we determine $\nel$ from IS generated by numerical simulation of the drift-diffusion model. }

{In order to  demonstrate how the model and the electronic ideality factor are related to conventional solar cell theory, we consider a three-layer cell in which there are no mobile ions present in the perovskite capable of forming interfacial Debye layers and screening the field  from the absorber. In this scenario the potential interfacial potential drops are all zero, that is $V_1 = V_2 = V_3 = V_4 = 0$. The potential profile no longer has the form depicted in Figure \ref{fig:V1-4}a but is instead similar to that typically  portrayed for an ideal `p-i-n' junction, where the built-in voltage and applied potential is dropped uniformly across the central intrinsic (\ie perovskite) layer. Hence, the uniform electric field in the perovskite (see equation \ref{eq:E(t)}) is given by
\be
E(t) = \frac{1}{b} \left( V_{bi} - V(t) \right).
\ee
Using the relation above and the fact that $V_{1-4} = 0$, equation (\ref{eq:j_rec_general}) for the recombination current simplifies to
\be
\jrec(t) = j_{R_i}^*\exp\left(\frac{V(t)}{\nel V_T}\right),
\ee
where we define $j_{R_i}^* = j_{R_i}\exp(-V_{bi}/(\nel V_T))$. Ignoring the contribution from the displacement current, the total current is given by
\be
J(t) = \jg - j_{R_i}^*\exp\left( \frac{V(t)}{\nel V_T}\right).
\ee
This can be compared to the classical non-ideal diode equation \cite{nelson2003physics}
\be
J(t) = \jg - j_{\rm g,therm}\exp\left( \frac{V(t)}{n_{id} V_T}\right), \label{eq:J(t)_non_ideal_diode}
\ee
where here $\jg=j_{\rm g,sol}+j_{\rm g,therm}$  incorporates both solar $j_{\rm g,sol}$ and thermal $j_{\rm g,therm}$ generation terms.
It is clear from this comparison that, in the absence of ions in the perovskite layer, the electronic ideality factor is exactly analogous to the ideality factor that appears in standard semiconductor diode theory.  Finally, by removing the ions in this way it is evident, from equation \eqref{eq:RandC_LF}, that the LF impedance response disappears, leaving only the HF semicircle described by equation \eqref{eq:RandC_HF}.}

{We have derived a new form of  ideality factor, namely the electronic ideality factor $\nel$, that is appropriate for analysing the behaviour of a PSC. In contrast to the apparent ideality factor $\nap$, which is commonly used to analyse PSC behaviour in the literature, the electronic ideality factor is not inherently voltage dependent and is not influenced by the distribution of potential drops $V_{1-4}$ across the cell. More specifically, it is a purely electronic parameter, which is not influenced by the physical behaviour of the ions in the perovskite material. In order to justify this assertion we note that $\nel$ is obtained using only the high frequency impedance measurements, via equation (\ref{eq:n_id}). At high frequencies, the cell is perturbed about its steady-state ionic configuration and the ions  are effectively immobile, because they move too slowly to respond to the voltage oscillations. As a result, the perturbed potential is only dropped across the interior of the perovskite layer, to produce an oscillating internal electric field which only modulates the electron and hole densities (this is illustrated in figure \ref{fig:HF_impedance_diagram}). Even at these high frequencies, the electron and hole concentrations remain in quasi-equilibrium. As such, only HF impedance is capable of probing the electronic properties of a PSC about a particular steady-state. Although the drift-diffusion model of a PSC leads us to conclude, at least where there is only a single source of recombination, that $\nel$ is independent of applied voltage it is, from a practical perspective,  probably best practice to determine $\nel$ from experiments conducted at the maximum power point as this provides the best picture of the cell working under typical operating conditions.}

%%%%%%%%%%%%%%%%%%%%%%%%%

\subsection{Qualitative behaviour of the IS response}
%The following sections provide a more thorough examination of the HF and LF capacitances and resistances.

\begin{figure}[htbp] % trim=left botm right top  \fbox{ whole include graphics bit for box around it to crop} 
\includegraphics[trim=0cm 0cm 0cm 0cm, clip, width=0.75\textwidth]{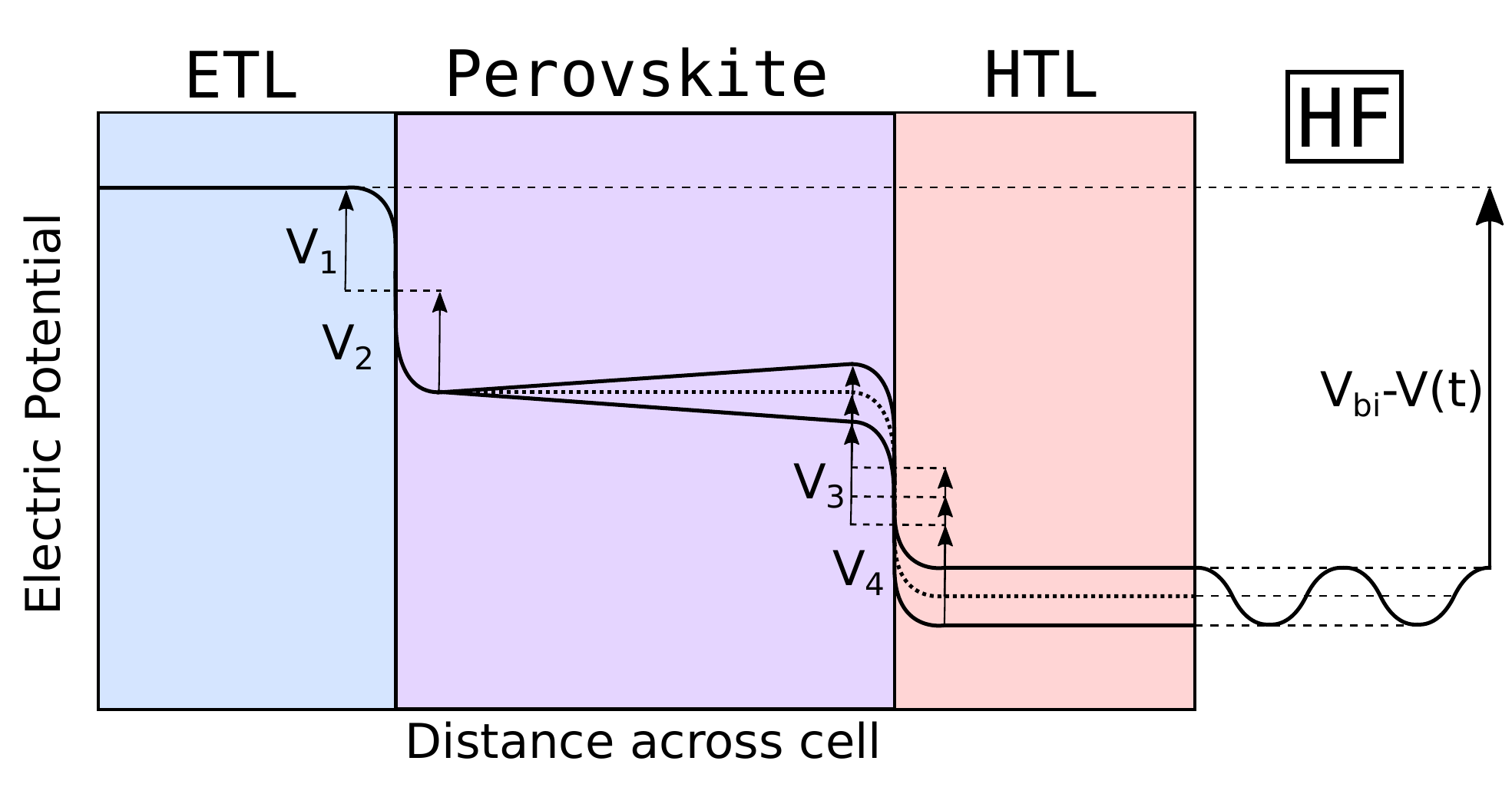}
\centering
\caption{Schematic of the potential distribution across a PSC during high frequency (above around 100 Hz) impedance measurements. The ion distribution is unable to adjust significantly over the short timescale of an impedance oscillation and so the potential drops $V_{1-4}$, across the
Debye layers, remain in their steady-state configuration. The time-dependent changes in applied potential thus lead to corresponding changes in electric field $E(t)$ within the perovskite layer.}
\label{fig:HF_impedance_diagram}
\end{figure}

\paragraph{High frequency feature.} \label{sec:HF}
The analytic and numerical results show a high frequency feature in the form of a semicircle above the axis on a Nyquist plot. This is consistent across all recombination types (light intensities and DC voltages) and matches that reported in experiment \cite{gonzalez2014general,pockett2017microseconds,contreras2019impedance,almora2018discerning}. A series resistance associated with the metal contacts and measurement apparatus is not observed \cite{bag2015kinetics}. 
{Unlike many other measurement techniques, high frequency impedance removes the transient effects of ion motion and leads to a response that is determined by the electronic properties of the PSC.} The typical evolution of the potential in a device, over a period of a high-frequency impedance measurement (greater than around 100Hz), is illustrated in figure \ref{fig:HF_impedance_diagram}. Further details of the HF response can be found in Riquelme \etal \cite{riquelme2020identification}.

The analytic model identifies the high frequency resistance (via \eqref{eq:RandC_HF}) as inversely proportional to the recombination current. This is in line with the usual interpretation of a recombination resistance \cite{pockett2015characterization, contreras2017origin}. 
In addition, it identifies the high frequency capacitance (see (\ref{eq:RandC_HF})) as a purely geometric capacitance \cite{almora2016mott,guerrero2016properties}, which is consistent with other reports found in the literature \cite{pockett2017microseconds,jacobs2018two,almora2016mott,pockett2015characterization, contreras2019impedance,zarazua2016surface, todinova2017towards,contreras2017origin}, and is a consequence of the contribution to the total current from the out-of-phase displacement current caused by polarisation of the perovskite layer, and given by (\ref{eq:j_d}).

\paragraph{Low frequency feature.} \label{sec:LF}
At low frequency, the possible impedance response of a PSC can be more varied. The analytic model shows that, depending upon the cell parameters, three different LF responses may be observed in the Nyquist plane. Either: (1) no visible LF feature; or (2) a `capacitive' semicircle above the axis; or (3) an `inductive' semicircle below the axis. These different LF features are displayed in Figures \ref{fig:R_combined_Nyquist_point1sun} and \ref{fig:R_combined_Nyquist_1Sun}. The time constant associated with the low frequency process is around 1-10 s, which is in line with experimental reports \cite{pockett2017microseconds,contreras2019impedance,wang2019kinetic}. 
%This is apparent on a Cole-Cole plot  plot 

\begin{figure}[htbp] % trim=left botm right top  \fbox{ whole include graphics bit for box around it to crop} 
\includegraphics[trim=0cm 0cm 0cm 0cm, clip, width=0.75\textwidth]{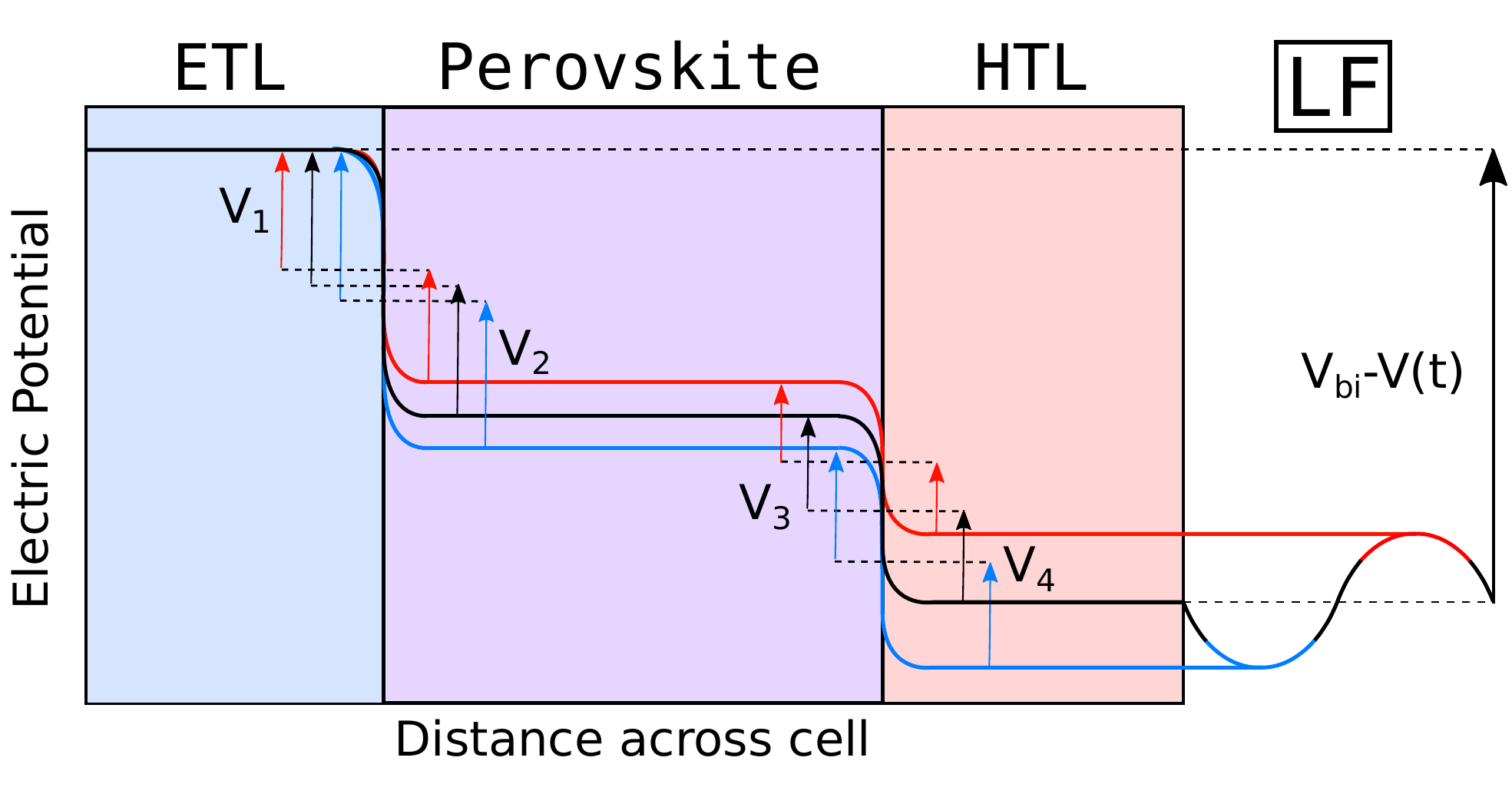}
\centering
\caption{Illustration of the effect of an oscillating potential difference on the distribution of the electric potential across a PSC during impedance measurements at ultra-low frequencies (below about 10 mHz). The slow modulation of applied voltage allows ionic charge to fill and deplete the perovskite Debye layers in-phase with the applied potential. The charging and discharging of Debye layers effectively screens the bulk electric field.}
\label{fig:LF_impedance_diagram}
\end{figure}

In the ultra-low frequency limit (below $\sim$ 10$^{-2}$ Hz), the modulation of the applied potential is so slow that the ionic distribution remains in approximate quasi-equilibrium throughout the perturbation and, as a result, the electric field within the perovskite layer is almost entirely screened (\ie $E(t) \approx 0$) from the interior of the perovskite layer, as illustrated in Figure \ref{fig:LF_impedance_diagram}. More generally, the LF response results in only partial screening of the electric field from the perovskite because the flow of charge, into and out of the Debye layers, lags behind the oscillating potential.  During a LF measurement the evolving potential drops across the Debye layers $V_{1-4}(t)$ modulate the recombination current (\ref{eq:j_rec_general}) and thus lead to an impedance response that is dependent on the properties of the ion motion, an interpretation which is in line with the discussions of IS in PSCs found in  \cite{pockett2017microseconds,jacobs2018two,moia2019ionic}. 

The forms of the relations for $R_{LF}$ and $C_{LF}$, given in (\ref{eq:RandC_LF}), provide insight into why certain features are observed in the IS response of a PSC. For example, when the recombination in a cell is dominated by bimolecular recombination, the low frequency resistance shrinks to zero (since in this scenario $\nap = \nel$ = 1), so that only a high frequency semicircle is observed in the Nyquist plot as, for example, shown in Figure \ref{fig:R_combined_Nyquist_point1sun}(c). While it is not realistic to entirely eliminate all other sources of recombination, this result is nevertheless interesting, since it demonstrates that the absence of a LF arc does not necessarily signify the absence of ion motion. Indeed the LF feature may also disappear where other types of recombination are present if $\nap = \nel$, see Table \ref{tab:negRandC_and_no_LF} for further details. 

The condition for the low frequency feature to appear `inductive' (\ie to lie below the axis on a Nyquist plot) is that both $R_{LF}<0$ and $C_{LF}<0$ and, as can be seen from (\ref{eq:RandC_LF}), this occurs where
\be
\nap < \nel. \label{eq:negRandCcondition}
\ee
More details about when such `inductive' arc can be expected to appear, for the particular types of recombination considered here, are provided in Table \ref{tab:negRandC_and_no_LF}. A notable result is that an `inductive' arc in the LF feature never occurs where the recombination occurs predominantly on one of the interfaces with the transport layers. Reframing this point, if a negative low frequency feature is observed in a Nyquist plot, we can infer from the model that the dominant recombination mechanism is bulk SRH within the perovskite layer.

\begin{table}[htbp]
\begin{center}
\begin{tabular}{ |c|c|c|c|}  
 \hline
  \textbf{Recombination} & \textbf{Conditions for no LF feature} &\textbf{ Can }$\boldsymbol{R_{LF}}$ \& $\boldsymbol{C_{LF}}$ \textbf{be negative?}  \\
 \hline 
  $R_b$: $R_{\textnormal{bulk}} = \beta np$ & always (since $\nap = \nel$) &  No ($R_{LF}=0, C_{LF}=\infty$) \\ 
 \hline
  $R_p$: $R_{\textnormal{bulk}} = p/\tau_p$  & if $V_1 + V_2 = V_3 + V_4$ &  Yes if $ V_1 + V_2 < V_3 + V_4$ \\ 
   \hline
  $R_n$:  $R_{\textnormal{bulk}} =  n/\tau_n$ & if $ V_1 + V_2 = V_3 + V_4$  &  Yes if $ V_3 + V_4 < V_1 + V_2$ \\ 
   \hline
  $R_l = v_{p_E}\pl$ & if $V_1 << V_2 + V_3 + V_4$ & No ($R_{LF} \geq 0, C_{LF}>0$) \\ 
   \hline
  $R_r = v_{n_H}\nr$ & if $V_4 << V_1 + V_2 + V_3$ &  No ($R_{LF} \geq 0, C_{LF}>0$)\\ 
  \hline
\end{tabular}
\end{center}
\caption{{Table showing the relationship between recombination mechanism and observed low frequency features. These conditions are derived using the inequality (\ref{eq:negRandCcondition}) and the expression for $\nap$ in \eqref{eq:ectypal_approx_def} and assume $\VDC$  is sufficiently close to $\Vbi$, see  \cite{courtier2020interpreting}. Table \ref{tab:negRandC_and_no_LF_gen} in the SI gives equivalent conditions where $\VDC$ lies further away from $\Vbi$.}} 
\label{tab:negRandC_and_no_LF}
\end{table}

{\subsection{General recombination mechanisms} \label{sec:general_recomb}}

{Up until now we have only considered scenarios where there is a single recombination mechanism. Whilst this is, in general, unrealistic there is usually a dominant form of recombination for any particular applied voltage $\VDC$, and therefore for any given impedance measurement. The results from one of these simple cases, with a single recombination mechanism, is therefore likely to give a good qualitative understanding of any particular impedance measurement conducted on a PSC. Nevertheless, it is possible to generalise our analytic model to cells with any combination of recombination mechanisms, the general expression for the HF and LF resistances and capacitances being given in \eqref{eq:RandC_HF_gen}-\eqref{eq:RandC_LF_gen}. We are therefore not restricted solely to the  forms found in Table \ref{tab:recomb}. However, we note that more general sets of recombination mechanisms lead to unwieldy equations for the HF and LF resistances and capacitances that are difficult to interpret.}

\begin{figure}[htbp] % trim=left botm right top  \fbox{ whole include graphics bit for box around it to crop}
\includegraphics[trim=3cm 11.2cm 4cm 11.5cm, clip,width=0.65\textwidth]{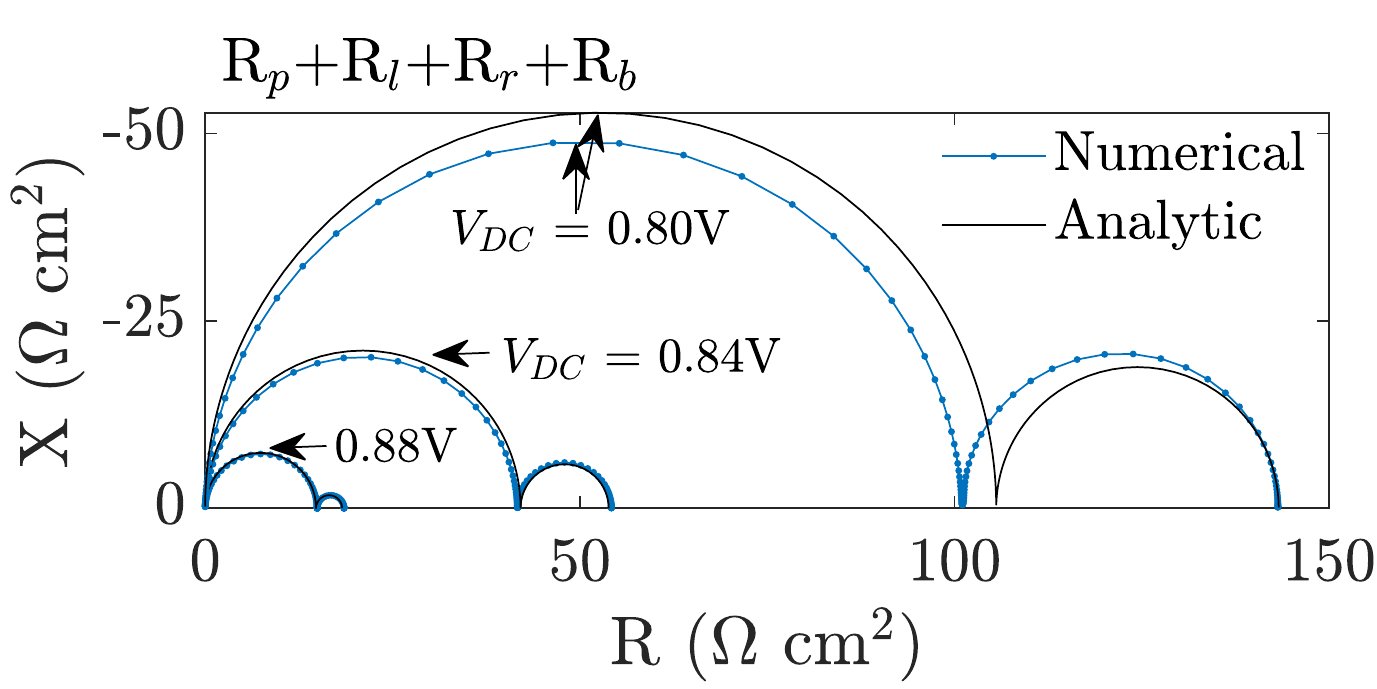}
\centering
\caption{Simulated impedance spectra for a PSC with four types of recombination under 0.1-Sun equivalent illumination. Spectra at three DC voltages are shown, including $\Voc$=0.88 V. Specifically, the recombination taking place is bimolecular ($R_b$) and hole-limited SRH ($R_p$) in the bulk and at both the ETL/perovskite ($R_l$) and perovskite/HTL ($R_r$) interfaces. Recombination parameters for each type are specified in Table \ref{tab:recomb} and the additional parameters are given in Table \ref{tab:params}. Figure \ref{fig:RallBode_point1sun} presents the corresponding frequency plot for this Nyquist diagram.}
\label{fig:RallNyquist_point1sun}
\end{figure}

{In order to illustrate the sort of behaviour that might be expected in a real cell we provide a representative example of a PSC with a combination of the recombination pathways given in Table \ref{tab:recomb}.  In particular, we assume the additive combination $R_b+R_p$ of bimolecular and hole-limited recombination in the perovskite, as given in Table  \ref{tab:recomb}, and the surface recombination pathways $R_l$ (on the ETL/perovskite interface) and $R_r$ (on the perovskite/HTL interface), again as given in  Table  \ref{tab:recomb}. The analytic results for this cell (as derived from equations \eqref{eq:RandC_HF_gen}-\eqref{eq:RandC_LF_gen}) are compared to the numerical solutions to the drift-diffusion model, in which the full recombination rates are used.  Figure \ref{fig:RallNyquist_point1sun} shows a comparison between the two approaches for this cell (\ie analytic model vs. drift-diffusion model) and demonstrates good agreement between the two approaches across the full frequency range. The corresponding frequency plots are  presented in Figure \ref{fig:RallBode_point1sun}. As expected, this combination of different recombination mechanisms leads to a decrease in open-circuit voltage ($\Voc$=0.88 V for $R_p+R_l+R_r+R_b$ as compared to $\Voc$=0.93 V with just $R_l$).} 

For a cell with multiple recombination pathways the interpretation of the high and low frequency features remains the same. However, interpretation of the electronic ideality factor and the apparent ideality factor, calculated using equations \eqref{eq:n_id} and \eqref{eq:n_ec_from_IS}, respectively, is a little more complicated. We show  in the SI, in \eqref{nel-mult}, that for a cell with multiple recombination pathways, the electronic ideality factor is given by
\be
\nel = \frac{2}{2-r_{\textnormal{SRH}}} \label{eq:n_id_general}
\ee
where $r_{\textnormal{SRH}}$ is the ratio of the SRH recombination current to the total recombination current. Therefore, the electronic ideality factor, calculated from an impedance spectrum using (\ref{eq:n_id}), lies close to 1 when SRH recombination is negligible relative to interfacial (or bimolecular) recombination. Correspondingly, a value for the electronic ideality factor that is close to 2 indicates that the dominant form of recombination is SRH in the perovskite. In the following section we compare the values of the electronic ideality factor to those of the apparent ideality factor for various types of recombination.

%%%%%%%%%%%%%%%%%%%%%%%%%%%%%%%%%%%%%%%%%%%%%%%%%%%%%%%%%%

\subsection{Comparison between electronic and apparent ideality factors} \label{sec:n_id_n_ec_comp}
{On referring to (\ref{eq:RandC_HF})-(\ref{eq:RandC_LF}) it is clear that apparent ideality factor can be written in the form
\be
 \nap(\VDC) = \frac{\jrec (\VDC)}{V_T}\bigg(R_{HF}(\VDC) + R_{LF}(\VDC)\bigg). \label{eq:n_ec_from_IS} \label{nap}
\ee
which gives a method for obtaining the apparent ideality factor $\nap$ from IS data without having to determine the gradient of a linear fitted function (as is required by the Suns-$\Voc$, dark-$JV$ and $R_{HF}-\Voc$ techniques). Notably, once $\nap$ has been determined, it can be used to estimate the potential barrier for recombination $F_i$ at steady state by inverting the formula \eqref{eq:ectypal_approx_def} to obtain
\be
F_i|_{\VDC} \approx \frac{\vbi-\VDC}{\nap(\VDC)},
\ee
see \cite{courtier2020interpreting} for further details.}

\begin{figure}[htbp] % trim=left botm right top  \fbox{ whole include graphics bit for box around it to crop}
\includegraphics[trim=4.0cm 10cm 4.5cm 10cm, clip, width=0.55\textwidth]{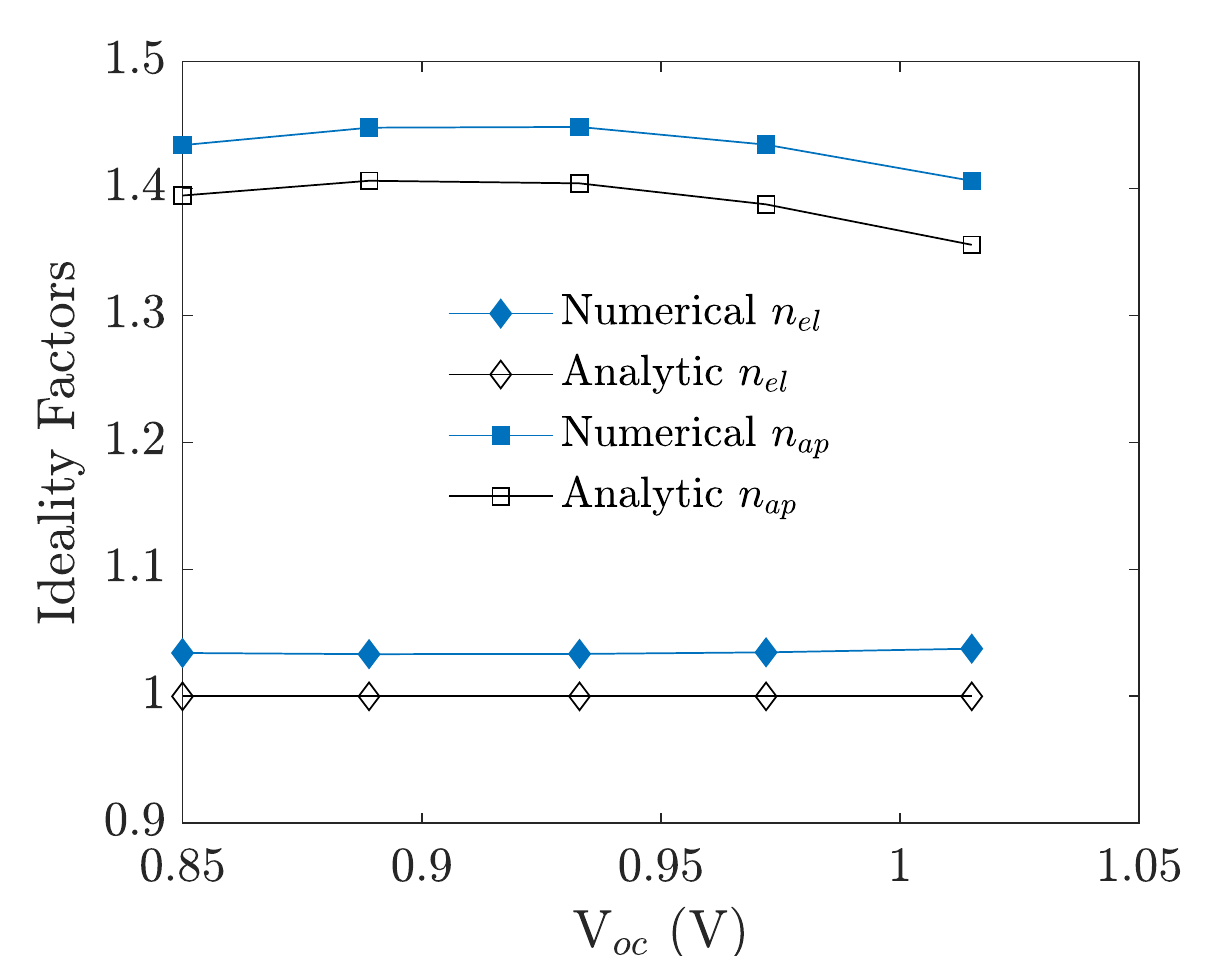}
\centering
\caption{Electronic ideality factor, $\nel$, and apparent ideality factor, $\nap$, calculated at different open-circuit voltages from impedance spectra obtained analytically and numerically. Calculated from the same spectra as used for Figure \ref{fig:V_oc_trends_combined}. Parameters used to calculate the numerical and analytic spectra are those from Table 1 for a cell with hole-limited interfacial recombination ($R_l$).} \label{fig:n_el_and_n_ec_trends}
\end{figure}

{Plots of the electronic ideality factor $\nel$ and the apparent ideality factor $\nap$ against open-circuit voltages $\voc$ (corresponding to different illumination intensities) are shown in Figure \ref{fig:n_el_and_n_ec_trends}. These are computed both from the analytic model (\ie from equations \eqref{eq:RandC_HF}-\eqref{eq:RandC_LF},  \eqref{nel} and \eqref{eq:n_ec_from_IS}) and from impedance spectra generated by numerical solutions of the PSC drift-diffusion model. The latter is accomplished by first obtaining $R_{LF}$, $R_{HF}$, $C_{LF}$ and $C_{HF}$ from the numerical spectra (as described in the discussion around equation \eqref{capac}), and $\jrec$ by using \eqref{eq:j_rec0_from_measurable_currents}, before using  \eqref{nel} to determine $\nel$, and \eqref{eq:n_ec_from_IS} to determine $\nap$. The results show the voltage independence of  $\nel$, for a single dominant source of recombination and, additionally, demonstrates that a non-integer value of $\nap$ is obtained, even in cells with a single (monomolecular) recombination mechanism. This supports the interpretation of $\nap$  as an apparent ideality factor rather than a true ideality factor.}

\begin{figure}[htbp] % trim=left botm right top  \fbox{ whole include graphics bit for box around it to crop}
\includegraphics[trim=2cm 9.5cm 2.5cm 9.4cm, clip, width=0.6\textwidth]{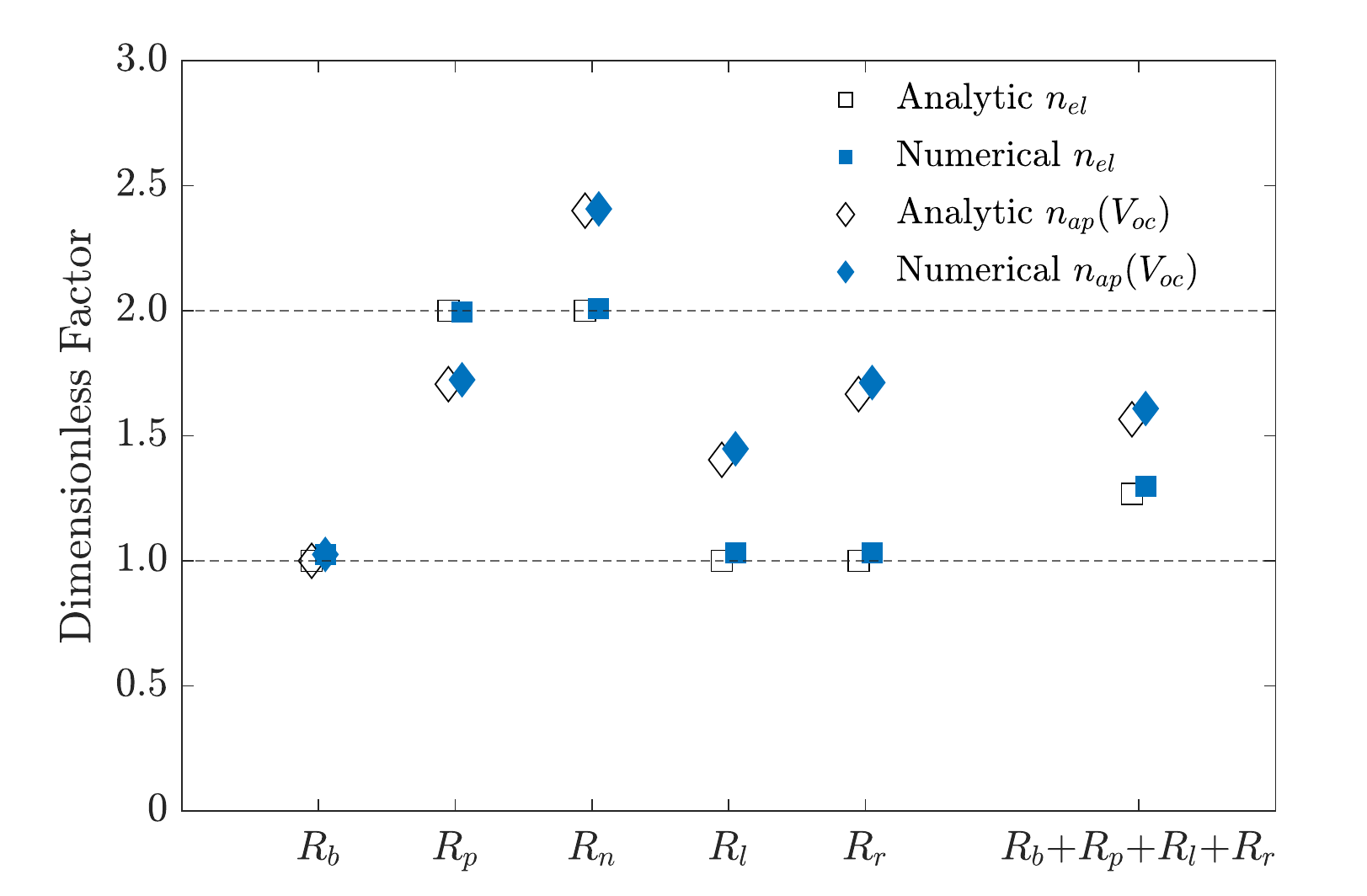}
\centering
\caption{Electronic ideality factor and the apparent ideality factor calculated from impedance spectra at open-circuit for the five recombination types considered in this work and cell parameters from Table \ref{tab:params}. The rightmost entry is for multiple recombination mechanisms, as displayed in Figure \ref{fig:RallNyquist_point1sun}.}
\label{fig:n_id_n_ec_scatter}
\end{figure}

{In order to highlight the difference between the properties of the apparent and electronic ideality factors we compute these quantities under 0.1-Suns illumination and at $\voc$. We compute these quantities both directly, from the drift-diffusion model, and indirectly, from the analytic model (via \eqref{eq:ectypal_approx_def} and \eqref{eq:n_id_general}), doing so for the five different forms of recombination detailed in Table \ref{tab:recomb} and for the cell described in \S\ref{sec:general_recomb} with multiple recombination mechanisms. The results of these computations are displayed in figure \ref{fig:n_id_n_ec_scatter}. We remark that while $\nap$ for this cell  lies within the range that typically would be expected for an ideality factor (\ie mainly between 1 and 2)  it is highly sensitive to device parameters (such as ion density or transport layer doping) and for certain cell parameter sets $\nap$ would lie well outside this range.} It is clear from the figure that the factors calculated from the drift-diffusion model closely match those predicted by the analytic model (at $\voc$). Even where only a single source of recombination is present the apparent ideality factor $\nap$ is non-integer (except for purely bimolecular recombination). This highlights the challenge of attributing  a particular form of recombination to a value of $\nap$. In contrast, however, where a single recombination mechanism dominates, the electronic ideality factor $\nel$ is an integer and its value can be used to distinguish between between bulk SRH recombination $\nel=2$ and interfacial recombination $\nel=1$. For multiple recombination mechanisms, the electronic ideality factor quantifies the proportion of bulk SRH to interfacial (and bimolecular) recombination via equation (\ref{eq:n_id_general}). If the size/proportion of some of the potentials $V_{1-4}$ are known it is then theoretically possible, by pairing this information about $\nel$ and $\nap$ (using equation \eqref{eq:ectypal_approx_def}),  to diagnose the exact form and location of the recombination which is limiting cell performance.

{Further information on interpretting impedance spectra is to be found in \S \ref{sec:reconstructingIS} with examples on how to apply the formulae (\ref{eq:RandC_HF})-(\ref{eq:RandC_LF}) in specific scenarios.}

%%%%%%%%%%%%%%%%%%%%%%%%%%%%%%%%%%%%%%%%%%%%%%%%%%%%%%%%%%%%%%%%%%%
%%%%%%%%%%%%%%%%%%%%%%%%%%%%%%%%%%%%%%%%%%%%%%%%%%%%%%%%%%%%%%%%%%%
%%%%%%%%%         CONCLUSIONS
%%%%%%%%%%%%%%%%%%%%%%%%%%%%%%%%%%%%%%%%%%%%%%%%%%%%%%%%%%%%%%%%%%%
%%%%%%%%%%%%%%%%%%%%%%%%%%%%%%%%%%%%%%%%%%%%%%%%%%%%%%%%%%%%%%%%%%%

\section{Conclusions}

In this work we have derived an approximate analytic model (\ie one based on a set of transcendental equations as opposed to a set of differential equations)  for the impedance response of a PSC from a commonly used drift-diffusion model of such devices, which includes the effects of halide ion vacancies  and charge carrier motion  (see for example, \cite{calado2016,calado2020,courtier2019transport,moia2019ionic,neukom2017,neukom2019,richardson2016can}). We have shown excellent agreement between the solutions of the analytic model and the drift-diffusion model for impedance simulations conducted on a physically realistic set of parameters at $\voc$ and very good agreement at the maximum power point. It is significant that no fitting is required in order to obtain this agreement between the analytic model and the drift-diffusion model; they only require that the same set of physical parameters are used in both. It is important to realise that there are some physical limitations to the analytic model that we have derived. The chief amongst these is that it is reliant on the charge carriers being close to quasi-equilibrium, so that in this particular instance they are close to being Boltzmann distributed. This approximation works well if there are sufficient carriers in the device to easily transport the current being extracted from it. In practice, this means that it works very well close to $\voc$ but breaks down as the device is brought towards short-circuit, where there is little resistance to flow of out of the perovskite absorber layer.

Simulations using both the analytic and the drift-diffusion model show good qualitative agreement with experimental IS studies and are able to predict some of the surprising features that are often seen in the literature, such as `inductive' low frequency arcs (which appear below the axis in Nyquist plots) and `giant low frequency capacitance`. The analytic model that we have derived has major advantages over the two standard approaches to fitting IS data when used as a tool to interpret IS experiments. These standard approaches  are to either: (A) fit to an equivalent circuit (\eg an RC-RC circuit) or (B) to fit directly to a drift-diffusion model \cite{calado2020,moia2019ionic,neukom2019}. The first of these approaches, (A), has the major disadvantage that the elements in the equivalent circuit cannot be related to real physical properties of the device, so that the fitting process becomes no more than an efficient way of capturing the shape of the data with a minimal number of parameters. Furthermore, the fitting often leads to physically infeasible circuit elements with, for example, negative resistances or capacitances. While the second approach, (B), is much more physically appealing, since it allows the results to be related to a physics-based model with parameters that can be understood in terms of the device's physics, it is nevertheless problematic: firstly, because the computational cost of simulating a single impedance experiment with a drift-diffusion model, with given  parameters, is non-trivial, so that the many simulations required to arrive at a set of physical parameters (for the drift-diffusion model) that fits even a single IS experiment is extremely costly; and secondly, because there are too many unknown parameters in the drift-diffusion model to obtain a unique fit to the impedance data from an IS experiment conducted at a specified voltage. In contrast, the analytic model that we have derived is physics-based, is easy to fit to data, and contains only five fitting parameters, which can be directly related to the physical parameters in the drift-diffusion model, although they typically vary with the steady-state applied voltage $\VDC$ about which the IS experiment is conducted. These five parameters are the high- and low-frequency resistances and capacitances  $R_{HF}$, $C_{HF}$, $R_{LF}$ and $C_{LF}$, measured by the IS experiment, and the recombination current $j_{rec}$.  From a practical perspective the impedance data is fitted to an RC-RC circuit to obtain  $R_{HF}(\VDC)$, $C_{HF}(\VDC)$, $R_{LF}(\VDC)$ and $C_{LF}(\VDC)$, while $j_{rec}(\VDC)$ is determined by fitting to the formula  \eqref{jrec}. Having determined these five quantities equations \eqref{eq:RandC_HF}-\eqref{eq:RandC_LF} can be used to obtain the following physical properties of the cell: $\nel$, the electronic ideality factor; $\nap(\VDC)$  the apparent ideality factor; $(1/G_+) (-d Q_+/d \VDC)$, which is the ionic capacitance of the cell (per unit area) divided by the ionic conductance of the perovskite layer (per unit area); and $\varepsilon_p/b$ the geometric capacitance of the cell (per unit area). The fifth physical property of the cell is the recombination current $\jrec$, which has already been obtained in the fitting process.

From a practical perspective probably the key result of this work is the identification of a new ideality factor within a PSC. This can be determined from the high frequency resistance and the recombination current via the relation (see \eqref{nel})
\be
\nel = \frac{R_{\textnormal{HF}}(\VDC)\jrec(\VDC)}{V_T},
\ee
and has the property that, provided the dominant source of recombination does not alter as the steady-state applied voltage $\VDC$ is changed, it is independent of $\VDC$. This is a consequence of it depending purely on electronic (as opposed to ionic) parameters. This is in stark contrast to the standard form of the ideality factor, which is determined from experiments such as Suns-$\voc$, and which we term here the apparent ideality factor $\nap$. This varies both with changes in $\VDC$, and (rather strongly) with changes in ionic parameters such as the ion concentration. Crucially, where there is a single dominant source of recombination loss within the cell, $\nec$ takes an integer value. In particular, if this dominant loss mechanism occurs via SRH recombination within the perovskite layer then $\nel=2$, whereas if the dominant loss mechanism occurs via  interfacial recombination on one of the interfaces between the perovskite and transport layers then $\nel=1$.

The analytic model is not capable of capturing the small `intermediate' frequency feature that is sometimes observed in both experiment and in solutions to the drift diffusion model. This, for example, takes the form of a bulge between the positive or negative LF feature and can be seen at higher illuminations in Figure \ref{fig:R_combined_Nyquist_1Sun}. Whilst not observed with the parameters used in this study, drift-diffusion simulations and experiments have also been reported that show a small loop in the Nyquist plane where the high and low frequency semicircles meet \cite{riquelme2020identification,jacobs2018two}. A wide range of IF features have been reported in the literature for experimental IS and while the intermediate frequency feature is interesting, it is not universally observed and even when it is, is much less significant than either the low or high frequency features \cite{contreras2019impedance,gonzalez2014general, pockett2017microseconds,almora2018discerning}.

Calculating the geometric capacitance from the perovskite permittivity and perovskite width returns capacitances up to 10-100 times less than those extracted experimentally. In order to reconcile this discrepancy between the geometric capacitance predicted by the thickness and permittivity of the perovskite absorber layer and experimentally measured value of $C_{HF}$ a roughness factor has been proposed which accounts for the non-planar (rough) nature of the perovskite transport layer interfaces \cite{pockett2015characterization, riquelme2020identification}. The low values of $C_{HF}$ obtained from our simulations (in both numerical and analytic impedance spectra) are reflected in the frequency plots, which show that the HF peak is shifted to slightly higher frequencies than is measured in experiment \cite{pockett2017microseconds,contreras2019impedance}. 

Finally, we remark on the generality of the approach that has been adopted to arrive at the analytic model. Although we have only applied this model to the standard perovskite drift-diffusion model of a PSC, with mobile charge carriers and a single mobile ions species, it is easily extended to other drift-diffusion type models that might be applied to such cells. For example, it is currently believed that a second very slow mobile ion species may play a significant role in PSC physics and, in \cite{domanski2017}, an appropriate PSC drift-diffusion model is formulated that encapsulates this mechanism. There is also speculation that high doping in the 
transport layers  may significantly alter the statistics of the charge carriers in these regions, so that rather than being Boltzmann distributed they may obey a Gauss-Fermi distribution (organic layers) or a Fermi-Dirac distribution (inorganic layers). In both of these scenarios the resulting drift-diffusion model of the PSC changes but it will still be possible to apply the techniques used here to arrive at an approximate analytic model for the IS response. Indeed it should be noted that the theory described here can either be applied directly, or in a slightly modified form, to any device based on mixed ionic/electronic semiconductors.

\paragraph{Acknowledgements}  LJB is supported by an EPSRC funded studentship from the CDT in New and Sustainable Photovoltaics, reference EP/L01551X/1. NEC was supported by an EPSRC Doctoral Prize (ref. EP/R513325/1). JAA thanks Ministerio de Ciencia e Innovación of Spain, Agencia Estatal de Investigación (AEI) and EU (FEDER) under grants PID2019-110430GB-C22 and
PCI2019-111839-2 (SCALEUP) and Junta de Andalucia under grant SOLARFORCE (UPO-1259175). AJR thanks the Spanish Ministry of Education, Culture and Sports for its supports via a PhD grant (FPU2017-03684).

\newpage

\bibliographystyle{abbrv}
\bibliography{IS_bibliog}

%%%%%%%%%%%%%%%%%%%%%%%%%%%%%%%%%%%%%%%%%%%%%%%%%%%%%%%%%%%%%%%%%%%
%%%%%%%%%%%%%%%%%%%%%%%%%%%%%%%%%%%%%%%%%%%%%%%%%%%%%%%%%%%%%%%%%%%
%%%%%%%%%%%%%%%%%%%%%%%%%%%%%%%%%%%%%%%%%%%%%%%%%%%%%%%%%%%%%%%%%%%
%%%%%%%%%%%%%%%%%%%%%%%%%%%%%%%%%%%%%%%%%%%%%%%%%%%%%%%%%%%%%%%%%%%
%%%%%%%%%%%%%%%%%%%%%%%%%%%%%%%%%%%%%%%%%%%%%%%%%%%%%%%%%%%%%%%%%%%
%%%%%%%%%%%%%%%%%%%%%%%%%%%%%%%%%%%%%%%%%%%%%%%%%%%%%%%%%%%%%%%%%%%
%%%%%%%%%         APPENDICES
%%%%%%%%%%%%%%%%%%%%%%%%%%%%%%%%%%%%%%%%%%%%%%%%%%%%%%%%%%%%%%%%%%%
%%%%%%%%%%%%%%%%%%%%%%%%%%%%%%%%%%%%%%%%%%%%%%%%%%%%%%%%%%%%%%%%%%%
%%%%%%%%%%%%%%%%%%%%%%%%%%%%%%%%%%%%%%%%%%%%%%%%%%%%%%%%%%%%%%%%%%%
%%%%%%%%%%%%%%%%%%%%%%%%%%%%%%%%%%%%%%%%%%%%%%%%%%%%%%%%%%%%%%%%%%%%%%%%%%%%%%%%%%%%%%%%%%%%%%%%%%%%%%%%%%%%%%%%%%%%%%%%%%%%%%%%%%%%%%
%%%%%%%%%%%%%%%%%%%%%%%%%%%%%%%%%%%%%%%%%%%%%%%%%%%%%%%%%%%%%%%%%%%

\appendix

% Below should be in sty file but we have it here for now. This is all to start equation numbering again and add the appendix label prefix for the equations and figures. Note the equation hyperlinks do not work when they point to the appendix
\setcounter{equation}{0}
\setcounter{figure}{0}
\renewcommand{\theequation}{\thesection.\arabic{equation}}
\renewcommand{\thefigure}{\thesection.\arabic{figure}}

% ---------------------------------------- SI ---------------------------------------- % 

\newpage

{\LARGE \textbf{Supplementary Information}}
\section{Symbols and acronyms} \label{sec:SI}

\small
\begin{center}
\begin{longtable}{|lll|}
\caption{List of symbols used in this work. Cell parameters and their values are displayed in Table \ref{tab:params} in the main text.} \label{tab:symbols} \\
\hline \multicolumn{1}{|l}{\textbf{Symbol}} & \multicolumn{1}{l}{\textbf{Definition}} & \multicolumn{1}{l|}{\textbf{Unit}} \\ \hline 
\endfirsthead

\multicolumn{3}{c}%
{{\bfseries \tablename\ \thetable{} -- continued from previous page}} \\
\hline \multicolumn{1}{|l}{\textbf{Symbol}} & \multicolumn{1}{l}{\textbf{Definition}} & \multicolumn{1}{l|}{\textbf{Unit}} \\ \hline 
\endhead

% \hline \multicolumn{3}{|r|}{{Continued on next page}} \\ \hline
% \endfoot

\hline %\hline
\endlastfoot
$A$         &  Cell area                                            & m\super{2}                \\
$B$         &  Susceptance                                          & $\Omega$\super{-1}m\super{-2} \\
$C_{HF}$    &  High frequency capacitance                           & Fcm\super{-2}             \\ 
$C_{LF}$    &  Low frequency capacitance                            & Fcm\super{-2}             \\   % Fcm$^{-2}$
$E$         &  Uniform bulk electric field                          & Vm\super{-1}              \\
$f$         &  Frequency                                            & s\super{-1} (Hz)          \\
$F_i$       &  Potential barrier to recombination                   & V                         \\
$F_T$       &  Total potential drop across the Debye layers         & V                         \\
$G$         &  Conductance                                          & $\Omega$\super{-1}m\super{-2} \\
$G_{bulk}$  &  Bulk generation rate                                 &  m\super{-3}s\super{-1}   \\ 
$G_+$       &  Ionic conductance per unit area                      & AV\super{-1}m\super{-2}   \\
%$i$         &  Imaginary unit                                       & Dimensionless             \\
$J$         &  Current density                                      & Am\super{-2}              \\
$j_{d}$     &  Displacement current density                         & Am\super{-2}              \\
$j_{n}$     &  Electron current density                             & Am\super{-2}              \\
$j_{p}$     &  Hole current density                                 & Am\super{-2}              \\
$j_{P}$     &  Ionic current density                                & Am\super{-2}              \\
$j_{rec}$   &  Recombination current density                        & Am\super{-2}              \\
$j_{R_i}$   &  Recombination current prefactor                      & Am\super{-2}              \\
$j_s$       &  Photocurrent density                                 & Am\super{-2}              \\
$k_{1-2}$   &  Deep trap constants for bulk SRH recombination       & sm\super{-3}              \\
$k_{3-4}$   &  Deep trap constants for interfacial recombination    & sm\super{-4}              \\
$k_E$       &  Ratio of electron density in the perovskite to that in the ETL& Dimensionless    \\
$k_H$       &  Ratio of hole density in the perovskite to that in the HTL& Dimensionless        \\
$L_D$       &  Debye length                                         & m                         \\
$n$         &  Electron density within the perovskite layer         & m\super{-3}               \\
$\nap$      &  Apparent ideality factor   ($\nap = \nmec$)          & Dimensionless             \\
$\nec$      &  True ectypal factor                                  & Dimensionless             \\
$\nmec$     &  Measured ectypal factor                              & Dimensionless             \\
$n_i$       &  Intrinsic carrier density within the perovskite      & m\super{-3}               \\
$\nel$      &  Electronic ideality factor                           & Dimensionless             \\
$\nl$       &  Electron density within the ETL at the ETL/perovskite interface   & m\super{-3}  \\
$\nr$       &  Electron density within the perovskite at perovskite/HTL interface & m\super{-3} \\
$p$         &  Hole density within the perovskite                   &  m\super{-3}              \\
$\pl$       &  Hole density within the perovskite at the ETL/perovskite interface  & m\super{-3}\\
$\pr$       &  Hole density within the HTL at the perovskite/HTL interface & m\super{-3}        \\
$Q$         &  Ionic surface charge density                         &  Cm\super{-2}             \\
$\QDC$      &  Steady-state or DC ionic surface charge density      &  Cm\super{-2}             \\
$R$         &  Resistance                                           & $\Omega$m\super{2}        \\
$R_b$       &  Bimolecular recombination rate                       &  m\super{-3}s\super{-1}   \\
$R_{bulk}$  &  Bulk recombination rate                              &  m\super{-3}s\super{-1}   \\ 
$R_l$       &  ETL/perovskite interfacial SRH recombination rate    &  m\super{-2}s\super{-1}   \\
$R_n$       &  Electron-limited bulk SRH recombination rate         &  m\super{-3}s\super{-1}   \\ 
$R_p$       &  Hole-limited bulk SRH recombination rate             &  m\super{-3}s\super{-1}   \\ 
$R_r$       &  Perovskite/HTL interfacial SRH recombination rate    &  m\super{-2}s\super{-1}   \\
$R_s$       &  Series resistance                                    & $\Omega$m\super{2}        \\
$R_{HF}$    &  High frequency resistance                            & $\Omega$m\super{2}        \\
$R_{LF}$    &  Low frequency resistance                             & $\Omega$m\super{2}        \\
$r_{\textnormal{SRH}}$ &  Ratio of SRH recombination current to total recombination current & Dimensionless   \\
$t$         &  Time                                                 & s                         \\
$V_{1}$     &  Potential drop across left DL within the ETL         & V                         \\
$V_{2}$     &  Potential drop across left DL within the Perovskite  & V                         \\
$V_{3}$     &  Potential drop across right DL within the Perovskite & V                         \\
$V_{4}$     &  Potential drop across right DL within the HTL        & V                         \\
$\Vbi$    &  Built-in voltage                                     & V                         \\
$\VDC$    &  Steady-state or DC voltage                           & V                         \\
$\Voc$    &  Open-circuit voltage                                 & V                         \\
$V_{p}$     &  Perturbation amplitude                               & V                         \\
$v_{n_E}$   &  Electron recombination velocity for SRH at ETL/perovskite interface& ms\super{-1}\\
$v_{n_H}$   &  Electron recombination velocity for SRH at perovskite/HTL interface& ms\super{-1}\\
$v_{p_E}$   &  Hole recombination velocity for SRH at ETL/perovskite interface    & ms\super{-1}\\
$v_{p_H}$   &  Hole recombination velocity for SRH at perovskite/HTL interface    & ms\super{-1}\\
$V(t)$      &  Applied voltage                                      & V                         \\
$x$         &  Spatial parameter                                    & m                         \\
$X$         &  Reactance                                            & $\Omega$m\super{2}        \\
$\beta$     &  Bimolecular recombination rate in the bulk           & m\super{3}s\super{-1}     \\
$\delta$    &  Ratio of perturbation amplitude to thermal voltage   & Dimensionless             \\
$\omega$    &  Angular frequency                                    & rad s\super{-1}           \\
$\omega_{HF}$ &  Characteristic high frequency                      & rad s\super{-1}           \\
$\omega_{LF}$ &  Characteristic low frequency                       & rad s\super{-1}           \\
$\Omega_E$  &  ETL doping and permittivity parameter from the SPM   & Dimensionless             \\
$\Omega_H$  &  HTL doping and permittivity parameter from the SPM   & Dimensionless             \\
$\tau_n$    &  Electron pseudo-lifetime for SRH recombination       & s                         \\
$\tau_p$    &  Hole pseudo-lifetime for SRH recombination           & s                         \\ 
$\mathcal{V}$& Capacitance relation from the surface polarisation model & V                     \\
\end{longtable}
\end{center}
\normalsize 

\newpage
\small
\begin{center}
\begin{longtable}{|ll|}
\caption{List of acronyms used in this work} \label{tab:acronyms} \\
\hline \multicolumn{1}{|l}{\textbf{Acronym}} & \multicolumn{1}{l|}{\textbf{Definition}} \\ \hline 
\endfirsthead
\multicolumn{2}{c}%
{{\bfseries \tablename\ \thetable{} -- continued from previous page}} \\
\hline \multicolumn{1}{|l}{\textbf{Acronym}} & \multicolumn{1}{l|}{\textbf{Definition}}  \\ \hline 
\endhead
% \hline \multicolumn{3}{|r|}{{Continued on next page}} \\ \hline
% \endfoot
\hline %\hline
\endlastfoot
AM          &  Air mass                                                                        \\
DC          &  Direct current                                                                  \\
DL          &  Debye layer                                                                     \\
ETL         &  Electron transport layer                                                        \\
HF          &  High frequency                                                                  \\
HTL         &  Hole transport layer                                                            \\
IS          &  Impedance spectroscopy                                                          \\
LF          &  Low frequency                                                                   \\
MA          &  Methylammonium                                                                  \\
MPP         &  Maximum power point                                                             \\
ODE         &  Ordinary differential equation                                                  \\
PSC         &  Perovskite solar cell                                                           \\
$r_{\textnormal{SRH}}$ &  Ratio of SRH recombination current to total recombination current    \\
Spiro       &  2,2',7,7'-Tetrakis[N,N-di(4-methoxyphenyl)amino]-9,9'-spirobifluorene           \\
SPM         &  Surface polarisation model                                                      \\
SRH         &  Shockley-Read-Hall                                                              \\
TL          &  Transport layer                                                                 \\
\end{longtable}
\end{center}
\normalsize

% ------------------------------------ Derivation ------------------------------------- % 
\newpage
\section{Full derivation}  \label{sec:derivation}
\setcounter{equation}{0}
\setcounter{figure}{0}

We present the full derivation of the analytic model. In this work, the chosen model for the ion dynamics is the \textit{surface polarisation model} \cite{courtier2019transport,courtier2018systematic,richardson2016can}. Here, we provide a brief description of the surface polarisation model, followed by the drift-diffusion equations within the perovskite bulk and the Boltzmann approximation to the carrier densities. The total and recombination current is then calculated. Lastly, these current equations are linearised and the impedance parameters extracted.

% Explain SPM and that it provides a good approx to DD equations
\paragraph{The surface polarisation model}
Courtier \textit{et al.} note that the unique properties of PSCs enable the following assumptions to be made: \cite{courtier2018systematic,courtier2019transport}

\begin{itemize}
    \setlength\itemsep{0.0em}
    \item The Debye lengths in the perovskite and transport layers are narrow with respect to their total widths. 
    \item The ionic vacancy density is high in relation to the electronic carrier density under normal operation (i.e up to and around 1 sun).
    \item The ionic vacancy distribution in the perovskite reaches equilibrium much slower than the electrons and holes in their respective transport layers.
\end{itemize}
With these assumptions, Courtier \textit{et al.} establish a \textit{surface polarisation model} in which they find: 
\begin{itemize}
    \setlength\itemsep{0.0em}
    \item The electronic carriers have a very limited effect on the electric potential. Therefore, to a good approximation, the ionic vacancies set the electric potential across the device.
    \item The electric potential is linear in the perovskite bulk and flat in the transport layers, except for the narrow Debye layers.
    \item The ionic charge $Q$ in the right perovskite Debye layer is balanced by an equal and opposite Debye layer charge (-$Q$) at the left perovskite Debye layer.
    \item The Debye layers within the selective contacts contain an equal and opposite charge to that contained within the adjacent perovskite Debye layer.    
    \item The potential drops $V_{1-4}$ across the device are set set via a non-linear capacitance relation. See equation eq.(\ref{eq:charge}) and Figure \ref{fig:V1-4}.
    \item The uniform electric field in the bulk is proportional to the rate of change of Debye layer charge.
\end{itemize}

The Debye layer (surface) charge $Q$ is given by the following ODE
\be \label{eq:dQdt3L}
\frac{dQ}{dt} = \frac{qD_+N_0}{V_Tb} \bigg(\Vbi - V(t) - V_1(Q(t)) - V_2(Q(t))  - V_3(Q(t))  - V_4(Q(t)) \bigg),
\ee
where % $\mathcal{V}_{1-4} = V_{1-4}/V_T$, and
\begin{align}
    V_1(Q) =  - \mathcal{V}(-\Omega_E Q), \hspace{2mm}   V_2(Q) =  - \mathcal{V}(- Q), \hspace{2mm} V_3(Q) =   \mathcal{V}( Q),  \hspace{2mm}   V_4(Q) =  - \mathcal{V}(-\Omega_H Q). \tag{\ref{eq:VsandOmegas} reprinted} % \label{eq:V4}
\end{align}

The potential drops across the four Debye layers of a PSC, V$_{1-4}$, are labelled in Figure \ref{fig:V1-4}.
The dimensionless parameters $\Omega_E$ and $\Omega_H$ set the relative magnitudes of $V_1$ and $V_4$ respectively and are given by 
\bet
\Omega_E = \sqrt{ \frac{\varepsilon_pN_0}{\varepsilon_Ed_E} }, \hspace{2mm} \Omega_H = \sqrt{ \frac{\varepsilon_pN_0}{\varepsilon_Hd_H} }.  \tag{\ref{eq:VsandOmegas} reprinted}
\eet
The capacitance relation, $\mathcal{V}(Q)$, is plotted in Figure \ref{fig:VofQ} and can be found in \cite{courtier2018systematic}. Here we quote its inverse which is given by 
\bet
Q(\mathcal{V}) = \sqrt{qN_0\varepsilon_p V_T}  \textnormal{ sign}(\mathcal{V})\sqrt{2}\left( e^{\mathcal{V}/V_T} - \mathcal{V}/V_T-1 \right)^{\frac{1}{2}},  \tag{\ref{eq:charge} reprinted} %\label{eq:charge}
\eet
The bulk electric field is related to the Debye layer charge via
\begin{equation} \label{eq:E=dQdt}
E(t) = \frac{V_T}{qD_+N_0} \frac{dQ}{dt}.  % = -\frac{\partial \phi}{\partial x}
\end{equation}
The perovskite bulk refers to the perovskite layer excluding its Debye layers. Within the perovskite bulk the electric field is uniform. For a specified applied potential $V(t)$, equation (\ref{eq:dQdt3L}) can be solved to determine the electric potential across the cell in time. It is reasonable to only consider the bulk of the perovskite as the generation and recombination in these layers is negligible compared to the bulk. The impact of the steep variation in potential across the Debye layers on charge carriers is accounted for via the boundary conditions.

\paragraph{Drift-diffusion equations}
The drift-diffusion equations within the perovskite bulk ($0 < x < b$) are given by
\begin{align}
- \frac{1}{q}\frac{\partial j_{n}}{\partial x} &= G_{\textnormal{bulk}}(x) - R_{\textnormal{bulk}}(n,p) , &  j_n &= qD_n\left(\frac{\partial n}{\partial x}+\frac{n}{V_T}E(t)\right), \label{eq:nDD}     \\
\frac{1}{q}\frac{\partial j_{p}}{\partial x} &= G_{\textnormal{bulk}}(x) - R_{\textnormal{bulk}}(n,p) , &  j_p &= -qD_p\left(\frac{\partial p}{\partial x}-\frac{p}{V_T}E(t)\right), \label{eq:pDD}
\end{align}
where $j_n$ and $j_p$ are the electron and hole current respectively. The Beer-Lambert generation is given by \begin{align} 
G_{\textnormal{bulk}}(x) &= F_{ph}\alpha e^{-\alpha x}, \label{eq:G(x)}
\end{align}
In this study, we consider ShockleyñReadñHall (SRH) and bimolecular recombination in the bulk and interfacial recombination at the perovskite/TL boundaries. These recombination mechanisms are detailed in Table \ref{tab:recomb}. At the ETL/perovskite interface:
\begin{align}
  n\big\rvert _{x=0^+} &=  k_E d_E \exp\left(-\frac{V_1+V_2}{V_T}\right), \label{eq:nBC}   \\
  j_p \big\rvert_{x=0^+} &= -qR_l \label{eq:jpatx=0},
\end{align} 
where $R_l$ is the recombination rate at the left boundary and the ratio between the electron densities across the boundary, $k_E$, is given by
\bet
k_E = \frac{g_c}{g_{cE}} \exp \left( \frac{E_{cE}-E_c}{V_T}\right).   \tag{\ref{eq:kE_kH} reprinted}
\eet
At the perovskite/HTL interface: 
\begin{align}
p\big\rvert _{x=b^-} &= k_H d_H \exp\left(-\frac{V_3+V_4}{V_T}\right), \label{eq:pBC}    \\
j_n \big\rvert_{x=b^-} &= -qR_r.
\end{align}
Here, $R_r$ is recombination rate at right interface and $k_H$ is the ratio between the hole densities across the interface, which is given by
\bet
k_H = \frac{g_v}{g_{vH}} \exp \left( \frac{E_{v}-E_{vH}}{V_T}\right).   \tag{\ref{eq:kE_kH} reprinted}
\eet
The total current output of the cell is given by
\be
J(t) = j_n(x,t) + j_p(x,t) + j_d(t) + j_{P}(t).
\label{eq:J(t)incj_P}
\ee
The displacement current $j_d$ is negligible under typical operation and current-voltage measurements. At high frequencies however, as is probed in impedance measurements, it is significant and is given by 
\begin{equation}
j_d(t) = \varepsilon_p \frac{d E}{d t}. \tag{\ref{eq:j_d} reprinted} 
\end{equation}
The ionic current, $j_{P}$, resulting from the motion of ions across the perovskite width, is at least four orders of magnitude less than the other current contributions. This is true across the full impedance spectrum and therefore it is reasonable to neglect this ionic current contribution.

\paragraph{The Boltzmann approximation}

This approximation was recently used by Courtier with their recent work on an ectypal diode theory for PSCs \cite{courtier2020interpreting}. Analysing the typical size of parameters, it becomes apparent that the following approximation can be made:
\be
\frac{\partial n}{\partial x}+\frac{n}{V_T}E(t) \approx 0, \hspace{5mm} \frac{\partial p}{\partial x}-\frac{p}{V_T}E(t)  \approx 0.
\ee
This approximation is accurate for cells operating at around open-circuit that have long carrier diffusion lengths with respect to the total perovskite layer width, as it is the case in state-of-the-art PSCs.
%\cite{stranks2013electron,dong2015electron}. 
Integrating and applying the boundary conditions (eq.(\ref{eq:nBC}) and eq.(\ref{eq:pBC})) one obtains
\begin{align} 
n(x,t) = k_Ed_E\exp\left(-\frac{V_1 + V_2 + xE}{V_T}\right), \hspace{2mm} p(x,t) = k_Hd_H\exp\left(-\frac{V_3 + V_4 + (b-x)E}{V_T}\right) \tag{\ref{eq:n_p_Boltz} reprinted}
\end{align}
Figures \ref{fig:npcomptoIM} and \ref{fig:npcomptoIM_atVMPP} show comparisons between the carrier densities calculated using the Boltzmann approximation and the full numerical solutions at open-circuit and maximum power point respectively. The parameters are those in Table \ref{tab:params} under 0.1-Sun equivalent illumination and with recombination at the ETL/perovskite interface ($R_l$).

\paragraph{The total current}

Integrating eq.(\ref{eq:pDD}) as follows
\be
\frac{1}{q}\int_0^b\frac{\partial j_{p}}{\partial x}dx = \int_0^b G_{\textnormal{bulk}}(x) - R_{\textnormal{bulk}}(n,p) dx,
\ee
one can write
\be
\frac{1}{q}j_{p}\big\rvert_{x=b} - \frac{1}{q}j_{p}\big\rvert_{x=0} = \int_0^1 G_{\textnormal{bulk}}(x) - R_{\textnormal{bulk}}(n,p) dx. \label{eq:jpatx=b}
\ee
Here, $j_{p}\big\rvert_{x=0}$ is defined via the boundary condition eq.(\ref{eq:jpatx=0}).   
Evaluating the right hand side of equation (\ref{eq:J(t)incj_P}) at $x=b$ we have
\be
J(t) = -qR_r + j_p\big\rvert_{x=b} + j_d(t).
\label{eq:Jforjp}
\ee
Therefore, using eq.(\ref{eq:jpatx=b}), one obtains an expression for the total current
\be
J = q\int_0^b G_{\textnormal{bulk}}(x) - R_{\textnormal{bulk}}(n,p) dx - qR_r - qR_l + j_d
\ee
Where one can define
\be
j_s = q\int_0^b G_{\textnormal{bulk}}(x) dx
\ee
\bet
\jrec = q\int_0^b R_{\textnormal{bulk}}(n,p) dx + qR_r + qR_l. \tag{ \ref{eq:jrec_int} reprinted}
\eet
Integrating the generation function given in equation (\ref{eq:G(x)}) gives the short circuit current density
\begin{equation}
j_s = qF_{ph}\left(1-e^{-\alpha b}\right) \tag{\ref{eq:js} reprinted}
\end{equation}
With the approximations above, one can obtain an equation for the total current
\be
J(t) = qI_sF_{ph}\left(1-e^{-\alpha b}\right) - j_{R_i}\exp\left(-\frac{F_i(V)}{V_T} - \frac{b}{\nel V_T}E\right)     +  \varepsilon_p \frac{d E}{d t}, \label{eq:J(t)full}
\ee
where $F_i(V)$ is the potential barrier to recombination and $\nel$ is the `electronic ideality factor', both of which are defined in Table \ref{tab:F_i(V)andjR_iforR_i}. The electronic ideality factor is detailed in Section \ref{sec:elec_id_fac}. The following section justifies writing the recombination current in this form.

\paragraph{Recombination current} 

We consider a PSC with a primary recombination mechanism that takes the form of one of the five mechanisms listed in Table \ref{tab:recomb}. The deep-level trap constants listed in the table for the full recombination rates are given by
\begin{align}
    k_{1,2} &= \left(\tau_n + \tau_p\right)n_i  \\
    k_{3} &= \left(\frac{1}{k_Ev_{p_E}} + \frac{1}{v_{n_E}} \right)n_i  \\
    k_{4} &= \left(\frac{1}{k_Hv_{n_H}} + \frac{1}{v_{p_H}} \right)n_i. \label{eq:k_1-4}
\end{align}
In this derivation for the analytic equations, the reduced forms of the recombination rates are chosen, which provide good approximations to the full rates. The recombination current for a cell with of multiple recombination pathways is derived at the end of this section. For each recombination type, the recombination current is obtained by substitution of the Boltzmann approximations to the carrier densities, (\ref{eq:n_p_Boltz}), into equation (\ref{eq:jrec_int}).
\\
\\
\textbf{Bimolecular Recombination:} For bimolecular recombination in the bulk, $R_{\textnormal{bulk}} = \beta np$ and $R_l$ = $R_r$ = 0. The recombination or equivalently \textit{dark} current is given by
\be
\jrec =  q\beta k_Ed_Ek_Hd_H \int_0^b \exp\left(-\frac{V_1(t) + V_2(t) + V_3(t) + V_4(t) + bE(t)}{V_T}\right) dx.
\ee
{Given that $V_{1-4}$ and $E(t)$ are not spatially dependent}
\be
\jrec =  qb\beta k_Ed_Ek_Hd_H \exp\left(-\frac{V_1(t) + V_2(t) + V_3(t) + V_4(t)+ bE}{V_T}\right).
\ee
Defining the following for bimolecular recombination:
\be
j_{R_b} =qb\beta k_Ed_Ek_Hd_H, \hspace{5mm} F_b(t) = V_1(t) + V_2(t) + V_3(t) + V_4(t), \hspace{5mm} \nel = 1,
\ee
one can write the recombination current in the form
\bet
\jrec(t) = j_{R_b}\exp\left(-\frac{F_b(V_1(t),V_2(t),V_3(t),V_4(t))}{V_T} - \frac{b}{\nel V_T}E(t)\right).   \tag{eq.(\ref{eq:j_rec_general}) for $R_b$} 
\eet
Table \ref{tab:F_i(V)andjR_iforR_i} details the definitions of $j_{R_b}$, $F_b(t)$ and the electronic ideality factor for bimolecular (and the other forms of) recombination.
\\
\\
\textbf{Hole-limited recombination:} Here, $R_{\textnormal{bulk}} =  p/\tau_p$ and $R_l$ = $R_r$ = 0. The recombination current is given by
\be
\jrec =   \frac{qk_Hd_H}{\tau_p} \int_0^b\exp\left(-\frac{V_3(t) + V_4(t) + (b-x)E(t)}{V_T}\right) dx.
\ee
Integrating this over the cell width, one obtains
\be
\jrec =   \frac{qk_Hd_H}{\tau_p}\exp\left( -\frac{ V_3(t) + V_4(t)}{V_T}\right)V_T\frac{1 - \exp(-\frac{bE(t)}{V_T})}{E}.
\ee
Making the substitution $\bar{E} = bE/V_T$, one can write
\be
\jrec =   \frac{qbk_Hd_H}{\tau_p}\exp\left( -\frac{ V_3(t) + V_4(t)}{V_T}\right)\frac{1 - \exp(-\bar{E})}{\bar{E}}.
\ee
Considering small perturbations to the bulk potential prop ($bE$) relative to the thermal voltage, as is reasonable for impedance measurements, one can expand about $\bar{E}=0$
\begin{align}
     \frac{1 - \exp(-\bar{E})}{\bar{E}} = 1- \frac{\bar{E}}{2} + \mathcal{O}(\bar{E}^2). \label{eq:tayloraboutEis0}
\end{align}
Neglecting terms of order $\mathcal{O}(\bar{E}^2)$ and noting that 
\begin{align}
     \exp(-\frac{\bar{E}}{2}) =  1- \frac{\bar{E}}{2} + \mathcal{O}(\bar{E}^2), \label{eq:exp(-E/2)approx}
\end{align}
one can rewrite the recombination current in the form
\be
\jrec =   \frac{qbk_Hd_H}{\tau_p}\exp\left( -\frac{ V_3(t) + V_4(t) }{V_T} - \frac{b}{2V_T}E(t)\right) + \mathcal{O}(E^2).
\ee
where $\bar{E}$ has been eliminated. 
Defining the following for hole-limited recombination:
\be
j_{R_p} = \frac{qbk_Hd_H}{\tau_p}, \hspace{5mm}  F_p(t) = V_3(t) + V_4(t) , \hspace{5mm} \nel = 2,
\ee
one can write the recombination current in the form of equation \eqref{eq:j_rec_general}.
\\
\\
\textbf{Electron-limited Recombination} Here $R_{\textnormal{bulk}}=n/\tau_n$ and $R_l$ = $R_r$ = 0. The recombination current is given by
\be
\jrec =  \frac{qk_Ed_E}{\tau_n} \int_0^b\exp\left(-\frac{V_1(t) + V_2(t) + xE(t)}{V_T}\right) dx.
\ee
Integrating one obtains
\be
\jrec =  \frac{qk_Ed_E}{\tau_n} \exp\left( -\frac{ V_1(t) + V_2(t)}{V_T} \right)V_T\frac{1 - \exp(-\frac{bE}{V_T})}{E(t)}.
\ee
Using the same approximations detailed in equations (\ref{eq:tayloraboutEis0}) and (\ref{eq:exp(-E/2)approx}), the recombination current can be written
\be
\jrec =  \frac{qbk_Ed_E}{\tau_n} \exp\left( -\frac{ V_1(t)+ V_2(t) }{V_T} - \frac{b}{2V_T}E(t)\right) + \mathcal{O}(E^2).
\ee
Defining the following for electron-limited bulk SRH recombination:
\be
j_{R_n} = \frac{qbk_Ed_E}{\tau_n}, \hspace{5mm} F_n(t) = V_1(t) + V_2(t), \hspace{5mm} \nel = 2, 
\ee
one can write the recombination current in the form of equation \eqref{eq:j_rec_general}.
\\
\\
\textbf{ETL/perovskite interfacial SRH recombination.}
Here $R_l= v_{p_E} \pl$. The recombination at the perovskite/HTL interface and in the bulk is set to zero. The hole density at the ETL/Perovskite interface is given by
\begin{align}
    \pl &= p\big\rvert _{x=0^+}  \exp\left( -\frac{V_2}{V_T} \right) \\
    \pl &= k_Hd_H\exp\left(-\frac{V_2(t) + V_3(t) + V_4(t) + bE(t)}{V_T}\right). \label{eq:p_l} 
\end{align}
Therefore the recombination current is given by
\begin{align}
\jrec &=  qk_Hd_Hv_{p_E} \exp\left(-\frac{ V_2 + V_3 + V_4 + bE}{V_T}\right).
\end{align}
Defining the following for ETL/Perovskite interfacial SRH recombination:
\be
j_{R_l} = qk_Hd_Hv_{p_E}, \hspace{5mm} F_l(t) = V_2(t) + V_3(t) + V_4(t) , \hspace{5mm} \nel = 1,
\ee
one can write the recombination current in the form of equation \eqref{eq:j_rec_general}.
\\
\\
\textbf{Perovskite/HTL interfacial SRH recombination.} Finally, we consider the case where $R_r= v_{n_H} \nr$ with zero recombination in the bulk and at the ETL/Perovskite interface. The electron density at the Perovskite/HTL interface is given by
\begin{align}
    \nr &= n\big\rvert _{x=b^-} \exp\left(-\frac{ V_3}{V_T}\right) \\
    \nr &= k_Ed_E\exp\left(-\frac{ V_1(t) + V_2(t) + V_3(t) + bE(t)}{V_T}\right). \label{eq:n_r}
\end{align}
Therefore the recombination current is given by
\begin{align}
\jrec &=  qv_{n_H}k_Ed_E \exp\left(-\frac{ V_1(t) + V_2(t) + V_3(t) + bE(t)}{V_T}\right)
\end{align}
Defining the following for Perovskite/HTL interfacial SRH recombination:
\be
j_{R_r} = qv_{n_H}k_Ed_E, \hspace{5mm} F_r(t) = V_1(t) + V_2(t) + V_3(t) , \hspace{5mm} \nel = 1 
\ee
one can write the recombination current in the form of equation \eqref{eq:j_rec_general}.

\paragraph{The general form for recombination current} 
As derived above, all forms of recombination result in recombination currents that can be written in the general form:
\begin{align}
        \jrec(t) = j_{R_i}\exp\left(-\frac{F_i(V_1(t),V_2(t),V_3(t),V_4(t))}{V_T} - \frac{b}{\nel V_T}E(t)\right)    \tag{\ref{eq:j_rec_general} reprinted}
\end{align}
where $F_i(V_1,V_2,V_3,V_4)$ is the potential barrier to recombination and $\nel$ is the electronic ideality factor, as defined in Table \ref{tab:F_i(V)andjR_iforR_i}. Note that for hole-limited and electron-limited bulk SRH recombination, eq.\eqref{eq:j_rec_general} is only correct up to terms of $\mathcal{O}(E^2)$.

\paragraph{Linearisation and impedance spectroscopy}

To measure the impedance of a solar cell, a `DC' (or constant) voltage is applied with an additional small sinusoidal signal to perturb the cell about a steady-state. The applied perturbation is small to induce a linear current response.
The phase and amplitude of the response is determined over a broad frequency range to generate an impedance spectrum. Impedance spectroscopy can be performed under illumination or in the dark and at any DC voltage. The general form of an applied voltage for an impedance measurement can be written
\bet
V(t) = \VDC + V_p\exp(i\omega t) % \tag{\ref{eq:V(t)_IS} reprinted}  % now the other one is V_psin()
\eet
where, $\VDC$ is the DC voltage, $\omega$ is the angular frequency of the sinusoidal perturbation and $V_p$ is its amplitude. Utilising the fact that the cell is perturbed about a steady-state, one can express the variables of the system as linear perturbations about this steady-state. We define the parameter
\be
\delta = V_p/V_T,
\ee 
where $V_T$ is the thermal voltage (which is around 25mV at room temperature). Impedance measurements require $\delta$ to be small for the analysis to be valid. {In practice, a perturbation amplitude of 20 mV suffices, even though this is not particularly small in comparison to the thermal voltage}. 

In experiment, the perturbation amplitude (typically 10-20 mV) is such that it's small enough to ensure a linear response, but large enough that the response is not negligible compared to noise. For the purposes of this work, we assume $\delta$ is small such that a linear response is observed and higher order terms can be reasonably neglected. The applied voltage can be written in the form
\be
V(t) = \VDC +\delta V^{(1)}(t) \hspace{5mm} \textnormal{where} \hspace{2mm} V^{(1)}(t) = V_T e^{i\omega t}
\ee
Now with this construction, the Debye layer charge, $Q$, and the bulk electric field can be written
\begin{align}
Q(t) &= \hspace{1.25mm} \QDC \hspace{0.5mm}+\hspace{0.5mm} \delta Q^{(1)}(t) + \mathcal{O}(\delta^2), \label{eq:Q(t)linear}\\
E(t) &= \underbrace{E_{DC}}_{ = \hspace{1mm} 0} + \hspace{1mm} \delta E^{(1)}(t) + \mathcal{O}(\delta^2).   
\end{align}
At steady-state, the charge contained within the left and right Debye layers is -$\QDC$ and $\QDC$ respectively with zero bulk electric field. Using eq.(\ref{eq:dQdt3L}) leading order (or steady-state) Debye layer charge $\QDC$ is given by solution to:
\bet
\VDC - \Vbi + V_1(\QDC) + V_2(\QDC) + V_3(\QDC) + V_4(\QDC) = 0.  \tag{\ref{eq:pot_drops} reprinted}  
\eet
To condense notation, we define the total potential drops across the Debye layers, $F_T(t)$
\be
F_T(t) = V_1(t) + V_2(t) + V_3(t) + V_4(t), \label{eq:F_T(t)} %\tag{\ref{eq:F_T(t)} reprinted}%\label{eq:G}
\ee
which, at steady state is simply
\begin{align}
F_T(\QDC) &= V_1(\QDC) + V_2(\QDC) + V_3(\QDC) + V_4(\QDC), \\
&= \Vbi - \VDC(\QDC)    
\end{align}
Owing to the complexity of the capacitance relations that set the potential drops $V_{1-4}$, it is not possible to determine an analytic solution for $\QDC$. Therefore, a numerical root finding algorithm must be used to determine $\QDC$ for a given applied potential $\VDC$. This is the only part of the model that requires numerical evaluation, the rest is analytic.
The equation for the leading order Debye layer charge, $Q^{(1)}(t)$, is given by 
\be
\frac{dQ^{(1)}(t)}{dt} = -\frac{qD_+N_0}{V_Tb} \Big( V^{(1)}(t) + Q^{(1)}(t)F_T'(\QDC)  \Big) 
\ee
where the prime denotes a derivative with respect to $Q$, hence
\be
F_T'(\QDC) = \frac{dV_1}{dQ}\bigg\rvert _{\QDC} + \frac{dV_2}{dQ}\bigg\rvert _{\QDC} + \frac{dV_3}{dQ}\bigg\rvert _{\QDC} +\frac{dV_4}{dQ}\bigg\rvert _{\QDC}.
\ee
Solving this first order ordinary differential equation one obtains
\begin{align}
    Q^{(1)}(t) &= G_+V_T\frac{-G_+F_T'(\QDC)+i\omega}{G_+^2F_T'(\QDC)^2+\omega^2}e^{i\omega t}, \label{eq:Q1oft}
\end{align}
where we have defined 
\bet 
    G_+ = \frac{qD_+N_0}{V_Tb}. \tag{\ref{eq:G_+} reprinted}
\eet 
Linearising the recombination current, one can write 
\begin{align}
    \jrec(t) = \jrec(\VDC) + \delta \jrec^{(1)}(t) + \mathcal{O}(\delta^2),
\end{align}
where 
\begin{align}
    \jrec(\VDC) &= j_{R_i}\exp\left(-\frac{F_i(\QDC)}{V_T}\right), \label{eq:jrec0} \\
    \jrec^{(1)}(t) &= Q^{(1)}(t)\frac{d\jrec}{dQ}\bigg\rvert_{Q=\QDC}.
\end{align}
We use the notation in Table \ref{tab:F_i(V)andjR_iforR_i} to keep this derivation general. Taking the derivative of eq.(\ref{eq:j_rec_general}), one obtains the first order recombination current
\be
\jrec^{(1)}(t) = Q^{(1)}(t)\jrec(\VDC)\left( - \frac{F'_i(\QDC)}{V_T} - \frac{b}{\nel V_T}E^{(1)\prime}\right),
\ee
where a prime denotes a derivative with respect to the Debye layer charge, $Q$. Similarly the short-circuit, displacement and total current can be expressed as
\begin{align}
    j_{s}    &= j_{s}(\VDC) + \delta\underbrace{j_{s}^{(1)}}_{ = 0 }, \\
    j_{d}(t) &= \underbrace{j_{d}(\VDC)}_{ = 0} + \delta j_{d}^{(1)}(t) + \mathcal{O}(\delta^2), \\
    J(t)     &= J(\VDC) + \delta J^{(1)}(t) + \mathcal{O}(\delta^2),
\end{align}
Using these linearised expressions and equation (\ref{eq:J(t)}), the leading and first order current is given by 
\begin{align}
    J(\VDC) &=  j_{s}(\VDC) - \jrec(\VDC),  \\
        &= qF_{ph}\left(1-e^{-\alpha b}\right) - j_{R_i}\exp\left(-\frac{F_i(\QDC)}{V_T}\right),  \\ % Removed I_s
    J^{(1)}(t) &=  - \jrec^{(1)}(t) + j_{d}^{(1)}(t), \\
    &= -Q^{(1)}(t)\jrec(\VDC)\left(- \frac{F'_i(\QDC)}{V_T}-\frac{b}{\nel V_T}E^{(1)\prime}\right) + \varepsilon_p \frac{d E^{(1)}}{d t}.
\end{align}

\paragraph{Impedance parameters}

Impedance is defined
\be
Z(\omega) = \frac{V^{(1)}}{J^{(1)}} = R + iX,
\ee
where $R$ is resistance and $X$ is reactance. For impedance measurements it is assumed the perturbation is small and hence the higher order terms can be neglected. When considering the impedance of a system, it is conventional to define current such that an increase in voltage results in an increase current. For the rest of this section, we will redefine the current to be consistent with this convention. This simply shifts the phase by 180$^{\circ}$. Hence
\be
Z(\omega) = \frac{V_Te^{i\omega t}}{Q^{(1)}(t)\jrec(\VDC)\left(- \frac{F'_i(\QDC)}{V_T}-\frac{b}{\nel V_T}E^{(1) \prime}\right) - \varepsilon_p \frac{d E^{(1)} }{d t}}.
\ee
The admittance $Y=1/Z$ is given by
\be
Y(\omega) = G + iB.
\ee
Hence, using the relations \eqref{eq:E=dQdt} and \eqref{eq:Q(t)linear}-\eqref{eq:Q1oft}, one obtains
\begin{align}
G(\omega) &= \frac{1}{G_+^2F_T'(\QDC)^2+\omega^2} \bigg( \frac{G_+^2}{V_T}\jrec(\VDC)F'_i(\QDC)F_T'(\QDC) \notag  \\
 &\hspace{59mm}+\omega^2\left( \frac{\jrec(\VDC)}{V_T\nel} - \frac{G_+\varepsilon_p}{b}F_T'(\QDC)\right)\bigg), \\
B(\omega) &=  \frac{\omega}{G_+^2F_T'(\QDC)^2+\omega^2} \left( \frac{G_+}{V_T}\jrec(\VDC)  \left( \frac{F_T'(\QDC)}{\nel} - F'_i(\QDC)  \right) +\frac{\varepsilon_p}{b}\omega^2 \right).
\end{align}
Now using 
\be
R = \frac{G}{G^2+B^2}, \hspace{5mm} X = - \frac{B}{G^2+B^2},
\ee
one obtains the resistance and reactance of a PSC as a function of frequency
\begin{align} 
\label{eq:Ranalytic}
R(\omega) & = \frac{G_+^2b\jrec(\VDC) F_i'(\QDC)F_T'(\QDC) + \omega^2\left( \frac{b\jrec(\VDC)}{\nel} - V_TG_+\varepsilon_pF_T'(\QDC) \right) }{\frac{b}{V_T}\Big(G_+\jrec\left(\VDC\right)F_i'\left(\QDC\right) \Big)^2 + \omega^2 \left( \frac{b\jrec(\VDC)^2}{V_T\nel^2} - 2G_+\varepsilon_p \jrec(\VDC) F_i'(\QDC) + \frac{V_T}{b}\varepsilon_p^2\omega^2 \right) }  \\
X(\omega) & = \frac{\omega \left( G_+b\jrec(\VDC) \left(F_i'(\QDC)-\frac{F_T'(\QDC)}{\nel} \right) -V_T\varepsilon_p\omega^2 \right) }{\frac{b}{V_T}\Big(G_+\jrec(\VDC)F_i'(\QDC) \Big)^2 + \omega^2 \left( \frac{b\jrec(\VDC)^2}{V_T\nel^2} - 2G_+\varepsilon_p \jrec(\VDC) F_i'(\QDC) + \frac{V_T}{b}\varepsilon_p^2\omega^2 \right) }. 
\label{eq:Xanalytic}
\end{align}
The Nyquist spectra generated by these relations consist of a high frequency semicircle above the axis and a low frequency semicircle that lies either above or below the axis. A spectrum that consists of two semicircular features can be reproduced by an equivalent circuit that is composed of two $RC$ elements in series. Each $RC$ element is simply a resistor and capacitor in parallel. This equivalent circuit model is presented in Figure \ref{fig:Nyq_Bode_labelled}. Comparing the frequency dependent relations, \ref{eq:Ranalytic} and \ref{eq:Xanalytic}, with the impedance response of two $RC$ elements in series it is found that the system is characterised by the following resistances and time constants:
\begin{align}
    R_{LF} &=  \frac{V_T}{\jrec(\VDC)}\left( \frac{F_T'(\QDC)}{F'_i(\QDC)} - \nel\right), \label{eq:R_LF} \\ 
    R_{HF} &=  \frac{V_T\nel}{\jrec(\VDC)} \label{eq:R_HF},\\
    \omega_{LF} &= G_+\nel F'_i(\QDC), \\
    \omega_{HF} &= \frac{\jrec(\VDC)b}{\varepsilon_p V_T \nel}. \tag{\ref{eq:omega_HF} reprinted}
\end{align}
See Figure \ref{fig:Nyq_Bode_labelled} for a Nyquist and frequency plot labelling these impedance parameters. The low frequency semicircle is above the axis when $R_{LF}>0$ and below the axis when $R_{LF}<0$.

\begin{figure}[htbp] % trim=left botm right top  \fbox{ whole include graphics bit for box around it to crop}
\includegraphics[trim=3.5cm 11.3cm 4cm 11.8cm, clip, width=0.65\textwidth]{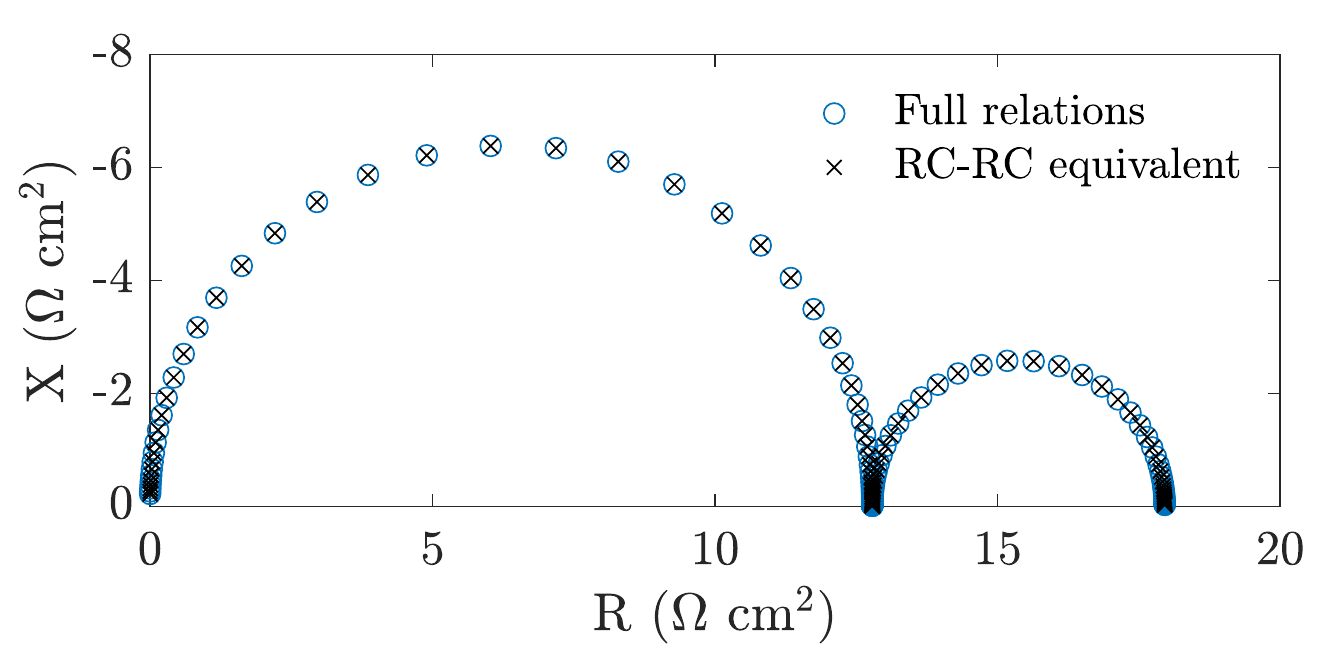}
\centering
\caption{Comparison between the full impedance relations, eq.(\ref{eq:Ranalytic}-\ref{eq:Xanalytic}), and an $RC$-$RC$ equivalent circuit approximation using the component values listed in eq.(\ref{eq:RandC_HF}-\ref{eq:RandC_LF}). The same cell parameters, from Table \ref{tab:params}, and simulation parameters have been used here as in Figure \ref{fig:RlNyquist} at open-circuit ($\VDC$ = 0.93 V). Specifically, recombination occurs at the ETL/perovskite interface ($R_l$) from Table \ref{tab:recomb} under 0.1-Sun equivalent illumination. Instead of the standard 256 frequencies, 128 frequencies are used here to more clearly show the data points.}
\label{fig:fullcomptoRCRC}
\end{figure}

Using the relations 
\be
C_{LF} = \frac{1}{R_{LF}\omega_{LF}}, \hspace{10mm} C_{HF} = \frac{1}{R_{HF}\omega_{HF}}, \label{eq:omega_from_RC}
\ee
the low and high frequency capacitances can be attributed to the spectrum generated from equations (\ref{eq:Ranalytic}) and (\ref{eq:Xanalytic})
\be
C_{LF} =  \frac{\jrec(\VDC)}{G_+V_T\nel\left( F_T'(\QDC) - \nel F'_i(\QDC) \right)}, \hspace{10mm} C_{HF} = \frac{\varepsilon_p}{b}. \label{eq:C_LFandC_HF}
\ee
Calculating the impedance using the theoretical impedance from an $RC$-$RC$ circuit with component values equal to equations (\ref{eq:RandC_HF}-\ref{eq:RandC_LF}), one obtains a spectra that matches the spectra of the full relations well. See Figure \ref{fig:fullcomptoRCRC} for this comparison. It should be noted that equations (\ref{eq:RandC_HF}) and (\ref{eq:RandC_LF}) do not exactly transform between the full expressions (\ref{eq:Ranalytic} and \ref{eq:Xanalytic}) and the equivalent circuit analogue. More complex frequency dependent parameterisations may be needed which are beyond the scope of this work. Figure \ref{fig:fullcomptoRCRC} shows how closely equations (\ref{eq:RandC_HF}) and (\ref{eq:RandC_LF}) as part of an $RC$-$RC$ circuit match the full impedance relations (\ref{eq:Ranalytic} and \ref{eq:Xanalytic}). This validates the use of equations (\ref{eq:RandC_HF}-\ref{eq:RandC_LF}) as equivalent circuit component values.

\paragraph{The apparent ideality factor}
In the main text the analytic relations are presented in a form that employ the apparent ideality factor rather than the potentials $F_i(V)$ and $F_T(V)$. This is because the apparent ideality factor is commonly measured (and mistaken for an ideality factor) and is therefore something that can easily be identified and determined from experimental measurements. This apparent ideality factor is equivalent to the ectypal factor, as derived in \cite{courtier2020interpreting}. We use the apparent ideality factor to refer to both the \textit{true} and the \textit{measured} ectypal factor. The measured ectypal factor, $\nmec$, is defined as \cite{courtier2020interpreting}
\be
\bar{n}_{ec}(\VDC) = \nec(\VDC)\left[ 1 - \frac{ \VDC - \Vbi }{\nec} \frac{d\nec}{dV}\bigg\rvert_{V=\VDC} \right]^{-1}, \label{eq:measured_n_ec}
\ee
where the true ectypal factor is given by
\be
\nec = \frac{\Vbi - V}{ F_i(V_1,V_2,V_3,V_4)}. \label{eq:n_ec}
\ee
The measured ectypal factor is related to the potential barrier to recombination, $F_i$, and the total potential, $F_T$ via
\be
\nmec(\QDC) = \frac{F'_T(\QDC)}{ F'_i(\QDC)}. \label{eq:nmec_to_F_i}
\ee
Note that
\be
F'_T(\QDC) = \frac{d}{dQ}\left( \Vbi - V\right)\big\rvert_{\QDC} = -\frac{d\VDC}{d\QDC}.
\ee
Substituting eq.\eqref{eq:nmec_to_F_i} in equations \eqref{eq:R_LF}-\eqref{eq:C_LFandC_HF} where $\nmec = \nap$ returns the equations found in the main text.
\newpage

\paragraph{Multiple recombination pathways and IS}
In the case of multiple recombination mechanisms, the recombination current can be composed as follows:
\be
\jrec = j_{\textnormal{rec}_\textnormal{b}} + j_{\textnormal{rec}_\textnormal{SRH}} +  j_{\textnormal{rec}_\textnormal{l}} + j_{\textnormal{rec}_\textnormal{r}}, \label{eq:j_rec_multiple_R_i's}
\ee 
where each component is the recombination current resulting from the specific form of recombination. SRH recombination is either electron-limited ($R_n$) or hole-limited ($R_p$). To reduce the length of the formula, we define the following 
\be
\jrec = \sum_{i=b,p,n,l,r} j_{\textnormal{rec}_\textnormal{i}}.
\ee 
The abbreviations used for the different types of recombination pathways are given in Table \ref{tab:recomb}. Additionally, we use $\neli$ and $n_{\textnormal{ap}_{i}}$ to distinguish the different values of the electronic ideality factor and the apparent ideality factor (measured ectypal factor) respectively for the corresponding recombination type. % corresponds to the recombination type $i$.
For example, for bimolecular recombination ($R_b$) and electron-limited bulk SRH recombination ($R_n$), the total recombination current is
\be
\jrec = j_{\textnormal{rec}_\textnormal{b}} + j_{\textnormal{rec}_\textnormal{n}} = \sum_{i=b,n} j_{\textnormal{rec}_\textnormal{i}},
\ee 
with
\be
n_{\textnormal{el}_b} = 1, \hspace{5mm} n_{\textnormal{el}_n} = 2.
\ee 
\be
n_{\textnormal{ap}_b}(\VDC) = 1, \hspace{5mm} n_{\textnormal{ap}_n}(\VDC) = \frac{F'_T(\QDC)}{ F'_n(\QDC)} = \frac{V'_1 + V'_2 + V'_3 + V'_4}{ V_1' + V_2'}.
\ee 
With this notation, the general form for the resistances and capacitances associated with the high and low frequency features can be written as
\begin{align} % WITH THE APPARENT IDEALITY FACTOR:
    R_{HF} &=  V_T \left(\sum_{i} \frac{j_{\textnormal{rec}_{\textnormal{i}}}(\VDC) }{ \neli } \right)^{-1}  &&C_{HF}= \frac{\varepsilon_p}{b}  \label{eq:RandC_LF_gen}\\
    R_{LF} &=  V_T \left(\sum_{i} \frac{j_{\textnormal{rec}_{\textnormal{i}}}(\VDC) }{ n_{ap_{i}} } \right)^{-1} - R_{HF}  &&C_{LF} =  \frac{\left(\sum_{i} j_{\textnormal{rec}_{\textnormal{i}}}(\VDC)/\neli \right)^2 \frac{d\QDC}{d\VDC} }{G_+V_T\sum_{i} j_{\textnormal{rec}_{\textnormal{i}}}(\VDC) \left( 1/n_{ap_{i}} - 1/\neli\right)}, \label{eq:RandC_HF_gen}
\end{align}
{where the sum over $i$ is the sum over recombination pathways, $i = b,p,n,l,r$ as defined in Table \ref{tab:recomb}.}

\paragraph{Multiple recombination pathways and the electronic ideality factor}
{Here, we consider what the electronic ideality factor is when there is more than one source of recombination. We start from the formula for $\nel$, found in (\ref{eq:n_id}) and, by writing (\ref{eq:j_rec_multiple_R_i's}) in the form
\be
\jrec = j_{\textnormal{rec}_\textnormal{SRH}} + j_{\textnormal{rec}_\textnormal{non-SRH}}
\ee 
where 
\be
j_{\textnormal{rec}_\textnormal{SRH}} =& j_{\textnormal{rec}_\textnormal{n}} \hspace{5mm} \textnormal{or} \hspace{5mm} j_{\textnormal{rec}_\textnormal{p}}  \\
j_{\textnormal{rec}_\textnormal{non-SRH}} =& \sum_{i=b,l,r} j_{\textnormal{rec}_\textnormal{i}},
\ee 
we see that the HF resistance (as defined in (\ref{eq:RandC_LF_gen})) can be written in the form
\be
R_{HF} =  V_T \left( \frac{j_{\textnormal{rec}_{\textnormal{SRH}}}(\VDC) }{ 2 } + j_{\textnormal{rec}_\textnormal{non-SRH}}(\VDC) \right)^{-1},
\ee
where appropriate values of $n_{el_{i}}$, the electronic ideality factor, are used for each recombination type. On defining $r_{\textnormal{SRH}}$ as the ratio of the SRH recombination current to the total recombination current, as follows
\be
r_{\textnormal{SRH}} = \frac{j_{\textnormal{rec}_\textnormal{SRH}}(\VDC)}{\jrec(\VDC),}
\ee
we find that the HF resistance is given by
\be
R_{HF}(\VDC) =  V_T \left( \jrec(\VDC) \left( \frac{r_{\textnormal{SRH}}}{2} + 1 - r_{\textnormal{SRH}}  \right)\right)^{-1}.
\ee
This simplifies to
\be
R_{HF}(\VDC) =  \frac{V_T}{\jrec(\VDC) \left( \frac{2-r_{\textnormal{SRH}}}{2} \right) }.
\ee
On comparison of the above with (\ref{eq:RandC_HF}), or by substituting it into the following 
\be
\nel = \frac{R_{\textnormal{HF}}(\VDC)\jrec(\VDC)}{V_T}, 
\ee
we see that the effective electronic ideality factor is given by
\be
\nel = \frac{2}{2-r_{\textnormal{SRH}}}     \label{nel-mult}
\ee
where $r_{\textnormal{SRH}}$ is the ratio of the SRH recombination current to the total recombination current.}

% ------------------------------------ Supplementary figures ------------------------------------- % 
\newpage
\section{Supplementary figures}  \label{sec:supfigs}
\setcounter{equation}{0}
\setcounter{figure}{0}

\subsection{Carrier Densities} \label{sec:npextrafigs}

\begin{figure}[htbp] % trim=left botm right top  \fbox{ whole include graphics bit for box around it to crop} 
\includegraphics[trim=1.1cm 8cm 2.0cm 7.5cm, clip, width=0.7\textwidth]{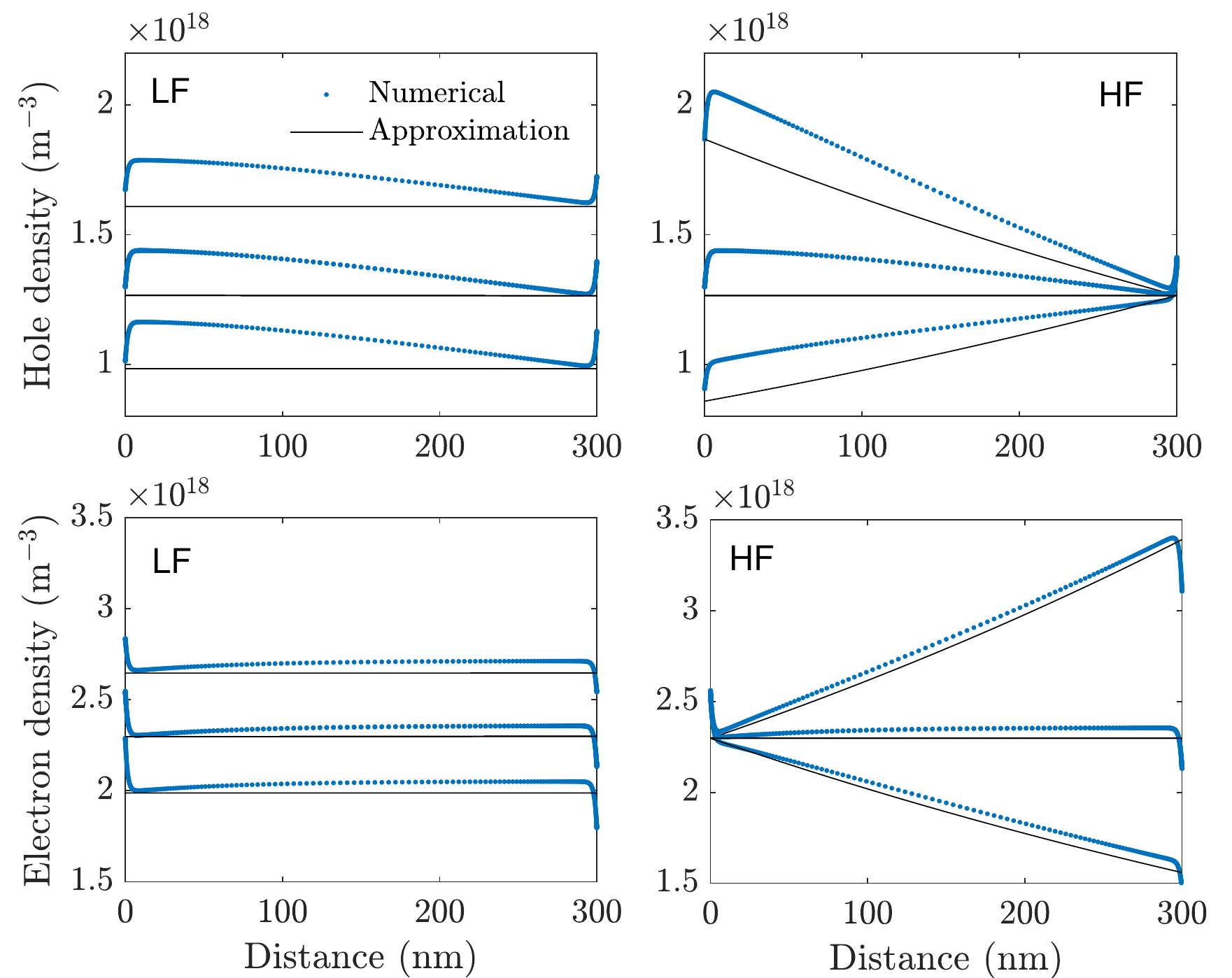}
\centering
\caption{Comparison between the Boltzmann approximation used in this work to the full numerical solutions at maximum power point. Left: carrier density at five equally spaced times over a low frequency period (1 mHz). The right is the equivalent but over an intermediate/high frequency period (25 kHz). The parameters used are detailed in Table \ref{tab:params}, under 0.1-sun equivalent illumination and with recombination at the ETL/perovskite interface ($R_l$) from Table \ref{tab:recomb}. Note that only three distinct distributions are apparent, due to the distributions closely overlapping at equilibrium.}
\label{fig:npcomptoIM_atVMPP}
\end{figure}

Figure \ref{fig:npcomptoIM_atVMPP} shows the numerical solutions for the carrier densities compared to the Boltzmann approximation (given by equations \eqref{eq:n_p_Boltz}-\eqref{eq:n_r_p_r}) across the perovskite layer at five equally spaced timepoints over low and high frequnecy perturbations. This is equivalent the Figure \ref{fig:npcomptoIM} but at the maximum power point. Specifically, $\VDC$ = 0.82 V for a simulated IS measurement under 0.1 sun equivalent illumination with recombination only occurring at the ETL/perovskite interface ($R_l$). Cell and recombination parameters are from Tables \ref{tab:params} and \ref{tab:recomb} respectively. It is clear that the Boltzmann approximation for the carriers is not as good at maximum power point (but still acceptable) compared to open-circuit (Figure \ref{fig:npcomptoIM}). Much below maximum power point the agreement is poor and the analytic model does not reproduce the results from the full numerical model. Therefore, we reiterate that this model is only suitable for interpreting IS measurements at or above the maximum power point.

\newpage
\subsection{Impedance spectra} \label{sec:ISextrafigs}

\begin{figure}[htbp] % trim=left botm right top  \fbox{ whole include graphics bit for box around it to crop}
\includegraphics[trim=3.2cm 11.4cm 4.5cm 10.2cm, clip, width=0.75\textwidth]{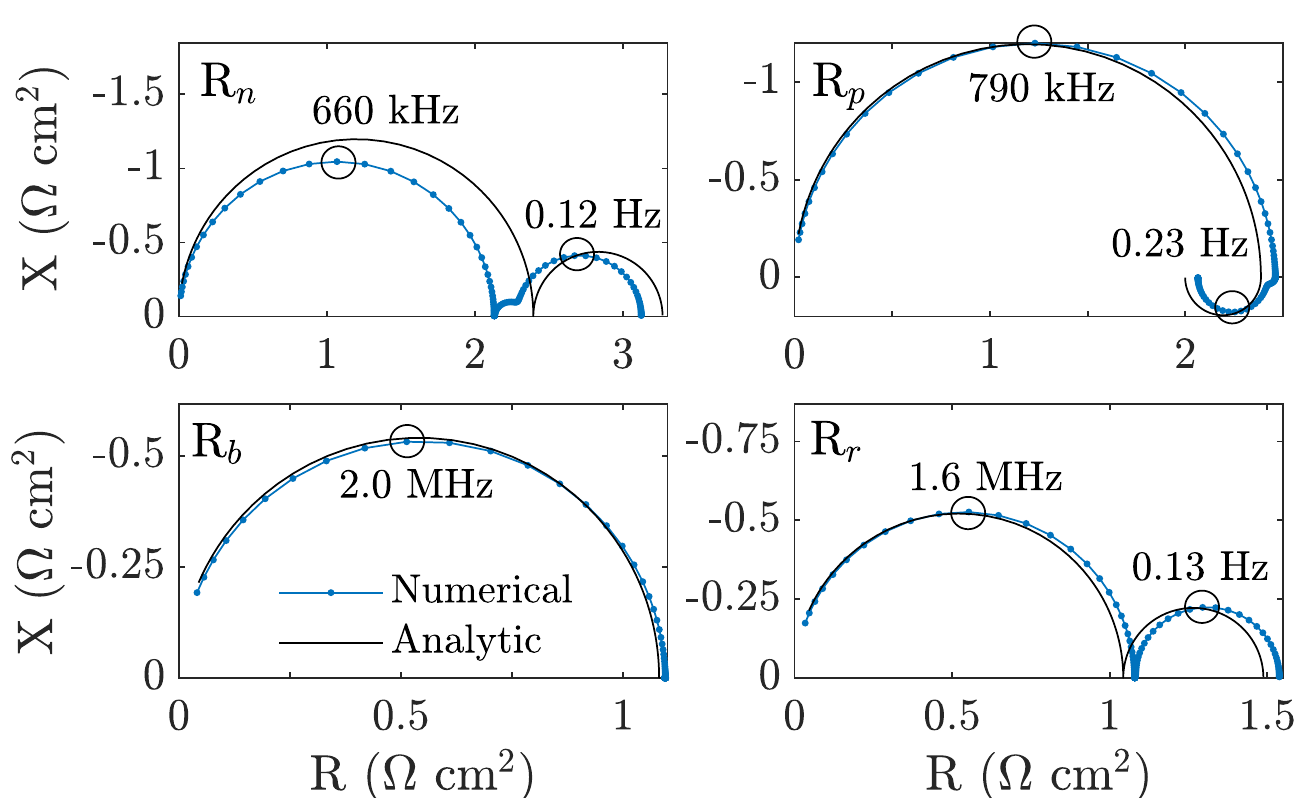}
\centering
\caption{Simulated impedance spectra at open-circuit with different recombination mechanisms. Clockwise from top left: electron-limited bulk SRH ($R_n$: V$_{oc}$=1.08 V), hole-limited bulk SRH ($R_p$: V$_{oc}$=1.02 V), perovskite/HTL interfacial ($R_r$: V$_{oc}$=1.04 V) and bimolecular bulk recombination ($R_b$: V$_{oc}$=1.01 V). Cell and recombination parameters are listed in Tables \ref{tab:params} and \ref{tab:recomb} respectively.}
\label{fig:R_combined_Nyquist_1Sun} % replaced by R_combined_Nyquist_1sun in appendix
\end{figure}

\begin{figure}[htbp] % trim=left botm right top  \fbox{ whole `includegraphics{}' bit for box around it to crop}
\includegraphics[trim=3.5cm 10.0cm 4cm 9.5cm, clip, width=0.70\textwidth]{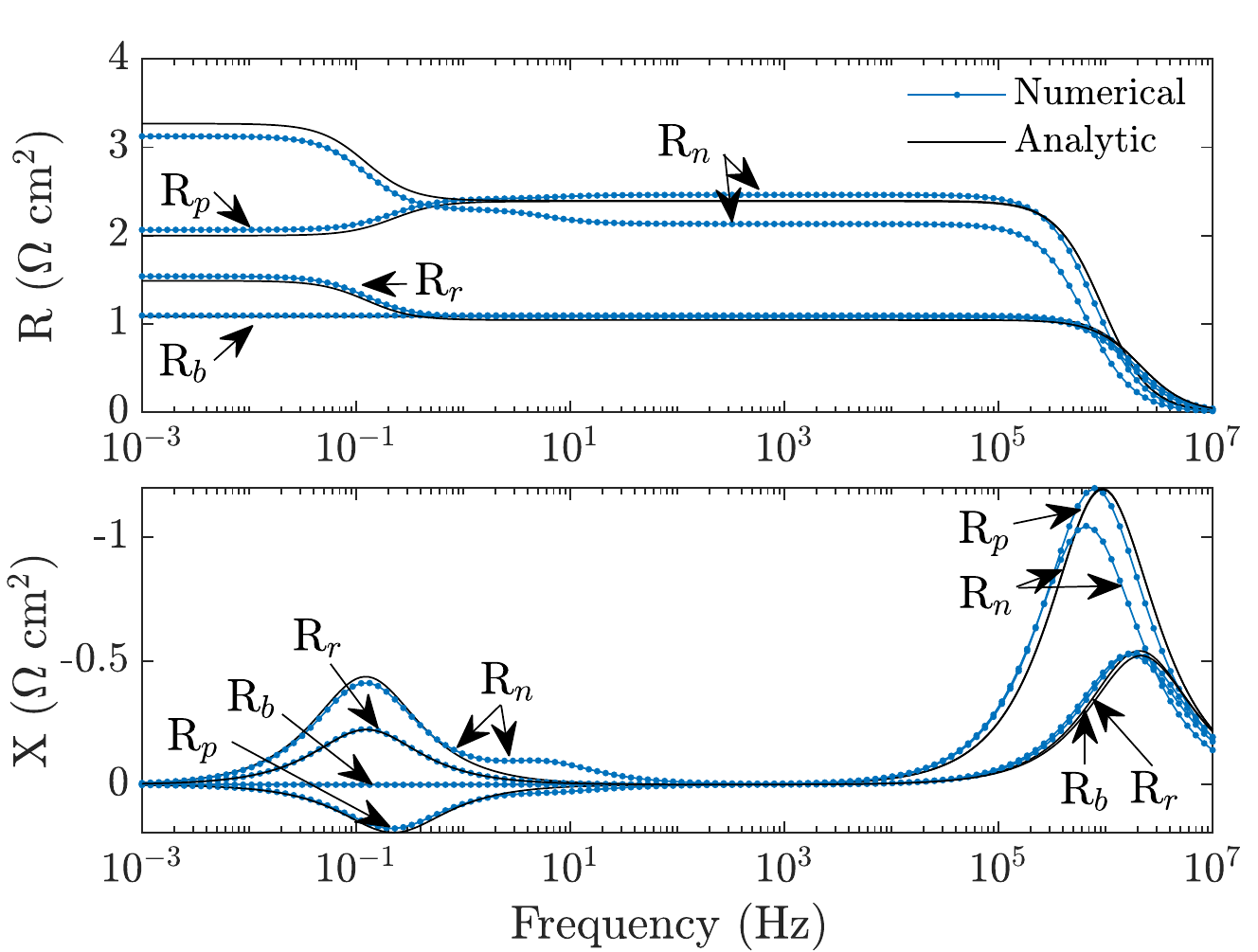}
\centering
\caption{Frequency plot of the spectra given in Figure \ref{fig:R_combined_Nyquist_1Sun}.}
\label{fig:R_combined_Bode_1Sun}
\end{figure}

All PSC spectra are simulated with parameters from Table \ref{tab:params} with a perturbation amplitude of 10 mV and over a frequency range of 10$^{-3}$-10$^7$ Hz. Details on the recombination types and parameters are specified in Table \ref{tab:recomb}. The numerical and analytic spectra are composed of 128 and 256 frequencies respectively. Spectra obtained numerically using IonMonger \cite{courtier2019ionmonger,riquelme2020identification} take under two minutes using a powerful laptop. Without optimisation, the analytic solutions are obtained in under 5 seconds. Figures \ref{fig:R_combined_Nyquist_1Sun} and \ref{fig:R_combined_Bode_1Sun} are equivalent to Figures \ref{fig:R_combined_Nyquist_point1sun} and \ref{fig:R_combined_Bode_point1sun} in the main text but under 1-Sun equivalent illumination. Numerical impedance spectra with electron-limited and hole-limited SRH recombination in the bulk show intermediate frequency features that are not reproduced by the analytic model. 

\begin{figure}[htbp] % trim=left botm right top  \fbox{ whole include graphics bit for box around it to crop}
\includegraphics[trim=3cm 9.25cm 3.5cm 9.25cm, clip, width=0.65\textwidth]{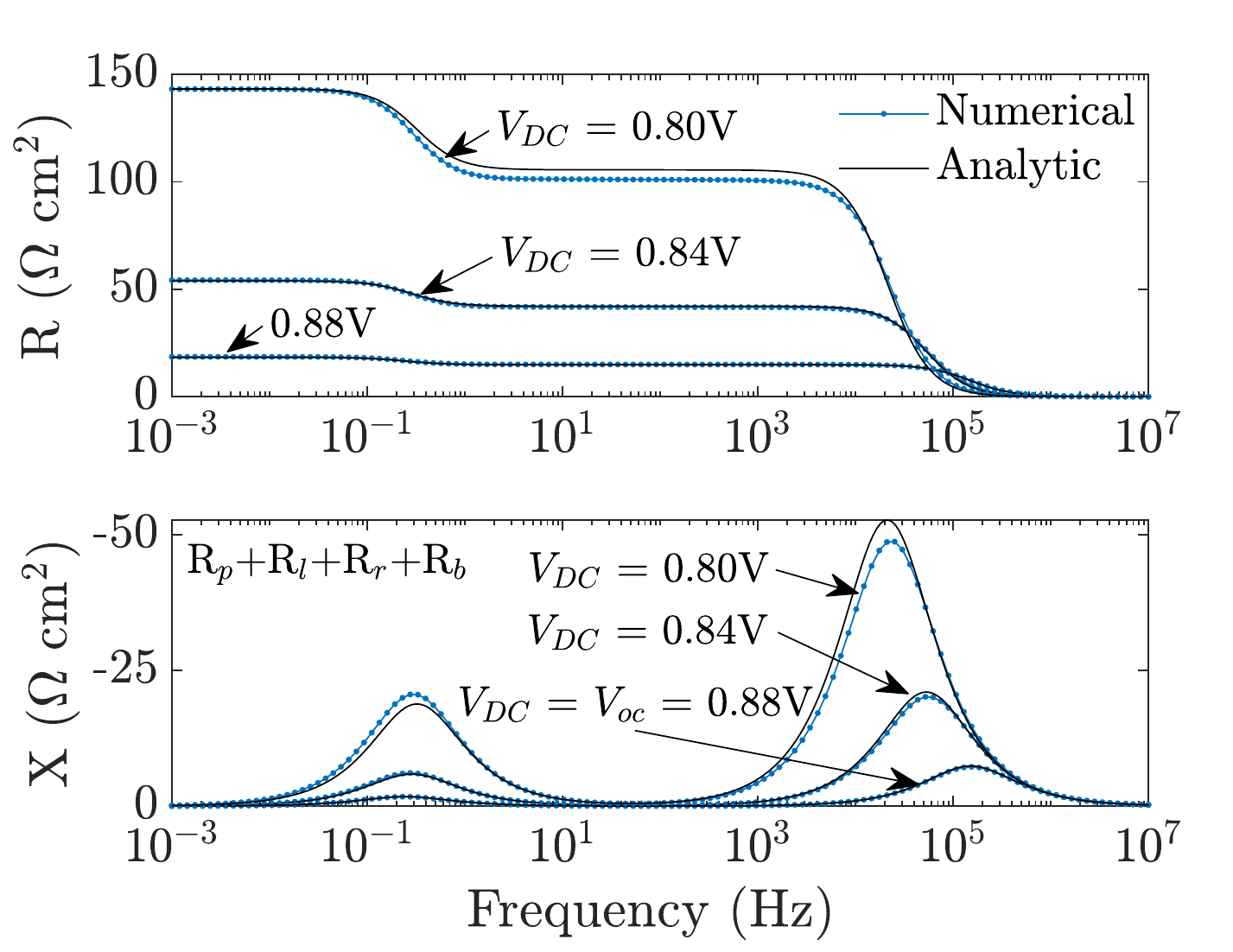}
\centering
\caption{Corresponding frequency plot for the Nyquist presented in Figure \ref{fig:RallNyquist_point1sun}. Simulated impedance spectra for a PSC under 0.1-Sun equivalent illumination with bimolecular ($R_b$) and hole-limited SRH ($R_p$) recombination in the bulk and recombination at both interfaces ($R_l$ and $R_r$).}
\label{fig:RallBode_point1sun}
\end{figure}

\begin{figure}[htbp]
\includegraphics[trim=3.2cm 11cm 3.5cm 10cm, clip, width=0.75\textwidth]{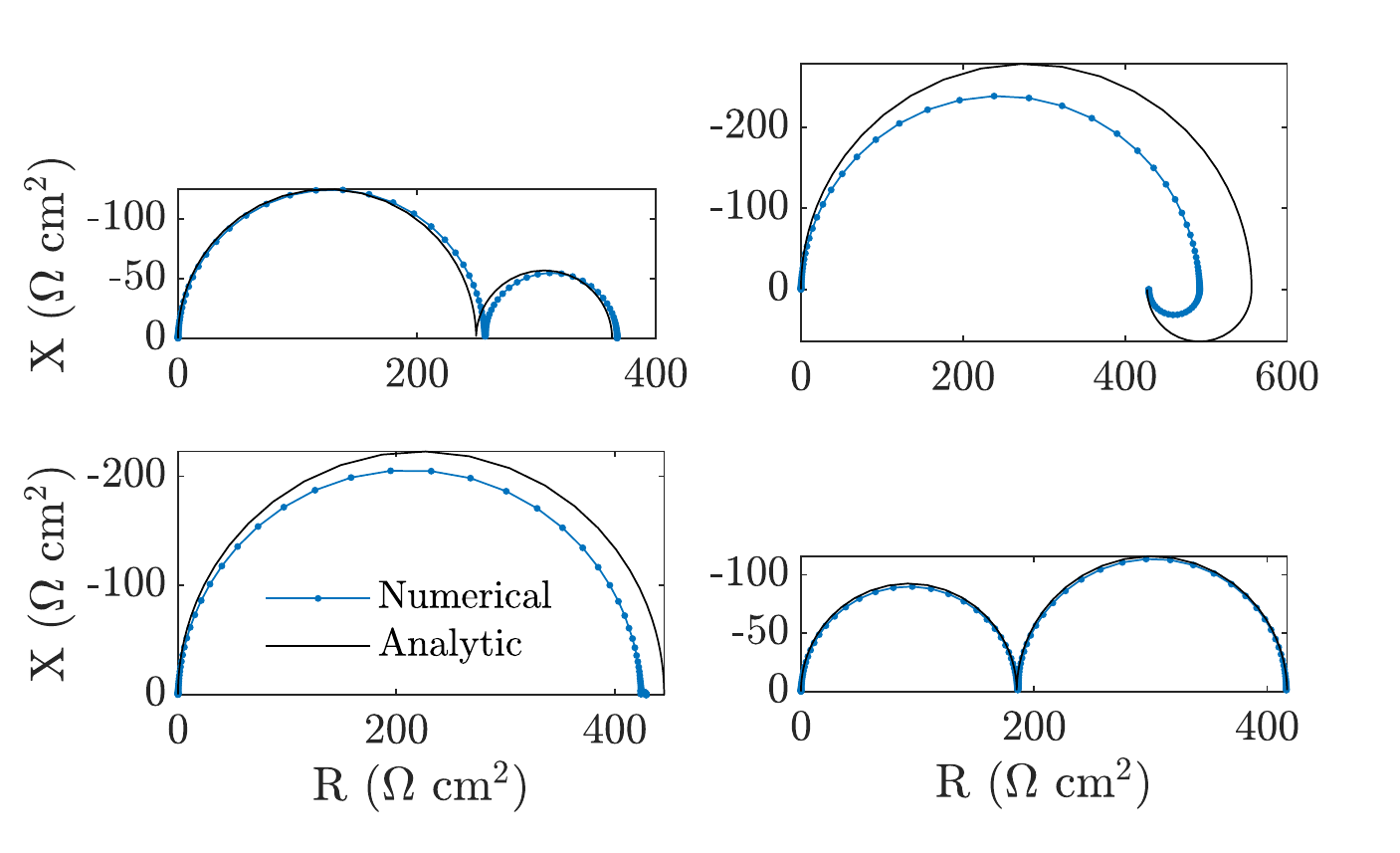}
\centering
\caption{Simulated impedance spectra at maximum power point with different recombination mechanisms. Clockwise from top left: electron-limited bulk SRH ($R_n$: $\VDC$=0.78 V), hole-limited bulk SRH ($R_p$: $\VDC$=0.79 V), perovskite/HTL interfacial ($R_r$: $\VDC$=0.81 V) and bimolecular bulk recombination ($R_b$: $\VDC$=0.86 V). Cell and recombination parameters are listed in Tables \ref{tab:params} and \ref{tab:recomb} respectively. This figure is equivalent to Figure \ref{fig:R_combined_Nyquist_point1sun} in the main paper but at maximum power point rather than at $\Voc$.}
\label{fig:Nyquists_at_MPP}
\end{figure}

Figure \ref{fig:RallBode_point1sun} is the corresponding frequency plot for the Nyquist presented in Figure \ref{fig:RallNyquist_point1sun}. This figure shows simulated impedance spectra for a PSC under 0.1-Sun equivalent illumination with multiple recombination mechanisms present. Specifically, there is bimolecular ($R_b$) and hole-limited SRH ($R_p$) recombination in the bulk and recombination at both interfaces ($R_l$ and $R_r$). Figure \ref{fig:Nyquists_at_MPP} shows four Nyquist plots, each with a different form of recombination, for a simulated PSC operating at maximum power point. This is equivalent to Figure \ref{fig:R_combined_Nyquist_point1sun} in the main text but at maximum power point rather than at $\Voc$.

\newpage 
\subsection{Open-circuit voltage trends} \label{sec:Voctrends}

\begin{figure}[htbp] % trim=left botm right top  \fbox{ whole include graphics bit for box around it to crop}
\includegraphics[trim=2.7cm 8.7cm 2.2cm 9cm, clip, width=0.80\textwidth]{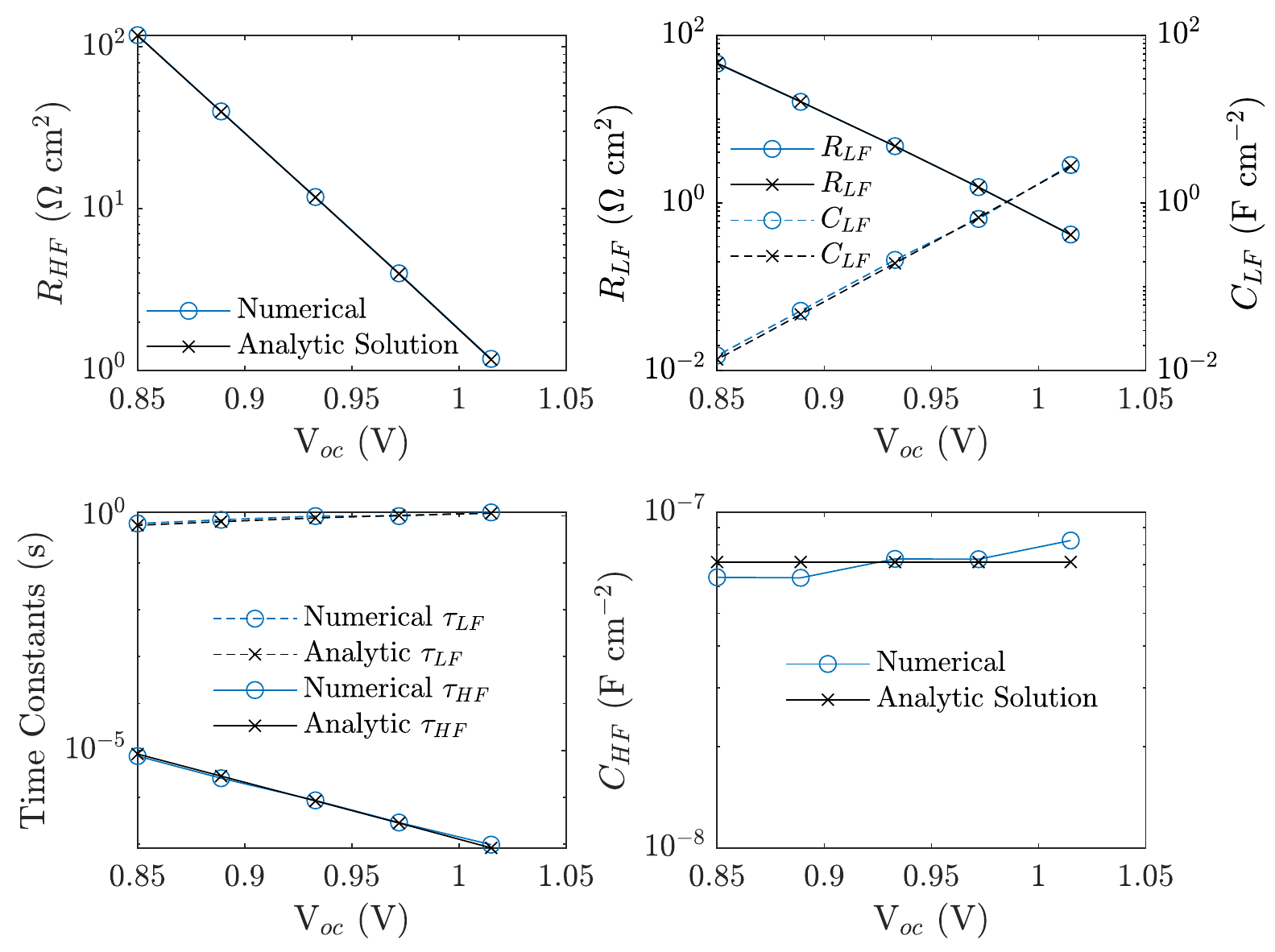}
\centering
\caption{High and low frequency resistances, capacitances and time constants, calculated at different open-circuit voltages from impedance spectra obtained analytically and numerically. Parameters calculated from numerical spectra are indicated by blue (solid or dotted) lines with circle markers and parameters calculated numerically are indicated by black (solid or dotted) lines with crosses. Parameters used to calculate the numerical and analytic spectra are those from Table \ref{tab:params} for a cell with hole-limited interfacial recombination ($R_l$). The time constants $\tau_f = 1/\omega_f$. The apparent ideality factor from the gradient of the HF resistance vs $\Voc$, eq.\eqref{eq:n_ec_from_R_HF_V_oc_trend}, is approximately 1.4. } 
\label{fig:V_oc_trends_combined}
\end{figure}

\begin{figure}[htbp] % trim=left botm right top  \fbox{ whole include graphics bit for box around it to crop}
\includegraphics[trim=4.0cm 10cm 4.5cm 10cm, clip, width=0.55\textwidth]{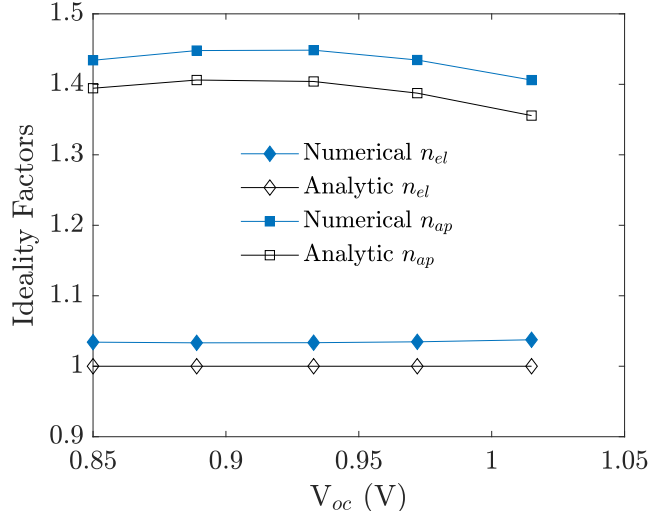}
\centering
\caption{Electronic ideality factor, $\nel$, and apparent ideality factor, $\nap$, calculated at different open-circuit voltages from impedance spectra obtained analytically and numerically. Calculated from the same spectra as used for Figure \ref{fig:V_oc_trends_combined}. Parameters used to calculate the numerical and analytic spectra are those from Table 1 for a cell with hole-limited interfacial recombination ($R_l$).} % Here, the `measured' \red{ectypal factor} approximates the true \red{ectypal factor} as $\Voc \approx \Vbi$.}
\label{fig:n_el_and_n_ec_trends2}
\end{figure}

Here we present the results showing the dependence of the resistances, capacitances, time constants (Figure \ref{fig:V_oc_trends_combined}) and ideality factors (Figure \ref{fig:n_el_and_n_ec_trends2}) on open-circuit voltage. For these simulations hole-limited SRH recombination at the ETL/perovskite interface ($R_l$) was chosen. The resistances and capacitances from the numerical spectra were extracted by fitting to an $RC$-$RC$ equivalent circuit. The analytic resistances and capacitances were calculated using eq.\eqref{eq:RandC_HF} and  eq.\eqref{eq:RandC_LF}. The electronic ideality factors and the apparent ideality factors were calculated from the impedance spectra using  eq.\eqref{eq:n_id} and eq.\eqref{eq:n_ec_from_IS} respectively.

\newpage
\subsection{Potentials and the surface polarisation model} \label{sec:SPMfigs}
 
Here, figures are provided to aid interpretation of the capacitance relation as well as the dependence of the potential drops $V_{1-4}$ on the Debye layer surface charge.

\begin{figure}[htbp] % trim=left botm right top  \fbox{ whole include graphics bit for box around it to crop}
\includegraphics[trim=3cm 9.3cm 3.9cm 9.5cm, clip, width=0.45\textwidth]{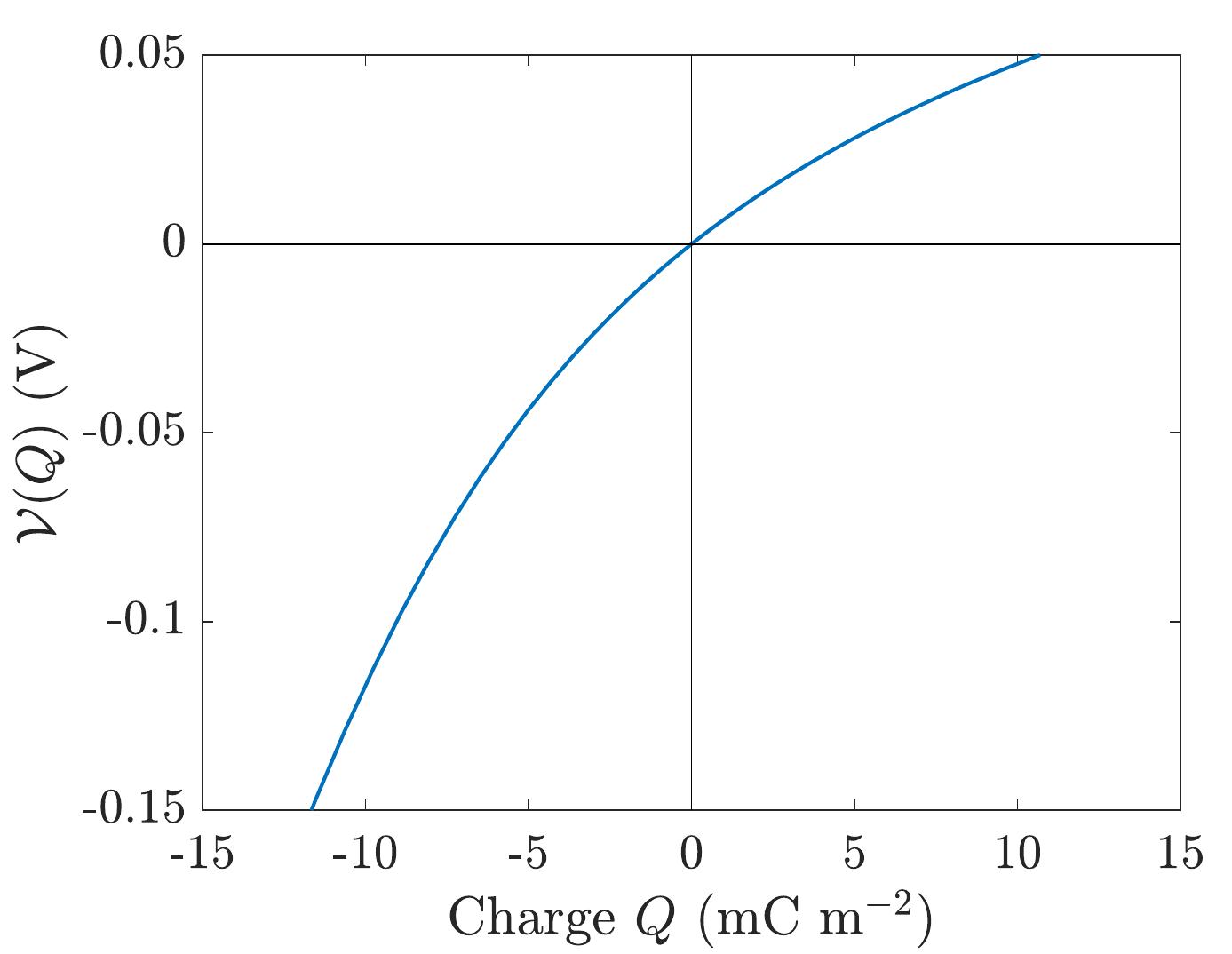}
\centering
\caption{Non-linear capacitance relation from the surface polarisation model \cite{courtier2019transport,courtier2018systematic,richardson2016can} given by equation the inverse of equation \eqref{eq:charge}.}
\label{fig:VofQ}
\end{figure}

\begin{figure}[htbp] % trim=left botm right top  \fbox{ whole include graphics bit for box around it to crop}
\includegraphics[trim=0.2cm 10.6cm 0.5cm 11cm, clip, width=0.7\textwidth]{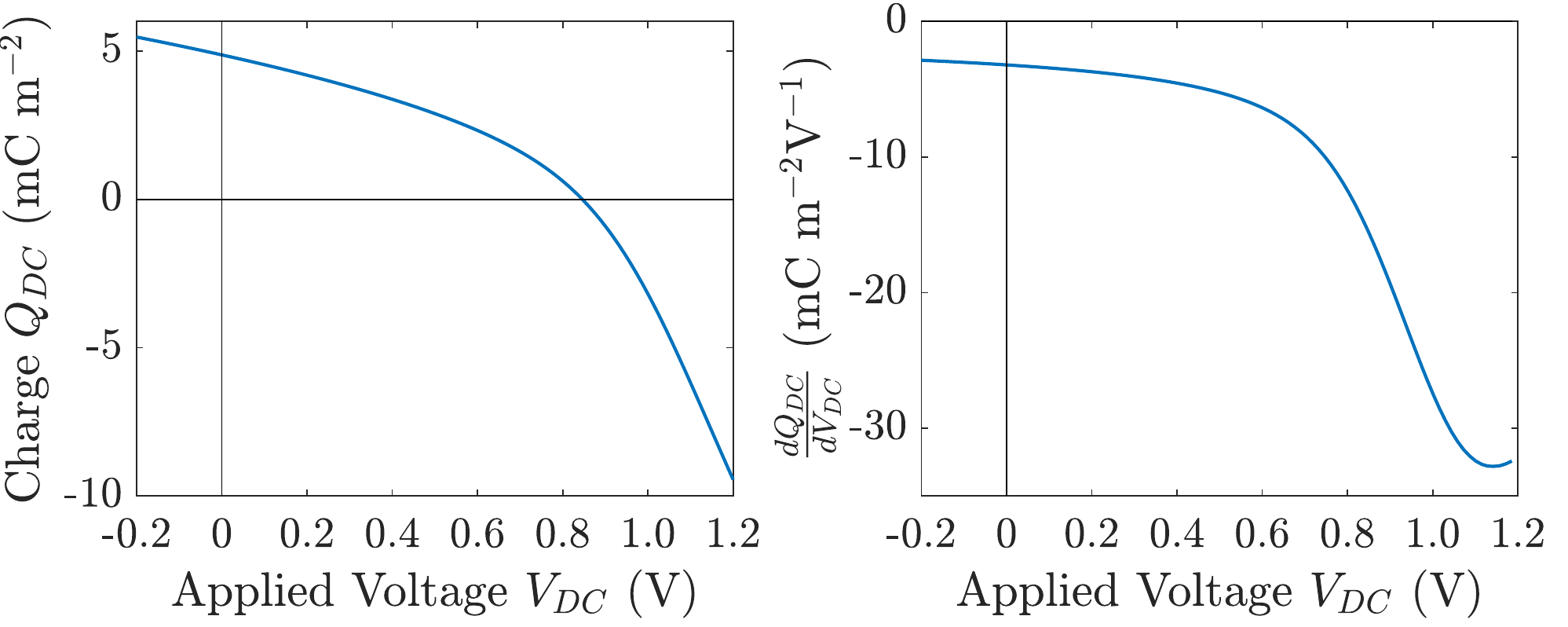}
\centering
\caption{Left: Steady-state charge within the Debye layers for for different applied voltages. Right: Derivative of the steady-state charge within the Debye layers with respect to the DC voltage.}
\label{fig:Q_0_vs_V_0_and_dQbydV}
\end{figure}

\begin{figure}[htbp] % trim=left botm right top  \fbox{ whole include graphics bit for box around it to crop}
\includegraphics[trim=0.1cm 10.6cm 0.5cm 11cm, clip, width=0.7\textwidth]{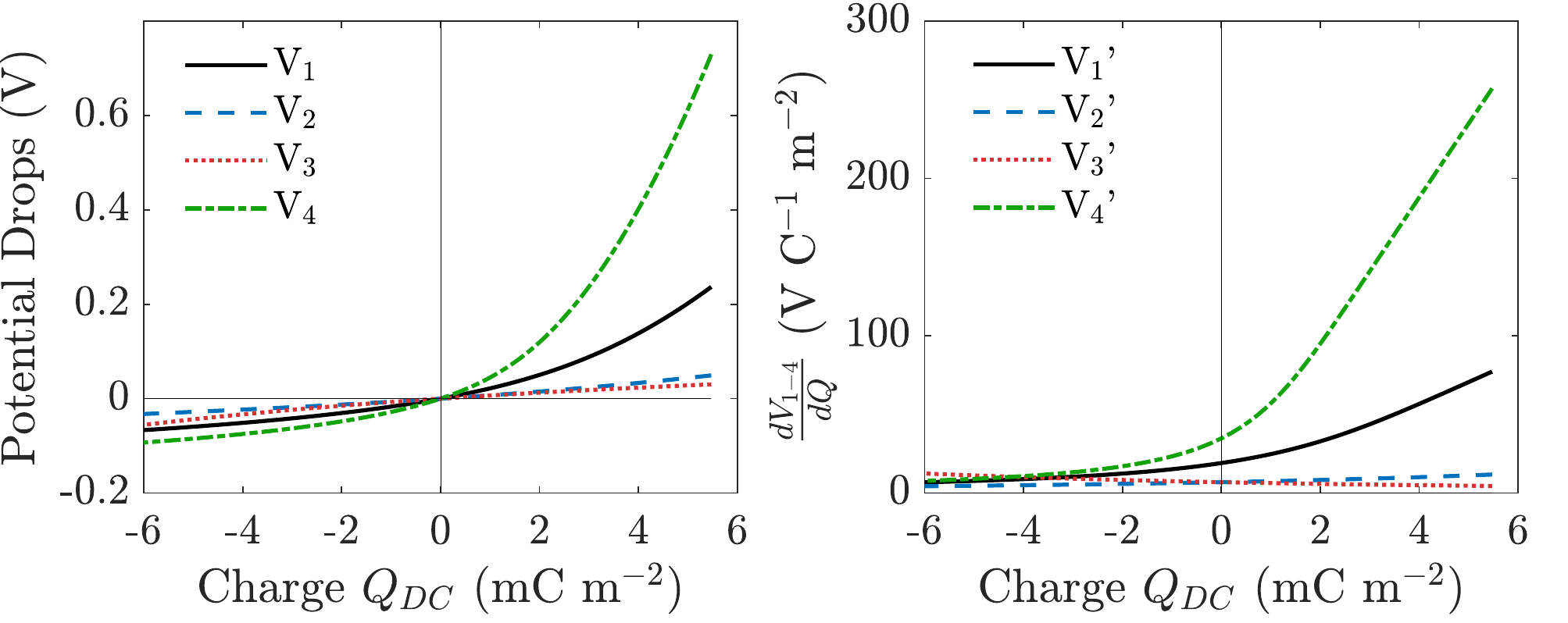}
\centering
\caption{Left: Potential drops across the interfaces of a PSC at steady-state with varying Debye layer charge. See Figure \ref{fig:V1-4} for an illustration of these potential drops across the cell. Right: Gradient of the potential drops across the interfaces of a PSC at steady-state with varying Debye layer charge.}
\label{fig:V_1_4_vs_Q_and_V_1_4_prime_vs_Q}
\end{figure}

%%%%%%%%%%%%%%%%%%%%%%%%%%%%%%%%%%%%%%%%%%%%%%%%%%%%%%%%%
%%%%%%% RECONSTRUCTING IS SPECTRA
%%%%%%%%%%%%%%%%%%%%%%%%%%%%%%%%%%%%%%%%%%%%%%%%%%%%%%%%%

\section{Reconstructing impedance spectra using (\ref{eq:RandC_HF})-(\ref{eq:RandC_LF})} \label{sec:reconstructingIS}
In this section we provide two examples in order to illustrate how the impedance data can be used to extract more information about the physics of the cell once the recombination mechanism has been deduced. In order to obtain  the apparent ideality (or ectypal) factor $n_{ap}$ at applied voltage $\VDC$ we first obtain $\QDC$, the surface charge density stored in the Debye layers at steady state, by solving \eqref{eq:pot_drops} for $\QDC$ as a function of $\VDC$ ({\it n.b.} $V'_{1-4}(\QDC)$ are, for the particular drift-diffusion model that we have chosen to simulate, given by \eqref{eq:VsandOmegas}-\eqref{eq:charge}). The apparent ideality factor is then given, in terms of $\QDC$,  by the formula (see \cite{courtier2020interpreting} for details)
\be
\nap(\QDC) = \frac{F'_T(\QDC)}{F'_i(\QDC)} \label{eq:n_trad_intermsof_F_i}
\ee
where $F_T(Q)$ = $V_1(Q) + V_2(Q) + V_3(Q) + V_4(Q)$ and the prime denotes a derivative with respect to the surface charge density $Q$. This enables (\ref{eq:RandC_HF})-(\ref{eq:RandC_LF}) to be expressed in terms of the potentials $V_{1-4}$ and their derivatives with respect to $Q$ at $\QDC$, $V'_{1-4}(\QDC)$. 

\paragraph{Recombination at the ETL/perovskite interface.} In this first example we consider a cell in which the only significant source of recombination is on the ETL/perovskite interface, $R_l$. From Table \ref{tab:F_i(V)andjR_iforR_i}, $\nel$ = 1 and $F'_i(\QDC) = V_2'(\QDC) + V_3'(\QDC) + V_4'(\QDC)$. Substituting these into equations (\ref{eq:RandC_HF})-(\ref{eq:RandC_LF}) and using \eqref{eq:n_trad_intermsof_F_i}, the analytic relations for the low and high frequency resistances and capacitances are given by
\begin{align}
    R_{LF} &=  \frac{V_T}{\jrec(\VDC)}\left( \frac{ V_{1}'(\QDC)}{V_2'(\QDC) + V_3'(\QDC) + V_4'(\QDC) }\right) \\ %\label{eq:R_LF} \\
    R_{HF} &=  \frac{V_T}{\jrec(\VDC)} \\ %\label{eq:R_HF} \\
    C_{LF} &=  \frac{\jrec  (\VDC)}{G_+V_T V_{1}'(\QDC)} \\
    C_{HF} &= \frac{\varepsilon_p}{b}
\end{align}
This shows how the low frequency feature is dependent on the specific potential drops, {which in turn depend on the steady-state ionic distribution}. Note that the recombination current is also dependent on the recombination mechanism (see Table \ref{tab:recomb}) however this is measurable from experiment using eq.\eqref{eq:j_rec0_from_measurable_currents}. 

\paragraph{Hole limited SRH recombination in the perovskite.}  In the second example we consider  a cell in which the only significant source of recombination is hole-limited SRH recombination in the perovskite, $R_p$. Table \ref{tab:F_i(V)andjR_iforR_i} define $\nel$ = 2 and $F'_i(\QDC) = V_3'(\QDC) + V_4'(\QDC)$. As such, the analytic relations for the low and high frequency resistances and capacitances are
\begin{align}
    R_{LF} &=  \frac{V_T}{\jrec(\VDC)}\left( \frac{ V_{1}'(\QDC) + V_{2}'(\QDC) - V_{3}'(\QDC) - V_{4}'(\QDC)}{V_3'(\QDC) + V_4'(\QDC) }\right) \\ %\label{eq:R_LF} \\ 
    R_{HF} &=  2\frac{V_T}{\jrec(\VDC)} \\ %\label{eq:R_HF} \\
    C_{LF} &=  \frac{\jrec  (\VDC)}{2G_+V_T \left( V_{1}'(\QDC) + V_{2}'(\QDC) - V_{3}'(\QDC) - V_{4}'(\QDC) \right)} \\
    C_{HF} &= \frac{\varepsilon_p}{b}
\end{align}

Both examples also illustrate how impedance measurements could be used to obtain the capacitance relations, used to determine the potentials $V_{1-4}$, from experiment in cases where the dominant recombination mechanism is known. This is a potentially valuable tool because, while the ionic model used in this case (equations \eqref{eq:VsandOmegas} and \eqref{eq:charge}) is consistent with solutions to the drift-diffusion model used to simulate the cell, it is possible that there are physical mechanisms that are not modelled by the drift-diffusion model that we use here but which still play a significant role in the operation of a real PSC. For example, mobile cation vacancies or degenerate statistics of the charge carriers in the transport layers may play an important role in the  behaviour of certain cells.

% --------------------------------------- Ectypal --------------------------------------- %

\clearpage
\newpage

\section{Additional information}  \label{sec:n_ec_and_n_id_fromIS}

\setcounter{equation}{0}
\setcounter{figure}{0}

\paragraph{Apparent ideality factor from HF Resistance Impedance Spectra.}
% \subsection{Obtaining the \red{ectypal factor} from HF resistance at different open-circuit voltages} 

Plotting $\ln (R_{HF})$ vs $\Voc$ has been shown to produce a straight line with a gradient proportional to an `ideality factor' \cite{pockett2017microseconds,riquelme2020identification}. Here we derive this result from eq.(\ref{eq:RandC_HF}) for $R_{HF}$ and the recombination current eq.(\ref{eq:j_rec_general}) to determine the correct interpretation for this ideality factor. Taking the log of the high frequency resistance at $\Voc$ one obtains
\be
\ln(R_{HF}(\Voc)) = \ln(V_T\nel) - \ln(\jrec(\Voc)).
\ee
Using the definition of the ectypal factor from \cite{courtier2020interpreting} (eq.\eqref{eq:ectypal_approx_def}), the recombination current at open-circuit can be written
\be
\jrec(\Voc) = j_{R_i}\exp\left(\frac{\Voc-\Vbi}{V_T\nec(\Voc)}\right),
\ee
where $\nap = \nec$. Therefore,
\be
\ln(R_{HF}(\Voc)) = \ln(V_T\nel) - \frac{\Voc - \Vbi}{V_T\nec(\Voc)} -\ln(j_{R_i}),
\ee
which can be rewritten
\be
\ln(R_{HF}(\Voc)) = -\frac{1}{V_T\nec}\Voc + C, \hspace{4mm} \textnormal{where} \hspace{2mm} C = \ln\left(\frac{V_T\nel}{j_{R_i}}\right) + \frac{\Vbi}{V_T\nec(\Voc)} \label{eq:n_ec_from_R_HF_V_oc_trend}
\ee
This shows that by plotting the log of the high frequency resistance vs open-circuit voltage, one obtains the ectypal factor from the gradient. A similar argument can be made using the fact that $R_{HF}$ is proportional to the derivative of the recombination current with respect to voltage. Unlike other measurement techniques, such as Sun-$\Voc$ and dark-$JV$ methods, this returns the value of the true ectypal factor, and not the measured ectypal factor. See \cite{courtier2020interpreting} for this distinction.

\paragraph{JV curves and impedance.}

\begin{figure}[htbp] % trim=left botm right top  \fbox{ whole include graphics bit for box around it to crop} 
\includegraphics[trim=0cm 0.1cm 0cm 0cm, clip, width=0.95\textwidth]{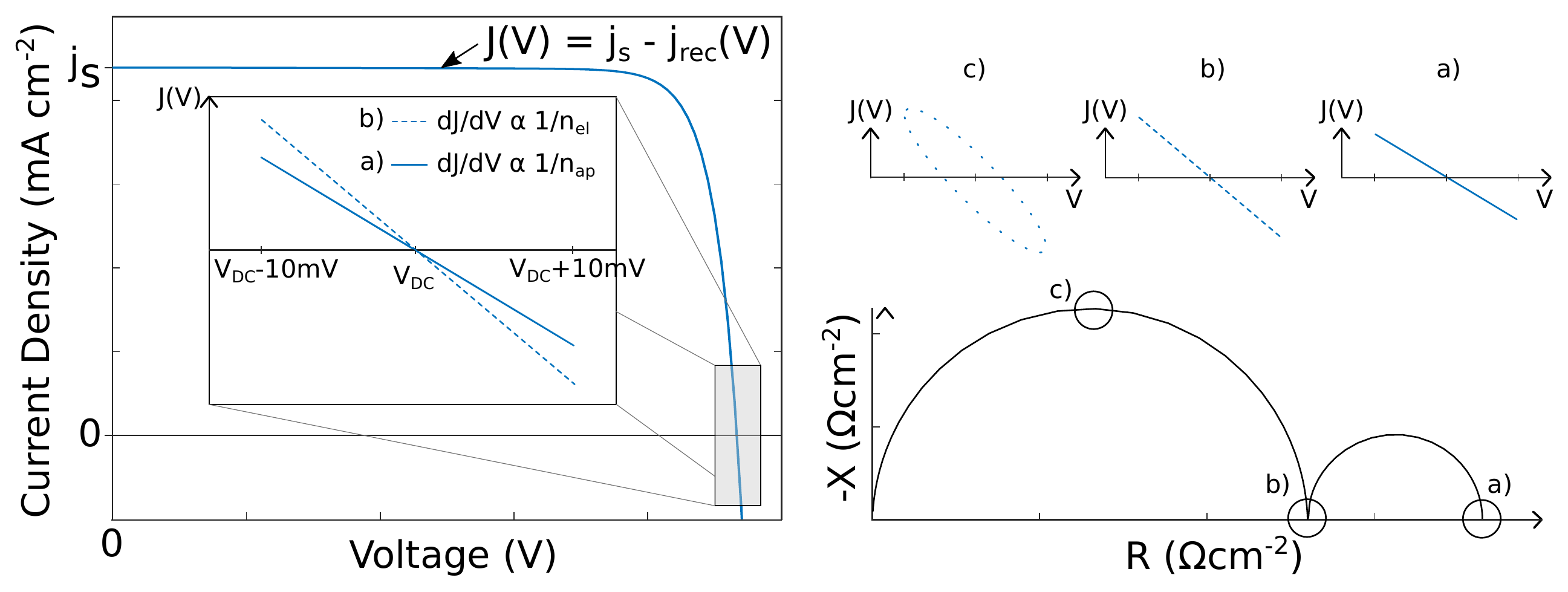}
\centering
\caption{Left: steady-state $JV$ curve with a Lissajous plot (J(t) versus V(t)) inset showing the gradient, as measured by impedance spectroscopy. Right: Lissajous plots illustrating the current response at points a), b) and c) indicated on the Nyquist plot. The gradients shown in a) and b) are proportional to the measured ectypal factor (or equivalently the apparent ideality factor) and the electronic ideality factor respectively. For simplicity, a series resistance has not been included.}
\label{fig:JV_and_impedance}
\end{figure}

Figure \ref{fig:JV_and_impedance} shows the relationship between time-dependent voltage-current responses and their positions on a Nyquist diagram. Additionally, the relationships between ideality factors and the gradient of the $JV$ curve is shown.  

\newpage 
\paragraph{The low frequency feature and recombination}
Here we provide the more general form of Table \ref{tab:negRandC_and_no_LF}, which does not assume that the applied voltage is close to the built-in voltage.

\begin{table}[htbp]
\begin{center}
\begin{tabular}{|c|c|c|c|}  
 \hline
  \textbf{Recombination} & \textbf{Can there be no LF feature?} &\textbf{ Can }$\boldsymbol{R_{LF}}$, $\boldsymbol{C_{LF}}$ \textbf{be negative?}  \\
 \hline 
  $R_b$: $R_{\textnormal{bulk}} = \beta np$ & Yes, always as $\nap = \nel$ &  No ($R_{LF}=0, C_{LF}=\infty$) \\ 
 \hline
  $R_p$: $R_{\textnormal{bulk}} = p/\tau_p$  & Yes if $V_1' + V_2' = V_3' + V_4'$ &  Yes if $ V_1' + V_2' < V_3' + V_4'$ \\ 
   \hline
  $R_n$:  $R_{\textnormal{bulk}} =  n/\tau_n$ & Yes if $ V_1' + V_2' = V_3' + V_4'$  &  Yes if $ V_3' + V_4' < V_1' + V_2'$ \\ 
   \hline
  $R_l = v_{p_E}\pl$ & Yes if $V_1' << V_2' + V_3' + V_4'$ & No ($R_{LF} \geq 0, C_{LF}>0$) \\ 
   \hline
  $R_r = v_{n_H}\nr$ & Yes if $V_4' << V_1' + V_2' + V_3'$ &  No ($R_{LF} \geq 0, C_{LF}>0$)\\ 
  \hline
\end{tabular}
\end{center}
\caption{A table to show the conditions required for no low frequency feature, or for it to be negative, for each recombination type considered in this study. These conditions are derived using the inequality defined by equation (\ref{eq:negRandCcondition}), the definition of the apparent ideality factor (ectypal factor) from \cite{tress2018interpretation} and the result that $V_{1-4}'(\QDC)>0$.} 
\label{tab:negRandC_and_no_LF_gen}
\end{table}

\end{document}